\documentclass[10pt]{article}
\usepackage{acl2015}
\usepackage{times}
\usepackage{url}
\usepackage{makecell}
\usepackage{latexsym}
\usepackage{physics}
\usepackage{amssymb}
\usepackage{tikz}
\usepackage{comment}
\usepackage[font=small,labelfont=bf]{caption}
\usepackage{caption}
\usepackage{subcaption}
\usepackage[hidelinks]{hyperref}
\usepackage{graphicx}
\usepackage[noend]{algpseudocode}
\usepackage{algorithm}
\usepackage{algpseudocode}
\usepackage{bbold}
\usepackage{changepage}
\usepackage[normalem]{ulem}
\usepackage[titletoc,toc,title]{appendix}
\usepackage{lipsum}
\usepackage{chngcntr}
\usepackage{afterpage}
\usepackage{tocloft}

\usetikzlibrary{quantikz2}
\usepackage[sorting = none, backend = bibtex, style=numeric-comp]{biblatex}
\newcommand{\braketmatrix}[3]{\left \langle #1 \middle| #2 \middle| #3 \right \rangle}
\algrenewcommand\algorithmicrequire{\textbf{Input:}}
\algrenewcommand\algorithmicensure{\textbf{Output:}}

\title{Compact Multi-Threshold Quantum Information Driven Ansatz For Strongly Interactive Lattice Spin Models}

\author{Fabio Tarocco$^{1}$, Davide Materia$^{1,2}$, Leonardo Ratini$^{1,2}$ and Leonardo Guidoni$^{2,*}$\\$^1$ Dipartimento di Ingegneria e Scienze dell'Informazione e Matematica,\\ Università degli Studi dell'Aquila, L'Aquila, Italy\\
$^2$Dipartimento di Scienze Fisiche e Chimiche,\\ Università degli Studi dell’Aquila,
L’Aquila, Italy\\
$^*$leonardo.guidoni@univaq.it}

\begin{document}

\maketitle
\begin{abstract}
Quantum algorithms based on the variational principle have found applications in diverse areas with a huge flexibility. But as the circuit size increases the variational landscapes become flattened, causing the so-called Barren plateau phenomena. This will lead to an increased difficulty in the optimization phase, due to the reduction of the cost function parameters gradient. One of the possible solutions is to employ shallower circuits or adaptive ans\"atze.

We introduce a systematic procedure for ansatz building based on approximate Quantum Mutual Information (QMI) with improvement on each layer based on the previous Quantum Information Driven Ansatz (QIDA) approach.
Our approach generates a layered-structured ansatz, where each layer's qubit pairs are selected based on their QMI values, resulting in more efficient state preparation and optimization routines. We benchmarked our approach on various configurations of the Heisenberg model Hamiltonian, demonstrating significant improvements in the accuracy of the ground state energy calculations compared to traditional heuristic ansatz methods. Our results show that the Multi-QIDA method reduces the computational complexity while maintaining high precision, making it a promising tool for quantum simulations in lattice spin models.

\end{abstract}

\section*{Introduction}
Quantum computers and algorithms represent a transformative leap in computational capabilities, with the potential to revolutionize various scientific disciplines \cite{Shor_1997,Banuls_2020,Hussain_2020}. Built on the principles of quantum mechanics, these technologies enable information manipulation and processing in ways fundamentally distinct from classical computers \cite{Arute2019-mb, Preskill_2018}, promising to drastically reduce the computational cost of specific tasks compared to their classical counterparts\cite{Feynman1982}. 
This breakthrough has generated wide public interest\cite{Seskir_2022, 2021_Bayer, bova_2021}, spanning applications from cryptography\cite{Shor_1997} to finance\cite{Egger_2020}, and particularly gaining traction in the academic realm of the natural sciences \cite{von_Burg_2021,Robert_2021} and applications to lattice models \cite{Kokail2019, bosse2021probinggroundstateproperties, Kattem_lle_2022, Cade2020}.
Simulations of quantum systems \cite{Barison_2021, Seskir_2022, Brown_2010} stand out as a particularly promising application of quantum algorithms, especially sparkling considerable enthusiasm due to its potential applications in tackling in the realm of quantum chemistry \cite{Whitfield_2011, Bauer_2020, Cao_2019, McArdle_2020}, aiming to address the electronic structure of large molecules where electron correlation plays a crucial role. 

A key method for near-term applications is the Variational Quantum Eigensolver (VQE), an hybrid quantum-classical algorithm designed to address challenges in quantum chemistry \cite{Peruzzo2014, McClean_2016, Kandala_2017, Cerezo_2021, fedorov2021vqe, Bharti_2022, Tilly_2022} aiming to determine the ground state energy of molecular systems. Its methodology involves employing a parameterized quantum ansatz, representing a trial wave function, and iteratively adjusting parameters to minimize energy through measurements on a quantum computer. This hybrid approach facilitates the exploration of the intricate quantum landscape inherent in molecular systems.

Despite its promise, VQE faces challenges in scaling to larger systems due to the expansion of the parameter space and the difficulties to implement long circuits describing the variational wavefunctions and operators on noisy quantum devices. To overcome these hurdles, various algorithms and strategies have been developed to enhance the performance of VQE
\cite{Barkoutsos_2018, Grimsley_2019,Yordanov_2021,benfenati2021improved}. Additionally, various approaches to constructing an ansatz have been developed, falling into two major classes. The first class involves translating classical Quantum Chemistry methods into the language of quantum computation. An example is the Unitary Coupled Cluster (UCC) method, which provides a unitary implementation of the classical Coupled Cluster method \cite{Hoffmann_1988,Cooper_2010,Evangelista_2011,romero2018strategies, Magoulas2023}. For strongly correlated lattice model like the Heisenberg Hamiltonian, specific ansatze have been developed to efficiently tackle the unique challenges presented by these systems. An example, is the Hamiltonian Variational Ansatz (HVA) which leverages the compact structure of lattice model Hamiltonians, using fermionic operators to exploit low Pauli weight encodings \cite{WeiHo2019,Wiersema2020,Cade2020}. This approach often results in shallower circuits compared to other ansatze like the Unitary Coupled Cluster Singles and Doubles (UCCSD).
In contrast, the second approach begins with wavefunctions directly constructed to leverage the characteristics of quantum hardware. This empirical approach, known as the Heuristic Ansatz\cite{Kandala_2017, Ganzhorn_2019,rattew2020domainagnostic, Tkachenko_2021, Tang_2021}, comprises repetitions of blocks composed of 1-qubit rotations and 2-qubits entangling gates. It is designed without relying on information about the physical system, focusing solely on exploiting the quantum hardware's capabilities, in fact they are also known as Hardware Efficient Ansatz (HEA). While the Heuristic Ansatz better utilizes quantum hardware, it comes at the cost of losing the physical meaning associated with the variational ansatz. Opting for the latter approach rather than the former allows for the consideration of shallower circuits, providing a potential avenue to address scalability concerns. 

Independently from the criterion according to which the ans\"atze is built, as the parameterized quantum circuits (PQC) start to get increasingly complex, the length and depth of the circuits increase too, not only enhancing the probability of accumulating errors, but also leading to the phenomenon of Barren Plateau \cite{larocca2024review} i.e. the exponentially flat parameter optimization landscape. While modification in the optimization methods or additional error mitigation are not strategies to avoid barren plateaus, shallow and variable structured ans\"atzes are efficient in the reduction of the phenomenon. Examples of these solutions are ADAPT-VQE ans\"atze \cite{Grimsley_2019}, which grows iteratively the complexity of the guess wavefunction one operator at time following information obtained by the simulated molecule. Exploiting a combined iterative optimization of the Hamiltonian and the wavefunction, the WAHTOR algorithm shows the capability of reducing the depth of the circuits and reducing the risk of falling into local minima while reaching a lower variational energy \cite{ratini_2022}.  Similarly to the adaptive approach, the Quantum Mutual Information Ansatz (QIDA) \cite{materia2023quantum} provides a quick and fast way to generate an initial guess with few CNOTs, following the correlation encoded in the molecular system, which can be used as an effective state preparation to guide the optimization in the right basin. 

The aim this work is to extend further the idea behind the QIDA approach building and optimizing ans\"atze layer by layer inspired by low-level preliminary classical calculations carried out on the system. The main focus of the QIDA method was to utilize Quantum Mutual Information (QMI) to define shallow-depth circuits. QMI, traditionally used in quantum chemistry to measure the correlation between orbitals \cite{Ding_2020}, has found applications in quantum computing such as optimizing qubit ordering \cite{Tkachenko_2021} and enhancing quantum algorithms \cite{Zhang_2021}. The circuits obtained by the QIDA approach represent compact and affordable initial starting states from which more precise and complex ans\"atze can be built.
Following the QIDA approach, our state-preparation method aims to recover the missing correlation by exploiting a layered-structured ans\"atz, where each layer is selected based on the QMI values. Multi-Threshold Quantum Information Driven Ansatz (Multi-QIDA) extends the single threshold QIDA approach by including qubit-pairs that still present mid-high QMI values, that are excluded by the original method. Combined with a repeated iterative VQE routine, Multi-QIDA allows us to obtain effective shallow circuits that surpass the generic ladder-fashion ansatz in terms of performance and convergence properties. In addition, we explore the usage of a different correlator, namely an $SO(4)$ 2-qubit gate, which has several advantages over the traditional CNOT.

We first begin with a brief introduction to concepts and methods behind our Multi-QIDA approach in Section  \ref{sec:theory}: Variational Quantum Eigensolver in Section \ref{ssec:vqe}, mutual-information measure in Section  \ref{ssec:mi_meas}, Tensor Networks, Matrix Product States, and the Density Matrix Renormalization Group in Section  \ref{ssec:tn_mps}. We then presented the system that has been studied in this work, the Heisenberg Model Hamiltonian, in Section  \ref{ssec:hamiltonians}. In Section \ref{ssec:so4},  the \textbf{SO}(4) gates used as an alternative to the standard combination of CNOTs and parametrized rotations are defined. 
The Quantum Information Driven Ansatz approach, is defined in Section  \ref{ssec:qida}, whereas Multi-QIDA, is introduced in Section \ref{sec:multi_qida}. The procedure to convert quantum mutual-information matrices into a sequence of entangling gates is defined in Section \ref{ssec:layer_builder}. Then, the construction of the parametrized quantum circuit representing the variational form introduced by our approach is explained in Section \ref{ssec:ansatz_comp}. The iterative optimization procedure combined with the multi-layer ansatz is defined in Section \ref{ssec:opt_routine}.
Computational details are presented in Section \ref{sec:comp_details}. In particular, the specific systems that have been tested are presented in Section \ref{ssec:tested_systems}, the metrics used to analyze the performance of our method compared to standard HEA in Section  \ref{ssec:metrics}, and for the simulation settings and details an explanation can be found in Section  \ref{ssec:sim_detail}. Finally, in Section \ref{sec:res} results for a specific system configuration are presented, with a particular focus on the whole workflow. Other systems are then collected in the Section \ref{ssec:complete_results}.
\section{Background}
\label{sec:theory}
In the following sections, all the principles and methods used in our work are briefly explained.
\subsection{Variational Quantum Eigensolver (VQE)}
\label{ssec:vqe}
The variational principle, used in the Variational Quantum Eigensolver (VQE) algorithm \cite{Peruzzo2014,McClean_2016}, states that the energy $E$ obtained by exploiting a parametrized wavefunction $\Psi({\theta})$ for a given system Hamiltonian $H$ is strictly lower bounded by the ground state energy $E_0$ of the system H:
\begin{equation}
    \label{eqn:var_principle}
    E_0 \leq\braketmatrix{\Psi(\theta)}{H}{\Psi(\theta)} = E.
\end{equation}
The VQE algorithm exploits the variational principle to variationally approximate the groundstate of the problem Hamiltonian by an iterative process of energy measurement of the parametrized quantum circuit, i.e. the parametrized wavefunction.
The VQE algorithm is a hybrid quantum-classical algorithm: the Quantum Processing Unit is responsible for the state preparation, the execution of the quantum circuit that generates the guess wavefunction $\Psi(\theta)$, and the measurement of the energy.
The classical CPU then composes the full energy measurement and feeds an optimizer which returns a newly computed set of parameters that will lower the energy of the guess wavefunction in the following iterations.

The routine starts with a random set of parameters as CPU input. The Quantum Processing Unit (QPU) is responsible for the problem-specific state preparation, for creating a parametrized quantum circuit (PQC) that must be executed on the device and the actual execution of the PQC. Once the execution is completed, a measurement is performed giving a sequence of bits. The output bit string is then passed to the classical processing unit.
The classical CPU then composes the full energy measurement and feeds an optimized which returns a newly computed set of parameters that will minimize the energy of the guess wavefunction in the following iterations. 
This routine is repeated until some optimization criteria are met.
\subsection{Quantum Mutual Information}
\label{ssec:mi_meas}
The Von-Neumann Quantum Mutual Information (QMI)~\cite{vonneumann1996} is one of the metrics that can be used to measure the correlation between two constituents of a quantum system.

Considering a system of $N$ elements, each one living in its own Hilbert space $\mathcal{H}_i$, for $i \in \{1,\dots,N\}$, the state of the system, which is the total composition of each subsystem combined, denoted with $\ket{\Psi}$ belongs to the composed Hilbert space, $\mathcal{H}_{tot}$, obtained by the tensor product of the single elements Hilbert space 
\begin{equation}
    \ket{\Psi} \in \mathcal{H}_{tot} = \bigotimes_{i=1}^{N}{\mathcal{H}_{i}}.
\end{equation}
Associated to a quantum state $\ket{\Psi}$, we can define its \textit{density matrix} (or \textit{density operator)}
\begin{equation}
    \rho = \ketbra{\Psi}.
\end{equation}
By selecting only a subset of $k$ elements, we can obtain the \textit{reduced density matrix} (RMD). The RMD of the $k$-qubits can be obtained from the density operator $\rho$ by tracing out the indices that are not the selected ones.
If $k \subset \mathbf{B}$, where $\mathbf{B}=\{1,\dots,N\}$ is the set of all the indices, then the RDM, $\rho_k$, is obtained by tracing out qubits with indices $i \in (\mathbf{B} - k)$
\begin{equation}
    \rho_k = \Tr_{_{\bar{k}}}\rho,
\end{equation}
where $\bar{k} = (\mathbf{B} - k)$.
Relatively to qubits, we define one-qubit and two-qubit RDMs as:
\begin{equation}
    \rho_i=\Tr_{_{\{\mathbf{B} - i\}}}{\rho},
\end{equation}
\begin{equation}
    \rho_{ij}=\Tr_{_{\{\mathbf{B} - (i,j)\}}}{\rho},
\end{equation}
where $i,j \in \{1\dots N\}$ and $\rho = \ketbra{\Psi}{\Psi}$.
To compute the QMI matrix, we need to compute the \textit{Von Neumann entropy}, $S$, for all the one-qubit and two-qubit RDM.
The entropy $S$ is defined as:
\begin{equation}
\label{eqn:vnentopy}
    S(\rho) = -\Tr(\rho \log{\rho}).
\end{equation}
Following Equation (\ref{eqn:vnentopy}), we can define:
\begin{equation}
    S_{i} = S(\rho_i) = -\Tr(\rho_i\log\rho_i) 
\end{equation}
and
\begin{equation}
    S_{ij} = S(\rho_{ij}) = -\Tr(\rho_{ij}\log\rho_{ij}),
\end{equation}
where $\forall i,j \in \{1,\dots,N\}$.
The QMI between qubits $i$ and $j$ is then obtained as:
\begin{equation}
\label{eqn:QMI}
\begin{split}
    I_{ij}&=(S_i + S_j - S_{ij})(1 - \delta_{ij})\\&=(S(\rho_i)+S(\rho_j)-S(\rho_{ij}))(1-\delta_{ij})
\end{split}
\end{equation}
where $i,j \in \{1,\dots,N\}$.
The QMI can be visualized with \textit{Quantum Mutual Information Maps} which are symmetric matrices with zero-valued entries on the diagonal by definition, shown in Figure \ref{appendix:fig_QMIs}. 

QMI found applications as a quantity to describe the electronic structure of a molecular system \cite{Ding_2020}. This measure does not solely quantify the entanglement a system includes, but it takes in account both quantum and classical correlation.
We can consider the values of the $I$ matrix as a measure of the \textit{total} correlation between subsystems of the total space, e.g. qubits or spin sites as used in this work. 

\subsection{Tensor Networks}
\label{ssec:tn_mps}
In this work, the QMI matrices are computed by exploiting \textit{Tensor Networks} (TN).
Tensor Networks are a tool used to represent variational wavefunction. In particular, they are used to approximate the wavefunction of a quantum system by providing a more manageable decomposition of the large tensor that represents the full wavefunction into smaller tensors.

A TN, as the name suggests, is a collection of correlated tensors. 
As TN wavefunction, we decided to work with Matrix Product States (MPS), in which the wavefunction is decomposed in a linear-fashion graph. 
An MPS quantum state wavefunction, $\ket{\Psi}$, composed by $N$ sites can be written as 
\begin{equation}
    \ket{\Psi} = \sum_{a_1\dots a_{N-1}}
A^{[1]}_{\sigma_1, a_1}A^{[2]}_{\sigma_2,a_1,a_2}\dots A^{[N]}_{\sigma_N,a_{N-1}}
\end{equation}
Each tensor $A^{[k]}$ has a \textit{physical} index, $\sigma_k$, which allows to iter over the $D$ \textit{degrees of freedom} of each element of the system and at most two \textit{virtual}, $a_k$, indices that connect one tensor to the neighbors. 
The level of \textit{expressibility} of an MPS wavefunction is fixed by a tunable parameter $\chi$, called \textit{bond dimension}, that corresponds to the maximum dimension allowed for the virtual indices.
The bond dimension sets an upper bound to the amount of the entanglement achievable by the approximated wavefunction, thus, if the bond dimension is large enough, the exact wavefunction can be recovered.

For the computation of the reference state from which the QMI matrices are created, we exploited the tensor network version of the density matrix renormalization group (DMRG)\cite{White1992, White93, Schollwock2005, SCHOLLWOCK2011}. TN-DMRG is a variational algorithm for optimizing an initial wavefunction written as MPS, such that after optimization the MPS is approximately the dominant eigenvector, which corresponds to the ground-state, of a problem Hamiltonian $H$ where the degree of approximation of the solution can be tuned by the $\chi$ parameter.
The problem matrix $H$ is usually assumed to be a Hermitian matrix written as a sum of local terms or in Matrix Product Operator (MPO) form, which is the tensor network version of an operator.
In physics and chemistry applications \cite{Baiardi_2020}, DMRG is mainly used to find ground states of Hamiltonians of many-body quantum systems.  
The overall complexity of the DMRG algorithm is $\mathcal{O}(ND\chi^3)$, where $N$ is the number of sites, $D$ is the degree of freedom of each site, and $\chi$ the maximum bond dimension imposed, if the bond dimension of the MPO is larger than the MPS bond dimension, the MPO one will be used to defined as $\chi$ in the above expression.

After running the DMRG algorithm, the converged wavefunction $T_{\sigma_1,\dots,\sigma_N}$, i.e. the best approximation of the ground-state at a fixed bond dimension, is obtained.
To compute the terms used in the QMI Equation \ref{eqn:QMI}, the RDMs for each couple of sites are required. The RDM can be seen as a partial trace.
 Given the matrix $A^{i'j'}_{i,j}$, which can also represent a density operator, the partial trace is obtained by summing over the indices that we want to trace out.
\begin{equation}
    \Tr_i(A^{i'j'}_{ij}) = \sum_{i} A^{ij'}_{ij} = A^{j'}_j
    \label{eqn:trace}
\end{equation}

To compute $\rho_{ij}$ which is define as a rank-4 tensor $\rho^{\sigma_i'\sigma_j'}_{\sigma_i\sigma_j}$, we need to trace out all the sites $\sigma_s \neq \{\sigma_i, \sigma_j\}$, which is equivalent to compute \begin{equation}
    \rho^{\sigma_i'\sigma_j'}_{\sigma_i\sigma_j}=\Tr_{\{\sigma_s \neq \sigma_i, \sigma_j\}}{\ketbra{\Psi}{\Psi}}
\end{equation}
As shown in Equation \ref{eqn:trace}, to trace out some sites, we need to sum over those indices, for the two-sites RDM, it is required to run over all the indices except $i$ and $j$. If the MPS representing the quantum state is defined as $\ket{\Psi} = T_{\sigma_1 \dots \sigma_N}$, its density operator is
\begin{equation}
    \rho = \ketbra{\Psi}{\Psi} = T_{\sigma_1 \dots \sigma_N}T^{\dagger}_{\sigma_1' \dots \sigma_N'} = \rho^{\sigma_1' \dots \sigma_N'}_{\sigma_1\dots \sigma_N}
\end{equation}
Now it is easy to see how to compute $\rho_{ij}$ starting from $\rho$ for each pair $(i,j)$:
\begin{equation}
\begin{split}
    \rho_{ij} &= \sum_{\sigma_s, s\neq i,j}{T_{\sigma_1 \dots \sigma_i\dots \sigma_j\dots\sigma_N}T^{\dagger}_{\sigma_1 \dots \sigma_i'\dots \sigma_j'\dots\sigma_N}}
    \\&=\rho_{\sigma_i\sigma_j}^{\sigma_i'\sigma_j'}.
\end{split}
\label{eqn:rho_ij}
\end{equation}
Finally, from each $\rho_{i,j}$ we can compute also the one-site RDMs $\rho_i$ and $\rho_j$ by tracing out the other index of two indices that resulted from the previous contraction (i.e. $j$ and $i$ respectively), completing all the necessary terms needed to compute the QMI value as defined in Equation \ref{eqn:QMI}.

\subsection{Heisenberg Model}
\label{ssec:hamiltonians}
A general Heisenberg model Hamiltonian is defined as follows:
\begin{equation}
    H = \sum_{\langle i,j\rangle}\mathbf{J}_{i,j}\mathbf{S}_i\cdot\mathbf{S}_j + \mathbf{h}_i\sum_i^N\mathbf{S}_i,
\end{equation}
where $\langle i,j\rangle$ indicates all nearest-neighbor spins pairs, $\mathbf{J}_{ij}$ is the interaction term between two adjacent spin sites, $\mathbf{S}_i$ and $\mathbf{S}_j$ are spins operators acting on sites $i$ and $j$, respectively, and $\mathbf{h}_i$ is the local magnetic field acting on the $i$-th site and this can be an external magnetic field or magnetic impurities.
Expanding each spin operator $\mathbf{S}_i$ into the three components along $x$, $y$ and $z$ axis, the Hamiltonian $H$ can be written as
\begin{equation}
\begin{split}
    H=&\sum_{\langle i,j\rangle}({J_{ij}^{xx}S_{i}^{x}S_{j}^{x}+J_{ij}^{yy}S_{i}^{y}S_{j}^{y}+J_{ij}^{zz}S_{i}^{z}S_{j}^{z}})\\
&+ h\sum_{i}^{N}{S_{i}^{z}},
\end{split}
\label{eqn:pauli_H_complete}
\end{equation}
where $\langle i,j\rangle$ indicates all nearest-neighbor spins in a 2D lattice with open boundary conditions, $J_{ij}^{\beta\beta}$ is the interaction term between two adjacent spin sites with $\beta \in \{x,y,z\}$, $h$ is the external magnetic field and $S_{i}^{\beta}$ is the $\beta$ component of the $\frac{1}{2}$-Spin operator acting on the $i$-th spin site.

The Hamiltonian is written directly on the qubit space by converting each $\frac{1}{2}$-Spin operator into Pauli operator following 
\begin{equation}
    S_{i}^{\beta} = \frac{\hbar}{2} \sigma_{i}^{\beta},
\end{equation}
where $\sigma_i^\beta \in \{\sigma^{x},\sigma^{y},\sigma^{z}\}$ i.e one of the three Pauli matrices.
The Hamiltonian to be measured on a quantum computer must be converted into a sum of Pauli strings. A Pauli string is defined as a tensor product of Pauli operators.
Each Pauli string corresponds to one of the coupling terms or the on site term in  Equation \ref{eqn:pauli_H_complete}.
For a general coupling term,  $J_{i,j}^{\beta\beta}$, the relative Pauli string is defined as
\begin{equation}
    I_0 I_1 \dots I_{i-1}\sigma^{\beta}_i I_{i+1}\dots I_{j-1}\sigma^{\beta}_j I_{j+1}\dots I_{N-1},
\end{equation} where $I_i$ are identity matrices.

\subsection{SO(4) Gates}
\label{ssec:so4}
Our approach has been also tested with a more general 2-qubit correlator, namely a fully parametrized two-qubit gates from \textbf{SO}(4).

At variance with CNOTs-based ansatz, the parametrized \textbf{SO}(4) gates offer a tunable correlation, that can be chosen by the optimization. 
With the right parametrization, the group of \textbf{SO}(4) transformations can also lead to the identity, and symmetry preserving gate, e.g. the fermionic swap Eqn.\ref{eqn:fswap}, which allows to imposing anti-symmetrization to the ansatz.
\begin{equation}
    fSWAP = \begin{pmatrix}
        1 & 0 & 0 & 0\\
        0 & 0 & 1 & 0 \\
        0 & 1 & 0 & 0 \\
        0 & 0 & 0 & -1 \\
    \end{pmatrix}
    \label{eqn:fswap}
\end{equation}

On top of the variety of intermediate transformations that can be obtained by parametrized \textbf{SO}(4) matrices, the decision of this type of gates is also related to the nature of the system that we decided to treat. Given the time reversibility of the Heisenberg model Hamiltonians thus we only need to include real transformations described by matrices in the \textbf{O}(4) group and \textbf{SO}(4).
In the context of \textbf{O}(4) and \textbf{SO}(4), the matrices consist of real numbers, and the operations involved in this group are based on real arithmetic, both can represent rotations in four-dimensional space.

Following Theorem 3 in \cite{Vatan_2004}, every two-qubit quantum gates in \textbf{SO}(4) can be realized by a circuit consisting of 12 elementary one-qubit gates and two CNOT gates.

A generic gate $U\in $\textbf{ SO}(4) is composed by two generic one-qubit rotations $A,B \in $ \textbf{SU}(2), four $S$ gates and two $R$ gates.
It is also known that every matrix $M \in$ \textbf{ SU}(2) can be written as a composition of $R_z(\alpha)R_y(\theta)R_z(\beta)$ for some $\alpha,\beta$ and $\theta$, while the $R$ gate is defined as $R_y(\pi/2)$ and the $S$ gate is obtained with $R_z(\pi/2)$. The $U$ gate is then parametrized using the two sets of three parameters of gates $A$ and $B$.
\begin{figure}[!htb]
    \centering
    \includegraphics[scale=0.3]{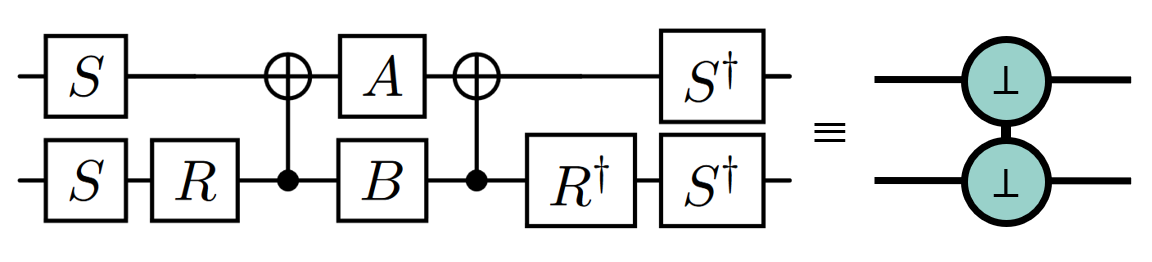}
    \caption{On the left, the circuit implementing a general $U \in$ \textbf{SO}(4). The gates A and B are general \textbf{SU}(2) parametrized gates. On the right, is the symbol we adopted in this work.}
    \label{fig:so4}
\end{figure}
\subsection{Quantum Information Driven Ansatz (QIDA)}
\label{ssec:qida}
Quantum Information Drive Ansatz (QIDA) \cite{materia2023quantum} is a heuristic ansatz approach used to design PQC with a topology that reflects the correlations that are contained in the quantum systems.
After the calculation of the QMI matrix $I$, coupling pairs are identified by fixing a threshold $\mu \in [0,1)$, splitting the qubits pairs into two separate sets:
\begin{equation}
\begin{cases}
(i,j)\text{ is selected}&  \text{if }I_{ij}\geq\mu,\\
(i,j) \text{ is discarded}& \text{if }I_{ij}<\mu.
    
\end{cases}
\end{equation}
The \textit{selected} qubit pairs are used to design the empirical ansatz.
In particular, the entangling block of the VQE is composed of two-qubit gates between each qubit couple $i,j \in selected$, i.e. the coupling with  QMI value greater or equal to the threshold chosen $I_{ij} > \mu$. In ref.\cite{materia2023quantum}, the entangling gate used, called \textit{correlator}, is the CNOT gate. Any other two-qubit quantum gate can be used to correlate the two qubits.

QIDA follows the idea that the tunable threshold, $\mu$, can allow provide the desired amount of correlation required to approximate the quantum state connecting only the spots which are more relevant for the approximate QMI reference. The algorithm does not necessarily connect all the qubits. 
Empirical ladder-fashion ansatz links all the qubits giving the possibility to span the whole Hilbert space at the price of having a larger entangling gate count, which will result in much more expensive quantum circuits.
Also, ladder-fashion ansatz does not carry any information about the physical problem that is going to be studied. On the contrary, With QIDA, the topology of the correlation in the quantum systems is obtained by a classical ground-state pre-computation from which the QMI map is defined.
\begin{figure}
    \centering
    \begin{quantikz}[column sep=1pt, row sep={14pt,between origins}, font=\tiny]
    &\gate{R_{x}(\theta_{0})}&\ctrl{1}&&&&&\gate{R_{x}(\theta_{6})}&\ctrl{1}&&&&&\gate{R_{x}(\theta_{11})}&\\
    &\gate{R_{x}(\theta_{1})}&\targ{-1}&\ctrl{1}&&&&\gate{R_{x}(\theta_{7})}&\targ{-1}&\ctrl{1}&&&&\gate{R_{x}(\theta_{12})}&\\
    &\gate{R_{x}(\theta_{2})}&&\targ{-1}&\ctrl{1}&&&\gate{R_{x}(\theta_{8})}&&\targ{-1}&\ctrl{1}&&&\gate{R_{x}(\theta_{13})}&\\
    &\gate{R_{x}(\theta_{3})}&&&\targ{{-1}}&\ctrl{1}&&\gate{R_{x}(\theta_{8})}&&&\targ{-1}&\ctrl{1}&&\gate{R_{x}(\theta_{14})}&\\
    &\gate{R_{x}(\theta_{4})}&&&&\targ{{-1}}&\ctrl{1}&\gate{R_{x}(\theta_{9})}&&&&\targ{-1}&\ctrl{1}&\gate{R_{x}(\theta_{15})}&\\
    &\gate{R_{x}(\theta_{5})}&&&&&\targ{{-1}}&\gate{R_{x}(\theta_{10})}&&&&&\targ{-1}&\gate{R_{x}(\theta_{16})}&\\
    \end{quantikz}
    \caption{Generic depth 2 heuristic ansatz with entangling map in a ladder configuration. }
    \label{fig:Ladder_heur_d_2}
\end{figure}
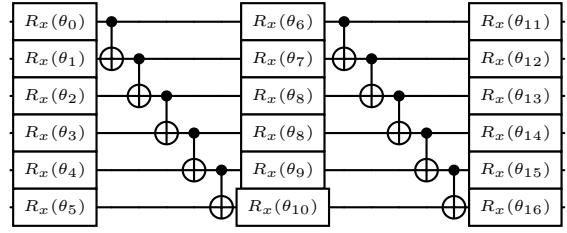
\section{Multi-QIDA}
\label{sec:multi_qida}
\begin{figure*}[!htb]
    \centering
    \includegraphics[scale=0.4]{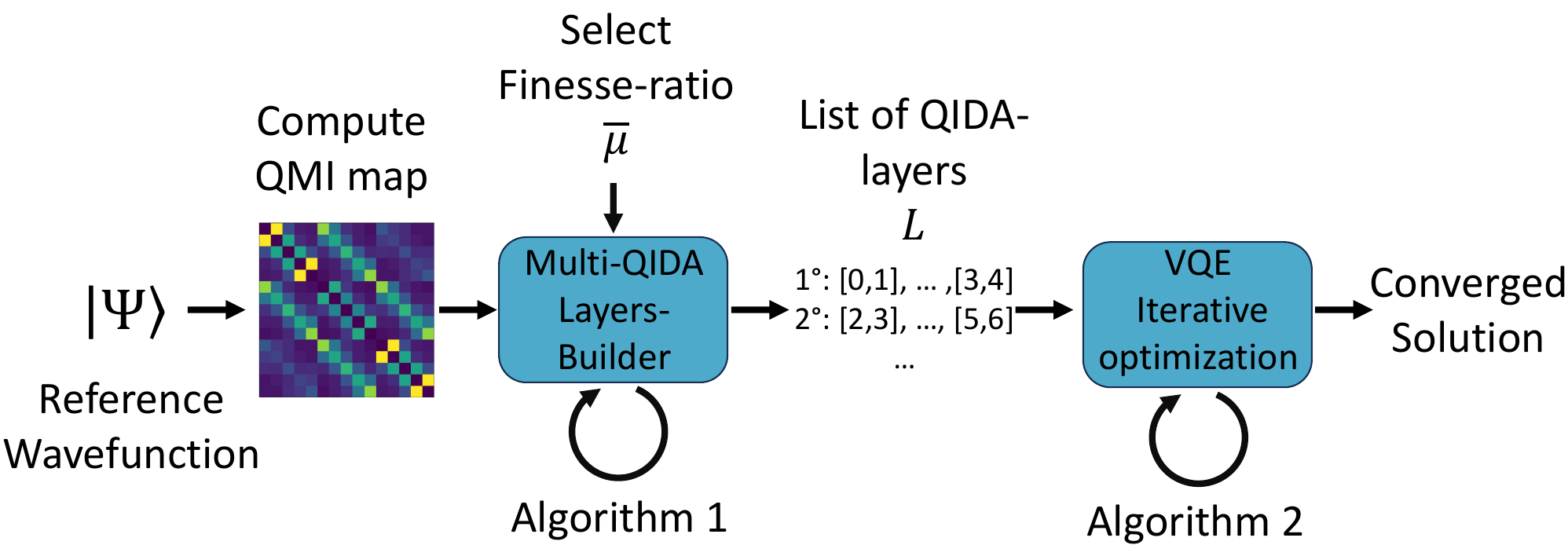}
    \caption{Pictorial representation of the workflows of the proposed Multi-QIDA algorithm. The reference wavefunction is used to build the QMI map on which is applied the Multi-QIDA Layers-Builder procedure as shown in Algorithm \ref{algo:sel_pseudocode}. The resulting collection of entangler maps is used to perform an Iterative-VQE, as defined in Algorithm \ref{algo:opt_layer}.}
    \label{fig:workflow}
\end{figure*}

The main focus of the Multi-QIDA approach is the building of a systematic ansatz adding, step by step, the missing correlation that is not caught using the first QIDA layer  \cite{materia2023quantum} as a standalone ansatz.
The implementation of Multi-Threshold Quantum Information Driven Anstaz (Multi-QIDA) is combination of three steps, which are described in the following sections.
The initial step is the \textit{Layers-Builder} procedure, a purely classical routine which, given an approximate QMI matrix, infers the adaptive layers of the VQE by sectioning the MI matrix by correlation levels, explained by in Section \ref{ssec:layer_builder} and with Algorithm \ref{algo:sel_pseudocode}.
Second, the \textit{Circuit-composition}, the circuit made of the different entangler maps for each layer is compiled, with details varying based on the entangler block chosen, $CNOT$ or \textbf{SO}(4).
Finally, we perform an \textit{iterative layered-VQE}, a varied version of the Layer-VQE~\cite{Liu_2022}, which converges to the full ansatz and defined by Algorithm \ref{algo:opt_layer} in Section \ref{ssec:opt_routine}. A general workflow of our Multi-QIDA approach is shown in Figure \ref{fig:workflow}.

\subsection{Layers-Builder}
\label{ssec:layer_builder}
To improve the QIDA method, we iteratively exploit QMI maps to obtain other additional layers. 
The selection is performed on the list of qubits-pairs ordered for decreasing QMI value from the QMI matrix $I_{i,j}$. Using a sequence of thresholds, $\bar\mu$, chunks of qubits-pairs that a layer must contain can be selected. This collection of thresholds $\bar\mu$ called \textit{finesse-ratio}, represents how the full QMI spectrum is split into different circuit layers.

\begin{algorithm}[!htb]
\caption{Schematic outline of the Multi-QIDA Layers-builder}
\begin{algorithmic}[1]
\Require QMI matrix $I_{ij}$, $\bar\mu$, $N_{qubits}>0$
\Ensure List of entangling map $L$
\State $L \gets$ empty list
\State $free \gets \{n | \forall n \in N_{qubits}\}$
\State $G \gets Disconnected\_Graph(free)$
\State $t\gets 1$
\State $all\_connected = false$
\While{$(free \neq \varnothing)\wedge \neg(all\_connected)$}
\State $l \gets$ empty list
\For{$q_i,q_j \in \{\forall I_{i,j}: \bar\mu[m]>I_{i,j}\geq \bar\mu[m+1]\}$}
\If{$\neg(G.connected(q_i,q_j))$}
\State $l.add([q_i,q_j])$
\State $G.add\_edge(q_i,q_j)$
\State $free.remove(q_i,q_j)$
\EndIf
\EndFor
\State $L.append(l)$
\State $t \gets t-\mu$
\State $all\_connected \gets G.is\_connected()$
\EndWhile
\State $L.append(\text{ladder\_ansatz}(N_{qubit}))$
\State \textbf{return} $L$

\end{algorithmic}
\label{algo:sel_pseudocode}
\end{algorithm}
Initially, all the qubits are in the set of \textit{free qubits}, meaning they have not been used in any entangling process. 
A QIDA entangling layer $l$-th is then composed by selecting the pairs $q_i, q_j$ whose correlation I$_{q_i, q_j}$ falls in the range $\bar\mu[l-1] > I_{q_i,q_j} \geq \bar\mu[l]$. 
Such pairs are then added to the candidate entangling list of the $l$-th layer, which represents the potential pairs to which then impose a correlator. The criterion according to which the entangling gates are added to the current layer $l$ is defined as follows: If the two qubits $q_0$ and $q_1$ are not reachable to each other through direct connection or exploiting previous layers qubit-pairs, they are selected. Thus, they are added to the circuit if they belong to distinct subsets of qubits. As a consequence, if at least one of the qubits was \textit{free} up to the current layer $l$, then this is clearly automatically satisfied, and the qubit is removed from the \textit{free} set.
This procedure is iterated again until the set of free qubits is empty and all the qubits belong to the same unique subset, i.e. they are all connected. 
In other terms, the layers are added until all the qubits in the lattice have been covered by at least one entangling layer and they are reachable to each other by exploiting a combination of correlators used amongst all the layers. 
After the last layer is defined with this procedure, and or more additional layers of entanglers in the ladder setup closes the Multi-QIDA ansatz..

At the end of this building procedure, we define the entire collection of layers by $L$, which is then passed to the ansatz construction procedure.

\subsection{Circuit ansatz composition}
\label{ssec:ansatz_comp}Until now, we only took care of retrieving how many correlators to impose on the quantum circuit and where to impose them, now, starting from the list of entangling maps $L$, we construct the ansatz in two ways. The first implementation follows the standard combination of parametrized single-qubit rotation gates and $CNOT$ as entanglers \cite{Kandala_2017}, and it has been included in the analysis as a comparison to the next type of ansatz. In the following Multi-Qida construction, this version of the Ansatz will be referred to as $QIDA^{CX}$. The second variant is obtained by entangling pairs of qubits using \textbf{SO}(4) gates, as in Fig.~\ref{fig:so4}, which, as we have seen, guarantees the most general description of real two-qubit gates, such as Multi-QIDA implementation is referred in the following as $QIDA^{SO4}$.

As already introduced in the first section, the phenomenon on the barren plateau is related to the \textit{circuit expressiveness}: the larger is the variational space that the ansatz has to explore, the greater is the probability, when initialized according to a random distribution of the parameters, to lead to initial states that are far from the right solutions. One of the possible solutions to avoid Barren Plateau is to employ short PQC \cite{larocca2024review}.
By variationally optimizing only a portion of the search space with the restricted parameters, in some scenarios, shallow circuits present no flattened potential surfaces, with the drawback that the optimal solution could also not be obtained inside this constrained space. 
Following this concept, similarly to the Adapt-VQE \cite{Grimsley_2019}, we decided to extend the shallow ansatz obtained by the QIDA method, adding layer-by-layer expressiveness to the PQC, without exploring the huge variational space at once. At each $l\neq1$, we want the new layer L$_l$ to be initialized to the identity to not disrupt all the previous $l-1$ rounds of optimizations once a new (yet) unoptimized layer is introduced.

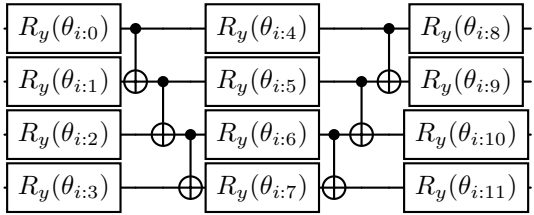
\begin{figure}[!htb]
    \centering
    \begin{quantikz}[column sep=1pt, row sep={20pt,between origins}]
    & \gate{R_y(\theta_{i:0})} & \ctrl{1} & \ghost{X}&\ghost{X}&\gate{R_y(\theta_{i:4})}&\ghost{X}&\ghost{X}&\ctrl{1}&\gate{R_y(\theta_{i:8})}&\\
    & \gate{R_y(\theta_{i:1})} & \targ{} &\ctrl{1}&\ghost{X}&\gate{R_y(\theta_{i:5})}&\ghost{X}&\ctrl{1}&\targ{}&\gate{R_y(\theta_{i:9})}&\\
    & \gate{R_y(\theta_{i:2})} & \ghost{X} &\targ{}&\ctrl{1}&\gate{R_y(\theta_{i:6})}&\ctrl{1}&\targ{}&\ghost{X}&\gate{R_y(\theta_{i:10})}&\\
    &\gate{R_y(\theta_{i:3})} & \ghost{X} &\ghost{X}&\targ{}&\gate{R_y(\theta_{i:7})}&\targ{}&\ghost{X}&\ghost{X}&\gate{R_y(\theta_{i:11})}&
    \end{quantikz}
    \caption{Circuit implementing a $V$-shape $i$-th additional layer with entangling map $[0,1]$,$[1,2]$,$[2,3]$. This circuit results into an identity, $\mathbb{1}$, when all the parameters $\theta_{i:j}$ are zero for all $j$.}
    \label{fig:V-shape}
\end{figure}

Specifically, for the combination of $R_y$ and CNOT, we can obtain the identity by imposing a $V$-shape structure to the entangling layer $l$, partly resembling the topology of a UCC excitation \cite{Hoffmann_1988,Cooper_2010,Evangelista_2011, Magoulas2023}. The $V$-shape is obtained by a series of $R_y$ rotations, the additional entangling layer, a central layer of $R_y$ rotations, again the additional entangling layer but with reversed-order, completed with a last series of $R_y$ rotations, as shown in Figure \ref{fig:V-shape}.

\begin{figure}[!htb]
    \centering
    \includegraphics[scale=0.45]{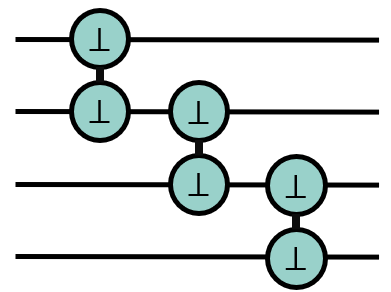}
    \caption{Additional layer with entangling map $[0,1]$,$[1,2]$,$[2,3]$ where each entangler is an  \textbf{SO}(4) 2-qubit gate as describer in Figure\ref{fig:so4}. This layer is initialized to identity, $\mathbb{1}$, at the beginning, with all the parameters of each entangler set to zero.}
    \label{fig:SO4-add_lh}
\end{figure}

If the additional layer is built with \textbf{SO}(4), only \textbf{SO}(4) gates are used in the newly added circuit, and since the identity gate is already included in the parameterization of a general \textbf{SO}(4), shown in Figure~\ref{fig:so4}, no further construction is needed to obtain it in this case. 
In summary, with both kinds of entangling maps (in Figure~\ref{fig:SO4-add_lh} and Figure~\ref{fig:V-shape}), setting the parameters to zero guarantees an identity gate.

The initialization of the parameters of the additional layers with zeros, or in general with a circuit that results in an identity, ensures that the cost function being optimized remains unchanged after introducing this new layer. The parameter optimization can therefore proceed step by step, avoiding the risk of barrel plateau due to large expressivity. As described in the next chapter this optimization can be also done gradually following a procedure that leads always to an improvement.
This structure is applied to all the entangling layers contained in $L$, except for the first one, as in that case there is yet no circuit to disrupt. Thus the first layer does not need to include an identity in its parametrization and as such is defined as a standard heuristic ansatz with random initial parameters.

\subsection{Iterative layered-VQE}
\label{ssec:opt_routine}
The selection of a suitable optimization routine is necessary when dealing with variational circuits in which the ansatz is built iteratively/adaptively and not optimizing directly the full wavefunction. As the name of this section suggests, we decided to perform the optimization of the full circuit iteratively along many steps, each of which will include two main phases: an optimization of the single layer $L_l$ at the $l$-th step, and a relaxation procedure in which a global optimization of all the previous layers $\{ 0 \dots, l-1 \}$ is performed. We remind the reader that each layer $L_l$ can be composed of two different kinds of entangler, as explained in Section \ref{ssec:ansatz_comp}, however, the procedure explained here is independent of this variable. Furthermore, as was specified before, any layer but the first one ($l>0$) needs to be initialized to the identity gate, which implies that the parameters of both $QIDA^{CX}$ and $QIDA^{SO4}$ ans\"atze are initialized to zero, for the first layer ($l=0$), the $M$ parameters are instead initialized randomly in the interval $[0,2\pi)$.

The starting parameters, denoted as $\bar{\theta}_l$, represent the unoptimized parameters associated with the $l$ layer. With the vector of parameters $\bar{\theta}_l$, it is possible to define the unitary transformation to bring $\ket{\Psi_l}$ to the $l+1$-th state $\ket{\Psi_{l+1}}= U(\bar{\theta}_l)\ket{\Psi_l}$.  
To optimize the $\bar{\theta}_{l}$ parameters we apply the VQE algorithm
by minimizing $\braketmatrix{\Psi_l|U^\dagger(\bar{\theta}_l)}{H}{U(\bar{\theta}_l)|\Psi_l}$. We mark the optimal parameter vector by $\bar{\theta}^*_l$. 

Now we may unfold the description of the total state $\ket{\Psi_l}$ by listing all the previously optimized layers with $\ket{\Psi_{l+1}}= U(\bar{\theta}^*_l)\ket{\Psi_{l}} = \prod_{j=0}^{j=l}U(\bar{\theta}^*_j)\ket{\Psi_{0}}$. After the completion of optimization of the layer, $L_{l}$, the algorithm proceeds with a new VQE. In this second phase of the iteration $l$ the algorithm encompasses all the layers $\{ 0 \dots, l \}$ in a general VQE that minimizes $\braketmatrix{\Psi_0|\prod_{j=l}^{j=0}U^\dagger(\bar{\theta}_j)}{H}{\prod_{j=0}^{j=l}U(\bar{\theta}_j)|\Psi_0}$.
 This optimization takes as a starting point the previous parameter vectors and gives as a result a new set of different parameter vectors. 

\begin{algorithm}[!htb]
\caption{Iterative (Re)-Optimization routine}
\begin{algorithmic}[1]
\Require $N_{qubits} > 0$, List of entangling map $L$
\Ensure Optimal parameters $\bar{\theta}^*$, Converged energy $E$
\State $QC_{empty} \gets $QuantumCircuit($N_{qubits}$)
\For{$l \in L$}
\State append($QC_{empty},l)$
\State $M \gets \text{n\_params}(QC_{empty})$
\If{$l$ is first layer}
    \State $\bar{\theta}_0 \gets \theta_{0:0},\dots,\theta_{0:M}\in_R[0,2\pi)$
    \State $\bar{\theta}_{l}^* \gets$VQE$(QC_{empty},\bar{\theta}_0)$  
\Else
    \State $\bar{\theta}_{l} \gets \theta_{l:0},\dots,\theta_{l:M}=0$
    \State $\bar{\theta}_{l}^* \gets $VQE$(QC_{prev}, \bar{\theta}_{l})$
    \State $\bar{\theta}_{tot} \gets \bar{\theta}_{prev}^* + \bar{\theta}_{l}^*$
    \State $E_{tot},\bar{\theta}_{tot}^* \gets $VQE$(QC_{empty}, \bar{\theta}_l)$
\EndIf
\State $\bar{\theta}_{prev}^* \gets \bar{\theta}_{tot}^*$
\State $QC_{prev} \gets $assign$(QC_{empty},\bar{\theta}_{prev}^*)$
\EndFor
\Return $E_{tot}, \bar{\theta}_{tot}^*$
\end{algorithmic}
\label{algo:opt_layer}
\end{algorithm}

We notice that the optimization procedure shown in Algorithm \ref{algo:opt_layer} is not limited to the use in combination with Multi-QIDA, but could also be used as an optimization method for other iterative ans\"atze, as it is similar procedure to the one used by the Adapt-VQE algorithm, as well as in Layer-VQE.

\section{Computational details}
\label{sec:comp_details}
\subsection{Spin Models}
\label{ssec:tested_systems}
To explore the validity of our MultiQIDA approach we have considered different spin systems for the following Heisenberg Hamiltonian: 
\begin{equation}
\begin{split}
    H&=\frac{J}{4}\sum_{\langle i,j\rangle}[\Delta(\sigma_{i}^{x}\sigma_{j}^{x}+ \sigma_{i}^{y}\sigma_{j}^{y} )+\sigma_{i}^{z}\sigma_{j}^{z}]-\\&-\frac{h}{2}\sum_{i}^{N}{\sigma_i^z}.
\end{split}
\label{eqn:sim_H}
\end{equation}
In Equation \ref{eqn:sim_H}, the operator is defined in terms of three parameters: $J=J_{ij}^{xx} =J_{ij}^{yy} =J_{ij}^{zz} > 0 , \forall $all $i,j \in \{1, \dots, N\}$, whereas $J$ is the spin coupling interaction, $\Delta$ as the Anisotropy term, and the external magnetic field $h$.
In Table \ref{tab:systems}, we collect all the information about the parameters used for each system configuration with relative exact energies, $E_{exact}$, as well as the energy of the classical  N\'eel state, $E_{\text{N\'eel}}$. $E_{exact}$ obtained from exact diagonalization of the Hamiltonian, while $E_{\text{N\'eel}}$ is obtained according to Equation \ref{eqn:neel_E}.
\begin{table}[!htb]
    \centering
    \begin{adjustwidth}{-0.2cm}{}
    \scalebox{.9}{
    \fontsize{10pt}{10pt}
    \setlength\tabcolsep{4pt}
    \begin{tabular}{|c|c|c|c|c|c|c|}
    \hline $Size$ & $\#Qubits$ & $J$ & $h$ & $\Delta$ & $E_{exact}$ &$E_{\text{N\'eel}}$\\ \hline\hline
        $3\times3$ & $9$ &$1.0$&$0.0$&$1.0$&$-4.749327$&$-3.00$\\
        $2\times6$ & $12$ &$1.0$&$0.0$&$1.0$&$-6.603472$&$-4.00$\\
        $3\times4$ & $12$ &$1.0$&$0.0$&$1.0$&$-6.691680$&$-4.25$\\
        $3\times4$ & $12$ &$1.0$&$2.0$&$1.0$&$-9.508473$&$-4.25$\\
        $3\times4$ & $12$ &$1.0$&$0.0$&$0.66$&$-5.338751$&$-4.25$\\
        $3\times4$ & $12$ &$1.0$&$0.0$&$0.1$&$-4.272670$&$-4.25$\\
       \hline
    \end{tabular}}
    \end{adjustwidth}
    \caption{Summary table containing all the configurations used in the simulations. The quantities $J$, $h$, and $\Delta$ have been defined in Section \ref{ssec:hamiltonians}, $E_{exact}$ is the energy obtained by exact diagonalization of the Hamiltonian of the system. The energy of the reference state, $E_{\text{N\'eel}}$, is obtained following Equation \ref{eqn:neel_E}.}
    \label{tab:systems}
\end{table}

\subsection{Metrics and measures}
\label{ssec:metrics}
We employed various metrics and measures to compare performances across different systems. In addition to the straightforward metric of the number of CNOTs, $\#CNOT$, other measures included are Absolute Quantum Energy, Relative Quantum Energy, Minimum Absolute Energy Deviation, and Minimum Relative Energy Deviation. 
While $\#CNOT$ is an unambiguous definition, other measures need a brief explanation:
\begin{itemize}
    \item \textit{Absolute Quantum Energy} ($AQE_i$) for the $i$-th VQE run, i.e. the energy obtained from the $i$-th VQE run and the one obtained from Exact Diagonalization, $E_{exact}$.
    \begin{equation}
    \label{eqn:AQE}
        AQE_{i} = \frac{E_{i}}{E_{exact}}*100.
    \end{equation}
    \item
\textit{Relative Quantum Energy} ($RQE_i$) i.e. the ratio between the difference of the $i$-th VQE run and the Néel energy and the difference between the ED energy and the Néel energy:
    \begin{equation}
    \label{eqn:RQE}
        RQE_i = \frac{|E_{i} - E_{\text{N\'eel}}|}{|E_{exact} - E_{\text{N\'eel}}|} * 100.
    \end{equation}
\end{itemize}

The Néel state represents the ground state of the corresponding classical system, and it is obtained by setting spin in an alternated configuration between spin-up, $\uparrow$, and spin-down, $\downarrow$. The energy associated with the Néel state, $E_{\text{N\'eel}}$, is a constant that scales with the number of pairings and the value of the coupling term $J_{zz}$, i.e. 
\begin{equation}
    E_{\text{N\'eel}} = \sum_{\langle i,j\rangle}J_{i,j}^{zz}.
    \label{eqn:neel_E}
\end{equation}
All the values relative to $E_{exact}$ and $E_{\text{N\'eel}}$ are shown in Table \ref{tab:systems}.
\begin{itemize}
    \item 
\textit{Minumum Absolute Energy Deviation} (MAED) represents the average deviation of the $AQE_i$ from the best achieved $AQE$ i.e. $AQE_{best}$, denoting the $AQE$ value of the top-performing run. It is computed by summing the differences between the $AQE$ of each VQE run and the $AQE_{best}$, then normalizing this sum by the total number of runs. It is expressed as:
\begin{equation}
    MAED = \frac{\sum^{\#VQE}_{i=1}{|AQE_i - AQE_{best}|}}{\#VQE}.
\end{equation}
\item
\textit{Mininum Relative Energy Deviation} (MRED) quantifies the average deviation of the relative quantum energy from the optimal relative energy $E_{best}$. Defined in the same way of the MAED but using $RQE$ values:
\begin{equation}
    MRED = \frac{\sum^{\#VQE}_{i=1}{|RQE_i - RQE_{best}|}}{\#VQE}.
\end{equation}
\item
\textit{Minimum Energy Deviation} (MED) defined as the two previous metrics, but using flat energies and not $\%$ values.
    \begin{equation}
        MED =\frac{\sum_{i = 1}^{\#VQE}|E_i - E_{best}|}{\#VQE}.
    \end{equation}
\end{itemize}

\subsection{Simulation details}
\label{ssec:sim_detail}
We split the workflow between pre-processing and simulation.
For the preprocessing part, which includes the DMRG calculation and QMI creation, we used \texttt{ITensors.jl} \cite{ITensor, ITensor-r0.3}, a supplementary Julia library for efficient tensor computations and tensor network calculations. 
For the Qubit-pairing selection and the framework for the simulations, we used Python, while for the quantum computing simulation and PQC creation, we used the Qiskit Python library\cite{Qiskit}.
We tested every Hamiltonian configuration with 50 VQE runs each. We employed noiseless statevector simulation. For the optimization procedure of the VQE, we decided to use the Broyden–Fletcher–Goldfarb–Shanno (BFGS)\cite{Flet87} algorithm with a convergence threshold set to $10^{-6}$.

\section{Results}
\label{sec:res}
In this section, we first present the detailed workflow and results obtained for the $3\times 4$ isotropic spin system. The complete workflow is shown in Figure \ref{figR:3x4_complete_workflow}. After the detailed description of this particular case, we will present the results for all the other system studied.

\subsection{Building the ansatz: QMI qubit-pairs and Layers-Builder procedure}
\label{ssec:qmi_res}
\begin{figure}[!htb]
    \centering
    \includegraphics[scale=0.4]{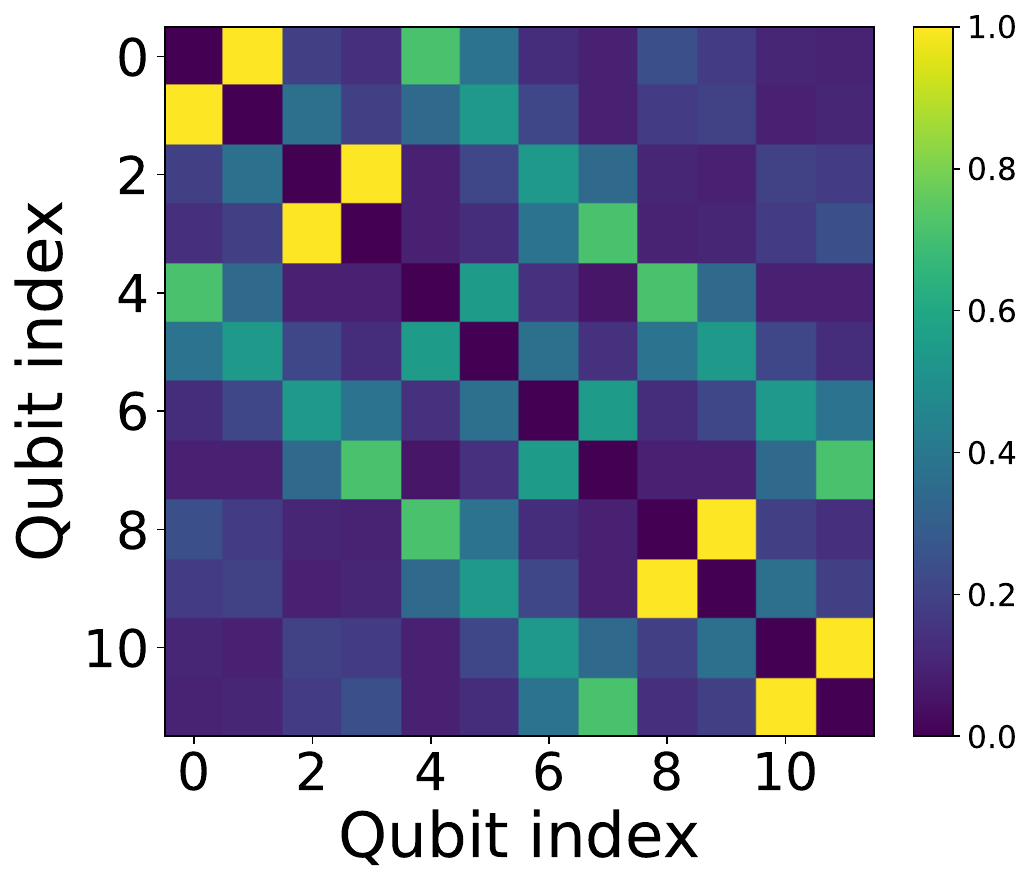}
    \caption{QMI map of $3\times4$ Isotropic Heisenberg model with $J=1.0$, $h=0$, and $\Delta = 1.0$. Reference wavefunction obtained from converged DMRG calculation.}
    \label{figR:3x4_qmi}
\end{figure}

\begin{figure*}[!htb]
    \centering
    \includegraphics[scale=0.4]{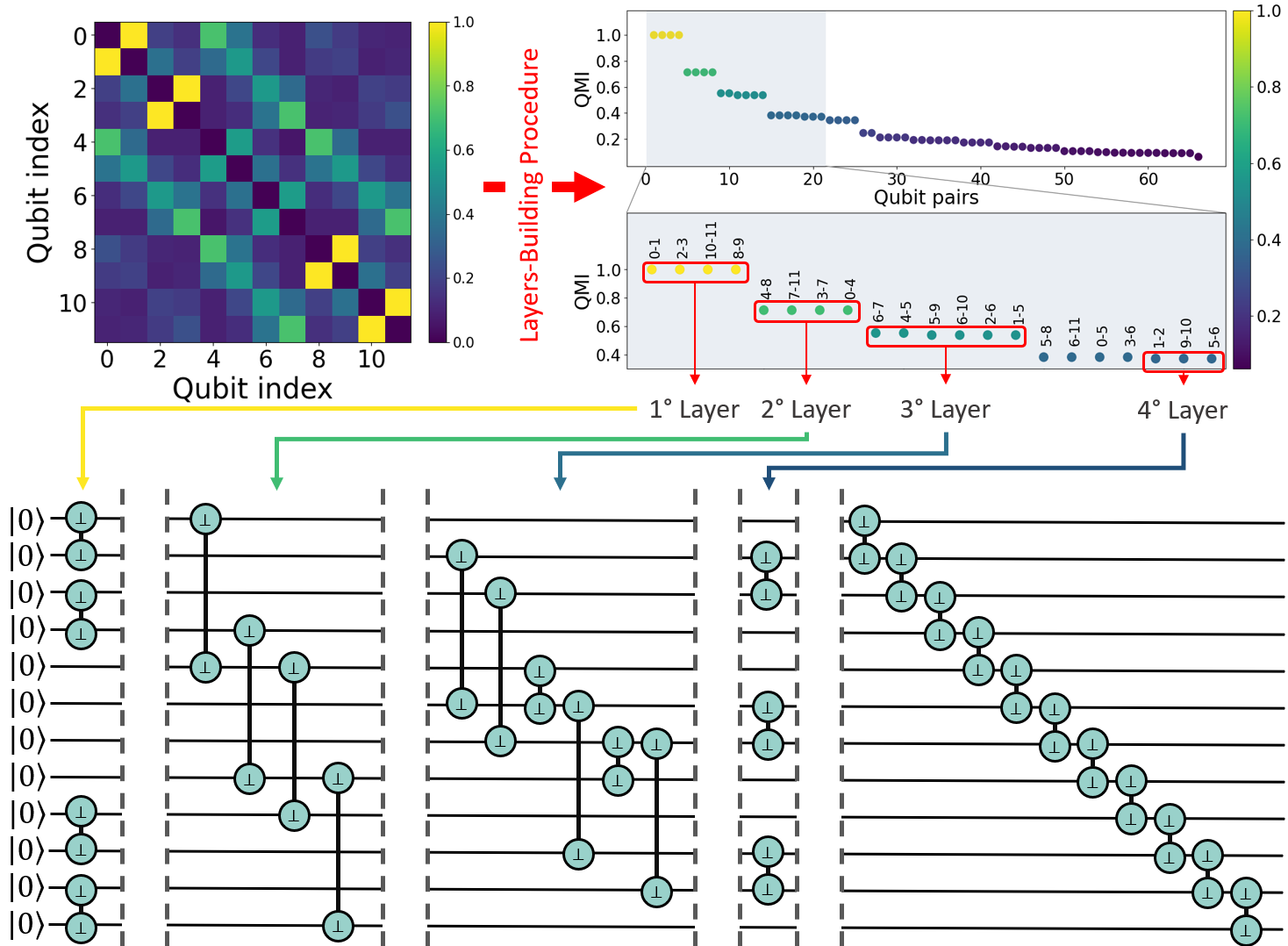}
    \caption{$3\times4$ Multi-QIDA complete workflow. In upper panel, the first two steps of the Multi-QIDA approach are displayed: (left) we start by building the QMI map from a reference wavefunction. (right) We then apply the Layer-building procedure to obtain the set of QIDA layers. The rounded boxes represent the qubit pairings selected for each QIDA-layer. The chunking is done according to the finesse-ratio defined in Section \ref{ssec:qmi_res}. In the lower panel, it is displayed the complete \textbf{SO}(4) circuit for $3\times4$ isotropic Heisenberg Hamiltonian system. From the Layers-Building procedure, we obtained 4 QIDA layers. The fifth and last entangling layer is the closing ladder.}
    \label{figR:3x4_complete_workflow}
\end{figure*}

Here, we describe the full procedure for obtaining the complete set of entangling maps that are needed to compose a single Multi-QIDA ansatz.
Starting from the DMRG reference MPS, we created the corresponding QMI map. We compute each term of Equation \ref{eqn:QMI} by exploiting properties of the MPS wavefunction that allow us to compute RMD matrices, defined in Equation \ref{eqn:rho_ij}, efficiently. 
In Figure \ref{figR:3x4_qmi} and in the upper panel of Figure \ref{figR:3x4_complete_workflow} (left), we show the QMI map obtained for the $3\times4$ isotropic Heisenberg Hamiltonian using DMRG.

\begin{figure}[!htb]
    \centering
    \includegraphics[scale=0.3]{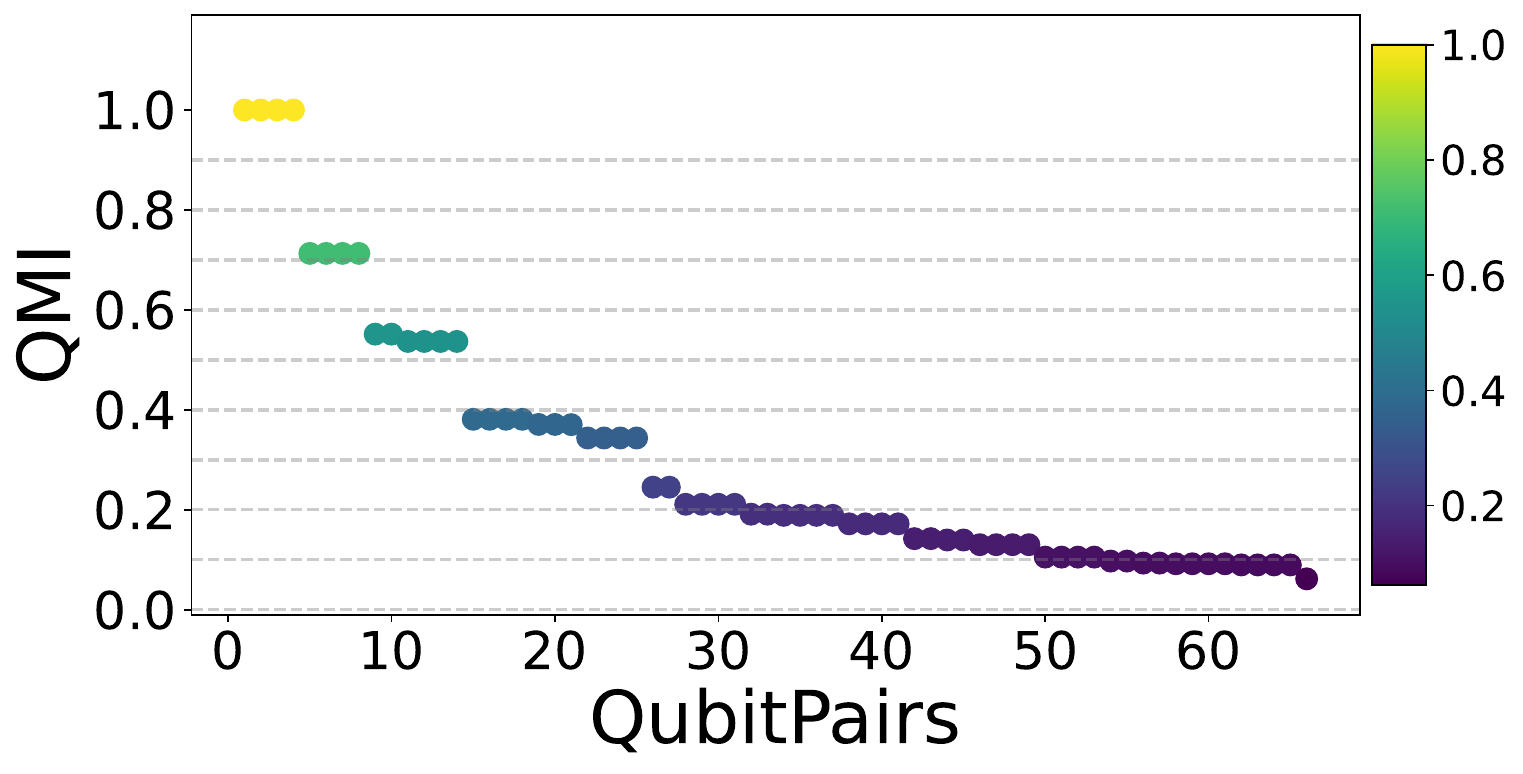}
    \caption{Qubit-Pairing for the $3\times4$ Isotropic Heisenberg model with $J=1.0$, $h=0$, and $\Delta = 1.0$. Dashed lines represent the values of the finesse-ratio $\bar\mu$ selected for this system.}
    \label{figR:3x4_close_up}
\end{figure}

From the QMI map, we have extracted the qubit-pairing and ordered them in descending mutual-information values. Then, we applied the layer-building procedure explained in Section \ref{ssec:layer_builder}, following Algorithm \ref{algo:sel_pseudocode}.
In Figure \ref{figR:3x4_close_up}, we show a pictorial representation of the ordered set of coupling obtained by a QMI map. All the qubit-pairs are represented, and the color of each point is chosen accordingly with the value of QMI. The most correlated couplings are the ones with a QMI value near 1, light-tone color, followed by decreasing values of QMI corresponding to a color gradient towards darker tone.

The first procedure of the Layers-builder procedure splits the correlators into QMI chunks, and this is done by imposing a finesse-ratio of $\bar\mu = [0.9, 0.8, 0.7, \dots]$. With this finesse-ratio, it is clear that not all the chunks will contain qubit pairing e.g. in the ranges $[0.9,0.8]$ or $[0.7,0.6]$. In Figure \ref{figR:3x4_close_up}, it is possible to see how the QMI values are chunked according to the finesse-ratio selected. The QMI pairs that fall inside two dotted lines are the candidate coupling that may be selected to be inserted inside a QIDA layer. Thus, it encloses also qubit-pairs that will be removed in the second phase of the Multi-QIDA layers-builder algorithm. These pairings are the ones that try to connect qubits that are already connected by the previous layer and thus, they are reachable to each other by exploiting the cross-correlation of the previous QIDA layers.

From the set of qubit-pairs split according to the QMI values and the finesse-ratio chosen, we retrieved the Multi-QIDA ansatz that we used as PQC applying Algorithm \ref{algo:sel_pseudocode}. 
This procedure is resumed in the upper pane (right) of Figure \ref{figR:3x4_complete_workflow} and the resulting circuit for the \textbf{SO}(4) implementation of the Multi-QIDA ansatz is shown in the lower pane of Figure \ref{figR:3x4_complete_workflow}.
From the QMI chunks obtained in the first procedure, we started building each QIDA layer by applying the selection criteria explained in Section \ref{ssec:layer_builder}. Algorithm \ref{algo:sel_pseudocode} stopped after the addition of the $7$-th range of QMI values, i.e. $30\%$ of correlation, with which covered and connected all the 12 qubits. 

The isotropic $3\times4$ model Hamiltonian required 17 of the 66 available qubit-pairs. The 17 correlators are split in the following layers :
\begin{itemize}
    \item First Layer (4 correlators):[0, 1], [2, 3], [8, 9], and [10, 11].
    \item Second Layer (4 correlators): [0, 4], [3, 7], [4, 8], and [7,11].
    \item Third Layer (6 correlators): [1, 5], [2, 6], [4, 5], [5, 9], [6, 7], and [6, 10].
    \item Fourth Layer (3 correlators):  [1, 2], [5, 6], and [9, 10],
\end{itemize}
and these pairings, i.e. the selected ones, are identified by the rounded boxed in the lower pane of Figure \ref{figR:3x4_complete_workflow}.
Following the above composition, we obtain a PQC that is composed of 52 CNOTs for $QIDA^{CX}$ and 56 CNOTs for $QIDA^{SO4}$ having in total 5 QIDA layers.
We notice that the $\#CNOT$ used in $QIDA^{CX}$ and $QIDA^{SO4}$ differs only for the number of correlators involved in the $1$-st layer. This is due to the fact that for the $QIDA^{CX}$ ansatz, all the layers except the first are composed of the $V$-shape configurations of the layer, which doubles the count of CNOTs, while $QIDA^{SO4}$ counts two CNOTs for each gate used, including the first layer.

We carried out the comparison between Multi-QIDA configuration and ladder Heuristic ansatz
defined by a layer of parametrized rotations $R_y(\theta)$, $N-1$ CNOTs in a top-down ordered configuration, followed by another set of parametrized rotations (see Figure \ref{fig:Ladder_heur_d_2}). 
In a situation in which the number of CNOTs of a multi-QIDA configuration can not be compared directly with the exact depth of the ladder circuit, we decided to compare ladders that are composed of the previous and next depths. Thus, if the ladder contains fewer CNOTs than our approach, and the ansatz with depth increased by one, contains a higher number of CNOTs, then we take both ladders as a comparison. 
The number of CNOTs for ladder fashion circuits is defined as $(N-1)*d$, where $N$ is the number of qubits, i.e. sites number, and $d$ is the number of repetition of each layer with ladder displacement of the entangling gate i.e. the depth. We refer to the ladder ans\"atze as $(L)_d^{CX}$.
For the $3\times4$ Isotropic Hamiltonian model, the number of sites is 12, which translates to 11 entangling gates per layer. The ladder ans\"atze we used as comparison contain $44$ CNOTs for $(L)_4^{CX}$, $55$ CNOTs for $(L)_5^{CX}$, and $66$ CNOTs for $(L)_6^{CX}$.

For the other system, the reference QMI maps are shown in Figure \ref{appendix:fig_QMIs}, while the relative qubit-pairing is available in Appendix \ref{appendix:fig_pairs_collected}. The resulting Multi-QIDA layers can be found in Appendix \ref{appendix:tab_multi_qida_conf} and the cost in terms of CNOTs in Appendix \ref{appendix:tab_CNOT_numbers}.

\subsection{Performance Analysis}
Here, we present the comparison between our Multi-QIDA approach and the heuristic ladder ans\"atze for one specific system that we have studied, the $3\times4$ Isotropic Heisenberg Model Hamiltonian. In Table \ref{tab:res_main}, we have collected all the results for this system configuration.
As already introduced in the previous section, we carried out the comparison between ladder fashion ans\"atze and Multi-QIDA method circuits, in both CNOTs and \textbf{SO}(4) parametrized gates.
\begin{itemize}
    \item \textit{$(L)_d^{CX}$ - Heuristic ladder-fashion ans\"atze}: For this type of ans\"atze, the results we obtained are very close between PQC at different depths. In particular, the two deeper circuits, with depths 5 and 6, present small variations in terms of mean result and best-performing run. The deeper circuit is the one that performs slightly better. For the absolute quantum error (AQE), the average is set to $84.37\%$, and the best-performing simulation reaches $90.36\%$, with a mean error from the best result of $5.99\%$. The relative quantum error (RQE) values are $57.17\%$, $73.58\%$, and $16.42\%$, corresponding to mean, best VQE, and deviation, in order. 
The ladder fashion circuits present on average a dispersion from the best-performing runs that is at least of $5.52\%$ for the absolute quantum energy and at least $15.10\%$ for the relative quantum energy. Looking at the mean energy deviation error (MEAD), we can see that the best-performing ladder incurs an error of $0.4$.
\end{itemize}

\begin{table*}[!htb]
\fontsize{8pt}{8pt}
    \centering
    \setlength\tabcolsep{2pt}
\begin{tabular}{|c|c|c|c|c|c|c|c|c|c|c|}
\hline
$Lattice$ & $Ansatz$& $E_{avg}$ & $E_{best}$ & $AQE_{avg}$ & $RQE_{avg}$ & $AQE_{best}$ & $RQE_{best}$ & $MED$ & $MAED$ & $MRED$ \\\hline\hline
$3\times4$ & $(L)^{CX}_4$ & $ -5.56(\pm24) $ & $ -5.924784 $ & $ 83.03 $ & $ 53.49  $ & $ 88.54  $ & $ 68.59 $ & $ 0.368738 $ & $ 5.51 $ & $ 15.10$ \\
$3\times4$ & $(L)^{CX}_5$ & $-5.65(\pm14)$ & $ -6.067645 $ & $ 84.36 $ & $ 57.14 $ & $ 90.67  $ & $ 74.44 $ & $ 0.422450  $ & $ 6.31 $ & $ 17.30$ \\
$3\times4$ & $(L)^{CX}_6$ & $-5.65(\pm25)$ & $ -6.046642  $ & $ 84.37  $ & $ 57.17  $ & $ 90.36  $ & $ 73.58  $ & $ 0.401  $ & $ 5.99  $ & $ 16.42$ \\ \hline
$3\times4$ & $(QIDA)^{CX}$ & $ -6.200065(\pm1)$ & $ -6.200066$ & $ 92.65$ & $ 79.87$ & $ 92.65$ & $ 79.87$ & $ \sim1\text{e-6} $ & $ \sim2\text{e-5} $ & $ \sim5\text{e-5}$ \\
$3\times4$ & $(QIDA)^{SO4}$ & $-6.3612(\pm42)$ & $ -6.362179 $ & $ 95.061$ & $ 86.46 $ & $ 95.08$ & $ 86.51 $ & $ 0.000998 $ & $ 0.014918 $ & $ 0.040885$ \\ \hline
\end{tabular}
    
    \caption{Simulations results for the $3\times4$ Isotropic Heisenberg model Hamiltonian divided by ansatz layout. Results were obtained with $\#VQE=50$ each, statevector simulation using Qiskit library, optimization algorithm BFGS with convergence tolerance 1e-6, and finesse ratio $\mu=0.1$ for the layer selection. $AQE_{avg}$ and $RQE_{avg}$ are defined as the average performance metrics $AQE_i$ and $RQE_i$ over $\#VQE$ simulations. For the $E_{avg}$, the value in the brackets is the standard deviation.}
    \label{tab:res_main}
\end{table*} 
\begin{itemize}
    \item \textit{$QIDA^{CX}$ - Multi-QIDA with CNOTS}: We increased the mean $AQE$ to $92.65\%$ and the best-performing run too, with a value of $92.65$. The mean absolute energy deviation that we commit using this ansatz is 2e-5$\%$. Consequently, the $RQE$ presents the same behavior, $79.87\%$ for the average $RQE$ and best VQE, and an error of 5e-5$\%$. For the $QIDA^{CX}$ ansatz, the $RQE_{avg}$ is increased by +$22.70\%$, the $RQE_{best}$ by +$6.29\%$, with increased precision of seven orders of magnitude. In terms of MAED, we obtain an energy deviation of $1e$-6.
\end{itemize}
\begin{itemize}
    \item \textit{$QIDA^{SO4}$ - Multi-QIDA with SO(4)}: We obtained $95.061\%$, $95.08\%$ and $0.149\%$, for average $AQE$, $AQE_{best}$ and deviation, in order. The $RQE$ results are instead $86.46\%$, $86.51$, and $0.04\%$ for mean $RQE$, best-simulation $RQE$ and minimum relative deviation.
In comparison to the ladders, for the $QIDA^{SO4}$, the enhancements that we obtained are +$29.29\%$ and +$12.93\%$, for $RQE_{avg}$ and $RQE_{best}$, respectively. The deviation from the minimum, $MRED$, is three orders of magnitude lower than the one obtained with the deeper heuristic ladder ansatz. We obtained a MEAD of $1e$-3 for this ans\"atze configuration.
\end{itemize}

\subsection{Convergence and precision}
\begin{figure}[!htb]
    \centering
    \includegraphics[scale=0.4]{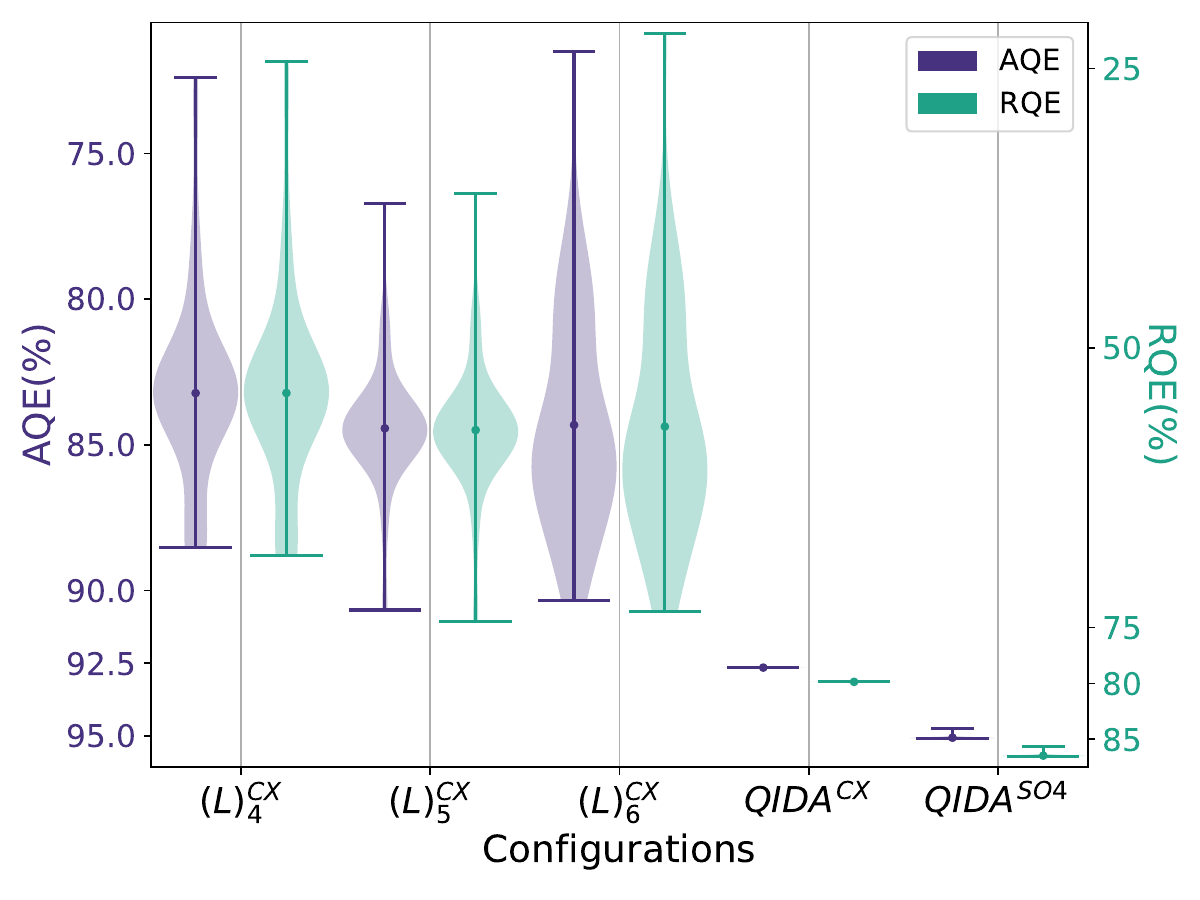}
    \caption{$3\times4$ isotropic Heisenberg Hamiltonian with $J=1.0$ VQE results. Comparison between AQE and RQE for the different ans\"atze settings. Starting from the left: HEA ladder depth 4 (44), depth 5 (55), and depth 6 (66), then,  Multi-QIDA CNOT(52) and Multi-QIDA (56) \textbf{SO}(4) ans\"atze. In round brackets, the cost in term of $\#CNOTS.$}
    \label{fig:3x4_violin}
\end{figure}

For a pictorial representation of the results shown in Table \ref{tab:res_main}, a series of resuming violin plots can be found in Figure \ref{fig:3x4_violin}. 

Each violin represents the $AQE$ and the $RQE$ for a different ans\"atze configuration for the same system, tested in both ladder and QIDA ansatz settings. The lower tick represents the best VQE, and the higher tick represents the worst VQE. The black dots are associated to the average value of both $AQE$ and $RQE$. The width of the violins represents the frequency of the simulation's outcomes and they can be thought of as a rotated and smoothed histogram of several optimization procedure obtained with different initial parameters. From these plots, it is possible to see the difference in the performance between general heuristic ans\"atze and Multi-QIDA variational forms. 

The ladder-fashion ans\"atze present average results that are generally distant from the minimum results, meaning that it is required to re-run the circuit different times before obtaining a satisfying result. Ladders can indeed encounter optimization paths that may converge to local minima that are far away from the best result that can be obtained. As you can see for the depth 4 ladder, only a few runs can reach similar results to the best-performing simulation for that configuration.

Multi-QIDA approaches instead guide the variational form in the right spot, providing good results with very high probability, thus, few independent optimizations are required to obtain these results.
From Table \ref{tab:res_main}, it is possible to see this behaviour noting that the difference between the average case and best-performing simulation is almost zero.
In terms of energy, $QIDA^{SO4}$ configuration provides the best choice. Looking at the errors committed by the two Multi-QIDA configurations, the $QIDA^{CX}$ ans\"atze maintain the optimal convergence and lower MRED. The $QIDA^{SO4}$ ans\"atze present a slightly larger inaccuracy with an error that is three orders of magnitude bigger than the CNOT implementation.

\begin{figure}[!htb]
    \centering
\includegraphics[scale=0.37]{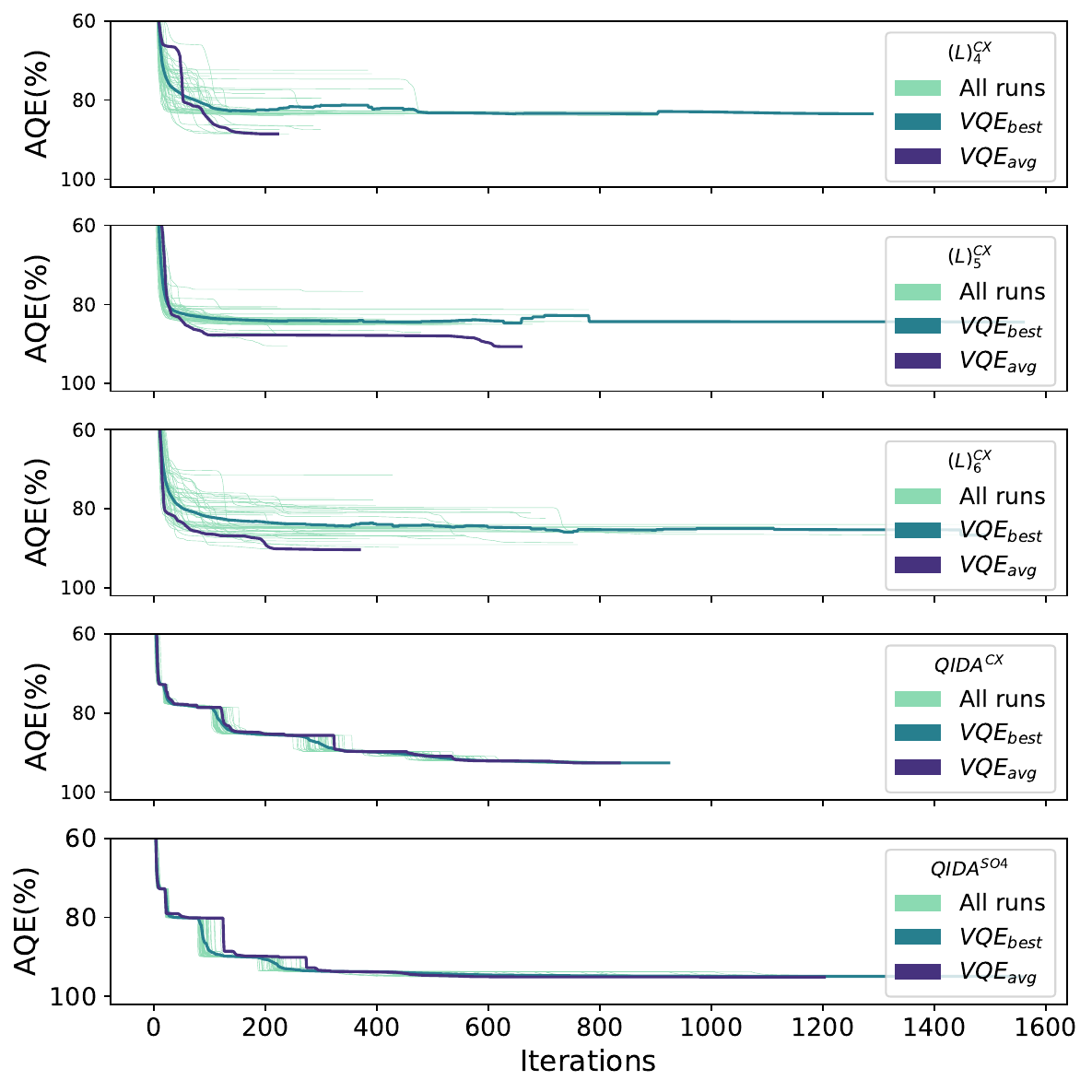}    \caption{Representation of the optimization trajectories plotted by each VQE in terms of AQE for the $3\times4$ isotropic Heisenberg Hamiltonian with $J=1.0$. From top to bottom: The first three plots represent ladder-fashion ans\"atze, then the $QIDA^{CX}$ circuit, and finally, the \textbf{SO}(4) implementation.}
\label{fig:traj_3x4_res}
\end{figure}

In Figure \ref{fig:traj_3x4_res}, we show the convergence plot for the $3\times4$ isotropic Heisenberg system. Here, we have collected all the optimization trajectories for each of the simulations. Each lighter-colored line represents one VQE simulation, while we plotted in a darker color, the best-performing VQE and average VQE values.
For these plots, it is possible to see how the VQEs simulations with Multi-QIDA ans\"atze tend to be more compact in terms of dispersion from the best results. Thus, we can say that with high precision Multi-QIDA-ans\"atzes drive almost all the VQE simulations to follow the same path as the best-performing VQE run.

In terms of the average number of iterations required for the different types of ansatz, it is clear that it is higher for the Multi-QIDA approach. This is due to the iterative optimization nature of the Multi-QIDA approach, against the standard VQE for the standard heuristic ladder ansatz which requires a single VQE routine.
For the $3\times4$ isotropic Heisenberg Hamiltonian system, we obtained that the number of optimization VQE steps for the three ladders ansatz is 338, 427, and 583, respectively for depth 3, 4, and 5. For our Multi-QIDA ansatz, we reached 825 evaluations for the CNOT version and 1073 for the \textbf{SO}(4) implementation. The mean number of iterations required by the Multi-QIDA ansatz is at least double the evaluations required by the heuristic ladder ans\"atze.

From Figure \ref{fig:traj_3x4_res}, it is possible to note that, for Multi-QIDA configuration, a not negligible time is spent during the relaxation phase, in which the optimization is trying to re-optimize already converged circuits. 
In this situation, we could have implemented a different halting criterion for the relaxation phase, which could have helped reduce the time in which the optimization is almost converged. Using a fixed number of evaluations for the re-optimization routine or an additional threshold on the $\Delta{E}$ could help speed up the whole procedure.

The convergence and the trajectories plots for the other tested systems can be found in Figures \ref{appendix:fig_violins} and \ref{appendix:fig_trajectories}, while the optimization iterations in Table \ref{appendix:tab_conv_time}.

\subsection{Complete results}
\label{ssec:complete_results}
In this section, we collect very briefly and concisely the results from all the systems configurations: 3x3, 2x6, and 3x4 isotropic Heisenberg model, 3x4 with $\Delta = 1/10$ and $2/3$ anisotropic term, and finally, the 3x4 isotropic with external magnetic field $h=2$.
In Table \ref{tab:res_isotropic} are presented the results for the three isotropic configurations, while in Table \ref{tab:res_other_3x4} are shown the results for the $3\times4$ variations.
 \begin{table*}[!hbt]
\fontsize{8pt}{8pt}
    \centering
    \setlength\tabcolsep{2pt}
\begin{tabular}{|c|c|c|c|c|c|c|c|c|c|c|}
\hline
$Lattice$ & $Ansatz$& $E_{avg}$ & $E_{best}$ & $AQE_{avg}$ & $RQE_{avg}$ & $AQE_{best}$ & $RQE_{best}$ & $MED$ & $MAED$ & $MRED$ \\ \hline\hline
$3\times3$ & $(L)^{CX}_4$ & $-4.22(\pm14) $ & $ -4.335326 $ & $ 88.81 $ & $ 69.61$ & $ 91.28 $ & $ 76.33 $ & $ 0.117578$ & $ 2.48 $ & $ 6.72$ \\
$3\times3$ & $(L)^{CX}_5$ & $ -4.30(\pm13)$ & $ -4.497591$ & $ 90.56$ & $ 74.37$ & $ 94.69 $ & $ 85.61$ & $ 0.196695$ & $ 4.14$ & $ 11.24$ \\ \hline
$3\times3$ & $(QIDA)^{CX}$ & $-4.333(\pm20)$ & $-4.336704$ & $ 91.25$ & $ 76.24$ & $ 91.31$ & $ 76.41$ & $ 0.003003$ & $ 0.063231$ & $ 0.171668$ \\
$3\times3$ & $(QIDA)^{SO4}$ & $ -4.6029(\pm11)$ & $ -4.603961$ & $ 96.92$ & $ 91.63$ & $ 96.94$ & $ 91.69$ & $ 0.001011$ & $ 0.021294 $ & $ 0.057811$ \\ \hline\hline
$2\times6$ & $(L)^{CX}_4$ & $ -5.42(\pm15)$ & $ -5.502422 $ & $ 82.07$ & $ 54.52$ & $ 83.33$ & $ 57.71$ & $ 0.083146$ & $ 1.26$ & $ 3.19$ \\
$2\times6$ & $(L)^{CX}_5$ & $-5.57(\pm11)$ & $ -5.662730$ & $ 84.31 $ & $ 60.21$ & $ 85.75$ & $ 63.87$ & $ 0.095113$ & $ 1.44$ & $ 3.65$ \\ \hline
$2\times6$ & $(QIDA)^{CX}$ & $ -6.23776(\pm31)$ & $ -6.238618$ & $ 94.46$ & $ 85.95$ & $ 94.48$ & $ 85.99$ & $ 0.000856$ & $ 0.012955$ & $ 0.032860$ \\
$2\times6$ & $(QIDA)^{SO4}$ & $-6.265514(\pm1)$ & $ -6.265514$ & $ 94.88$ & $ 87.02$ & $ 94.88$ & $ 87.02$ & $ \sim5\text{e-7} $ & $ \sim8\text{e-6} $ & $ \sim2\text{e-5}$ \\ \hline\hline
$3\times4$ & $(L)^{CX}_4$ & $ -5.56(\pm24) $ & $ -5.924784 $ & $ 83.03 $ & $ 53.49  $ & $ 88.54  $ & $ 68.59 $ & $ 0.368738 $ & $ 5.51 $ & $ 15.10$ \\
$3\times4$ & $(L)^{CX}_5$ & $-5.65(\pm14)$ & $ -6.067645 $ & $ 84.36 $ & $ 57.14 $ & $ 90.67  $ & $ 74.44 $ & $ 0.422450  $ & $ 6.31 $ & $ 17.30$ \\
$3\times4$ & $(L)^{CX}_6$ & $-5.65(\pm25)$ & $ -6.046642  $ & $ 84.37  $ & $ 57.17  $ & $ 90.36  $ & $ 73.58  $ & $ 0.401  $ & $ 5.99  $ & $ 16.42$ \\ \hline
$3\times4$ & $(QIDA)^{CX}$ & $ -6.200065(\pm1)$ & $ -6.200066$ & $ 92.65$ & $ 79.87$ & $ 92.65$ & $ 79.87$ & $ \sim1\text{e-6} $ & $ \sim2\text{e-5} $ & $ \sim5\text{e-5}$ \\
$3\times4$ & $(QIDA)^{SO4}$ & $-6.3612(\pm42)$ & $ -6.362179 $ & $ 95.061$ & $ 86.46 $ & $ 95.08$ & $ 86.51 $ & $ 0.000998 $ & $ 0.014918 $ & $ 0.040885$ \\ \hline
\end{tabular}
    
    \caption{Simulation results divided by ansatz layout and system configurations for all the Isotropic Heisenberg Hamiltonian lattices. For the $E_{avg}$, the value in the brackets is the standard deviation.}
    \label{tab:res_isotropic}
\end{table*} 
\begin{table*}[!htb]
\fontsize{8pt}{8pt}
    \centering
    \setlength\tabcolsep{2pt}
    \begin{tabular}{|c|c||c|c|c|c|c|c|c|c|c|}\hline $Lattice$ & $Ansatz$ & $E_{avg}$ & $E_{best}$ & $AQE_{avg}$ & $RQE_{avg}$ & $AQE_{best}$ & $RQE_{best}$ & $MED$ & $MAED$ & $MRED$ \\\hline\hline
$h=2$ & $(L)^{CX}_4$ &$ -9.066(\pm92)$&$ -9.184634$&$ 95.35 $&$ 91.59 $&$ 96.59 $&$ 93.84 $&$ 0.118396$&$ 1.25$&$ 2.25$ \\
$h=2$ & $(L)^{CX}_5$ &$ -9.068(\pm87)  $&$ -9.189169  $&$ 95.37  $&$ 91.62 $&$ 96.64  $&$ 93.93  $&$ 0.121285  $&$ 1.28  $&$ 2.31$ \\\hline
$h=2$ & $(QIDA)^{CX}$ & $-9.286827(\pm3\text{e-4})  $&$ -9.286827 $&$ 97.67 $&$ 95.79 $&$ 97.67 $&$ 95.76 $&$ \sim3\text{e-10} $&$ \sim3\text{e-9} $&$ \sim5\text{e-9}$ \\
$h=2$ & $(QIDA)^{SO4}$ &$ -9.4144(\pm61) $&$ -9.426608 $&$ 99.01 $&$ 98.21  $&$ 99.14  $&$ 98.44  $&$ 0.012195  $&$ 0.128254$&$ 0.231911 $ \\\hline\hline
$\Delta=\frac{2}{3}$ & $(L)^{CX}_4$ &$ -4.80(\pm16)$&$ -5.091959 $&$ 89.93 $&$ 50.61 $&$ 95.38 $&$ 77.33$&$ 0.290964  $&$ 5.45  $&$ 26.72 $ \\
$\Delta=\frac{2}{3}$ & $(L)^{CX}_5$ & $-4.866(\pm49)$&$ -4.934131  $&$ 91.09  $&$ 56.35   $&$ 92.42 $&$ 62.84   $&$ 0.070614 $&$ 1.32  $&$ 6.49 $ \\
$\Delta=\frac{2}{3}$ & $(L)^{CX}_6$ &$ -4.89(\pm17)$&$ -5.135422 $&$ 91.52  $&$ 58.42 $&$ 96.19 $&$ 81.33 $&$ 0.249399 $&$ 4.67 $&$ 22.91$ \\\hline
$\Delta=\frac{2}{3}$ & $(QIDA)^{CX}$ & $-5.1777(\pm25)  $&$ -5.181318  $&$ 96.98  $&$ 85.21  $&$ 97.05  $&$ 85.54  $&$ 0.003624 $&$ 0.067884 $&$ 0.332874 $ \\
$\Delta=\frac{2}{3}$ & $(QIDA)^{SO4}$ & $-5.232248(\pm1\text{e-2})  $&$ -5.232248 $&$ 98.01 $&$ 90.22 $&$ 98.01  $&$ 90.22  $&$ \sim4\text{e-9} $&$ \sim8\text{e-8} $&$ 4\text{e-7}$ \\\hline\hline
$\Delta=\frac{1}{10}$ & $(L)^{CX}_4$ &$ -4.2599(\pm40)  $&$ -4.272528  $&$ 99.70  $&$ 43.64  $&$ 99.99 $&$ 99.37 $&$ 0.013 $&$ 0.295712 $&$ 55.73 $ \\
$\Delta=\frac{1}{10}$ & $(L)^{CX}_5$ & $-4.2629(\pm37)$&$ -4.266284$&$ 99.77 $&$ 56.71$&$ 99.85$&$ 71.83$&$ 0.003428$&$ 0.080234 $&$ 15.12$ \\\hline
$\Delta=\frac{1}{10}$ & $(QIDA)^{CX}$ & $-4.272574(\pm3) $&$ -4.272579 $&$ 99.99 $&$ 99.57 $&$ 99.99$&$ 99.60 $&$ \sim6\text{e-6} $&$ 0.000141$&$ 0.026570$ \\
$\Delta=\frac{1}{10}$ & $(QIDA)^{SO4}$ & $-4.272645(\pm4)  $&$ -4.272656  $&$ 99.99  $&$ 99.89$&$ 99.99 $&$ 99.94$&$ \sim1\text{e-5} $&$ 0.000249$&$ 0.046940 $\\\hline
\end{tabular}
    
    \caption{Simulation results for the $3\times4$ Heisenberg model variations. Starting from the top: Isotropic with External magnetic field $h=2$, Anisotropic with $\Delta = \frac{2}{3}$, and Anisotropic with $\Delta=\frac{1}{10}$. For the $E_{avg}$, the value in the brackets is the standard deviation. }
    \label{tab:res_other_3x4}
\end{table*}
For the Isotropic configurations remaining the $3\times3$ and the $2\times6$, our approach in the \textbf{SO}(4) settings performs better than the standard HEA in both the lattices. For the $2\times6$, both the Multi-QIDA configurations performed significantly better than the deeper ladder circuit $(L)_5^{CX}$ with an increase in the average $RQE$ of +$25.74\%$ and +$26.81\%$ for CNOT and \textbf{SO}(4) Multi-QIDA ans\"atze, respectively (shown in Figure \ref{appendix:fig_2x6_violin}). The $RQE_{best}$ is increased by +$22.12\%$ by the $QIDA^{CX}$ ansatz and +$23.15\%$ for the $QIDA^{SO4}$. For the $3\times3$ lattice instead,  the $QIDA^{CX}$ has the same performance of the $(L)_4^{CX}$ in terms of best-run VQE and worse performance for the $(L)_5^{CX}$ (Figure \ref{appendix:fig_3x3_violin}). On this lattice, the \textbf{SO}(4) implementation performs better than the deeper ladder ansatz, $(L)_5^{CX}$, obtaining an enhanced $RQE_{avg}$ by +$17.26\%$ and a best $RQE$ of +$6.08\%$. For the variants of the $3\times4$ lattice, the enhancement in terms of best-performing result is not very pronounced as the improvement in the average case. For the Anisotropic $\Delta=2/3$ configuration, the $RQE_{avg}$ is increased by +$26.79\%$ and the best runs by +$4.21\%$ for the CNOT ansatz. $QIDA^{SO4}$ instead lead to a +$31.80\%$ for the average case and +$8.89\%$ for the best-performing run (Figure \ref{appendix:fig_3x4_0_66}). For the $\Delta=1/10$ Anisotropic system, the CNOT Multi-QIDA ansatz increases the average case by +$42.86\%$ and the best one of +$27.77\%$. The \textbf{SO}(4) results are close to the previous one, changing only by a few decimals.
Finally, for the Isotropic system with external magnetic field $h=2$, $QIDA^{CX}$ obtains an increment in the average case of +$4.13\%$ and +$1.83\%$ for the best, and similar results of +$6.56\%$ and +$4.51\%$ for the \textbf{SO}(4) version. 

The preprocessing information as QMI maps and qubit-pairings can be consulted in Figures \ref{appendix:fig_QMIs} and \ref{appendix:fig_pairs_collected}, respectively. The composition and cost of each Multi-QIDA configuration are shown in Tables \ref{appendix:tab_multi_qida_conf} and \ref{appendix:tab_CNOT_numbers}.
Convergence violin plots are collected in Figure \ref{appendix:fig_violins}, while the trajectories of the optimizations are shown in Figure \ref{appendix:fig_trajectories}. The cost in terms of VQE evaluations are displayed in Table \ref{appendix:tab_conv_time}.

\section{Discussion and Conclusion}
In this work, we introduced the Multi-Threshold Quantum Information Driven Ansatz (Multi-QIDA), a novel approach to enhance quantum simulations of strongly interactive lattice spin models. By extending the single-threshold QIDA method, Multi-QIDA leverages multiple thresholds of Quantum Mutual Information (QMI) to construct more efficient and compact quantum circuits. Our approach utilizes a layered structure, where each layer is built based on the QMI values of qubit pairs, enabling the inclusion of mid-high QMI value pairs and thereby capturing more correlations within the system.

The effectiveness of Multi-QIDA was validated through extensive benchmarking on various configurations of the Heisenberg model Hamiltonian. Our results demonstrated that Multi-QIDA outperforms traditional heuristic ansatz methods, achieving higher precision in ground-state energy calculations while reducing computational complexity. Specifically, Multi-QIDA circuits showed a significant improvement in accuracy and convergence, with fewer simulations needed to try to obtain optimal solutions compared to ladder-fashion heuristic ans\"atzes. 

One crucial aspect of constructing a layered ansatz is the ability to add layers that can also act as identities for specific parameter sets. The group of \textbf{SO}(4) transformations possesses this characteristic, and its introduction has proven to be highly effective. This approach systematically improves energy efficiency and significantly alleviates the barrel plateau problem, provided the parameters are correctly set and managed as outlined in the described procedure.
The iterative optimization routine of Multi-QIDA, which combines shallow and adaptable quantum circuits, indeed proved to be an effective strategy in navigating the parameter space and mitigating issues such as barren plateaus. This iterative approach allowed for a more targeted exploration of the quantum landscape, leading to more reliable and consistent results.
Our approach, although slightly more costly in terms of the number of evaluations of the cost function, has the advantage of bringing the average errors on a single optimization at least two orders of magnitude lower than using the standard heuristic ladder approach.

Overall, the Multi-QIDA method presents a promising advancement in the field of quantum simulations for lattice spin models. Its ability to efficiently utilize quantum hardware and deliver high-precision results positions it as a valuable tool for future research in quantum computing and quantum chemistry. Further exploration and adaptation of this method to other quantum systems could potentially unlock new possibilities and applications in the realm of quantum simulations.

\section*{Acknowledgments}

The authors acknowledge funding from the MoQS program, founded by the European Union’s Horizon 2020 research and innovation under the Marie Skłodowska-Curie grant agreement number 955479.
The authors acknowledge funding from Ministero dell’Istruzione dell’Università e della Ricerca (PON R \& I 2014-2020).
The authors also acknowledge funding from the National Centre for HPC, Big Data and Quantum Computing - PNRR Project, funded by the European Union - Next Generation EU.\\
L.G. acknowledges funding from the Ministero dell'Università e della Ricerca (MUR) under Project PRIN 2022 number 2022W9W423 through the European Union Next Generation EU.

\clearpage
\printbibliography

@Article{Peruzzo2014,
author={Peruzzo, Alberto
and McClean, Jarrod
and Shadbolt, Peter
and Yung, Man-Hong
and Zhou, Xiao-Qi
and Love, Peter J.
and Aspuru-Guzik, Al{\'a}n
and O'Brien, Jeremy L.},
title={A variational eigenvalue solver on a photonic quantum processor},
journal={Nature Communications},
year={2014},
day={23},
volume={5},
number={1},
pages={4213},
issn={2041-1723},
doi={10.1038/ncomms5213},
url={https://doi.org/10.1038/ncomms5213}
}

@article{McClean_2016,
   title={The theory of variational hybrid quantum-classical algorithms},
   volume={18},
   ISSN={1367-2630},
   url={http://dx.doi.org/10.1088/1367-2630/18/2/023023},
   DOI={10.1088/1367-2630/18/2/023023},
   number={2},
   journal={New Journal of Physics},
   publisher={IOP Publishing},
   author={McClean, Jarrod R and Romero, Jonathan and Babbush, Ryan and Aspuru-Guzik, Alán},
   year={2016}, pages={023023} }

@misc{materia2023quantum,
      title={Quantum Information Driven Ansatz (QIDA): shallow-depth empirical quantum circuits from Quantum Chemistry}, 
      author={Davide Materia and Leonardo Ratini and Celestino Angeli and Leonardo Guidoni},
      year={2023},
      eprint={2309.15287},
      archivePrefix={arXiv},
      primaryClass={quant-ph}
}

@article{Vatan_2004,
   title={Optimal quantum circuits for general two-qubit gates},
   volume={69},
   ISSN={1094-1622},
   url={http://dx.doi.org/10.1103/PhysRevA.69.032315},
   DOI={10.1103/physreva.69.032315},
   number={3},
   journal={Physical Review A},
   publisher={American Physical Society (APS)},
   author={Vatan, Farrokh and Williams, Colin},
   year={2004}}

@article{Shor_1997,
   title={Polynomial-Time Algorithms for Prime Factorization and Discrete Logarithms on a Quantum Computer},
   volume={26},
   ISSN={1095-7111},
   url={http://dx.doi.org/10.1137/S0097539795293172},
   DOI={10.1137/s0097539795293172},
   number={5},
   journal={SIAM Journal on Computing},
   publisher={Society for Industrial & Applied Mathematics (SIAM)},
   author={Shor, Peter W.},
   year={1997},pages={1484–1509} }

@article{Egger_2020,
   title={Quantum Computing for Finance: State-of-the-Art and Future Prospects},
   volume={1},
   ISSN={2689-1808},
   url={http://dx.doi.org/10.1109/TQE.2020.3030314},
   DOI={10.1109/tqe.2020.3030314},
   journal={IEEE Transactions on Quantum Engineering},
   publisher={Institute of Electrical and Electronics Engineers (IEEE)},
   author={Egger, Daniel J. and Gambella, Claudio and Marecek, Jakub and McFaddin, Scott and Mevissen, Martin and Raymond, Rudy and Simonetto, Andrea and Woerner, Stefan and Yndurain, Elena},
   year={2020},
   pages={1–24} }

@article{Seskir_2022,
   title={The landscape of the quantum start-up ecosystem},
   volume={9},
   ISSN={2196-0763},
   url={http://dx.doi.org/10.1140/epjqt/s40507-022-00146-x},
   DOI={10.1140/epjqt/s40507-022-00146-x},
   number={1},
   journal={EPJ Quantum Technology},
   publisher={Springer Science and Business Media LLC},
   author={Seskir, Zeki Can and Korkmaz, Ramis and Aydinoglu, Arsev Umur},
   year={2022}}

@article{von_Burg_2021,
   title={Quantum computing enhanced computational catalysis},
   volume={3},
   ISSN={2643-1564},
   url={http://dx.doi.org/10.1103/PhysRevResearch.3.033055},
   DOI={10.1103/physrevresearch.3.033055},
   number={3},
   journal={Physical Review Research},
   publisher={American Physical Society (APS)},
   author={von Burg, Vera and Low, Guang Hao and Häner, Thomas and Steiger, Damian S. and Reiher, Markus and Roetteler, Martin and Troyer, Matthias},
   year={2021}}

@article{Robert_2021,
   title={Resource-efficient quantum algorithm for protein folding},
   volume={7},
   ISSN={2056-6387},
   url={http://dx.doi.org/10.1038/s41534-021-00368-4},
   DOI={10.1038/s41534-021-00368-4},
   number={1},
   journal={npj Quantum Information},
   publisher={Springer Science and Business Media LLC},
   author={Robert, Anton and Barkoutsos, Panagiotis Kl. and Woerner, Stefan and Tavernelli, Ivano},
   year={2021}}

@Article{Feynman1982,
author={Feynman, Richard P.},
title={Simulating physics with computers},
journal={International Journal of Theoretical Physics},
year={1982},
day={01},
volume={21},
number={6},
pages={467-488},
issn={1572-9575},
doi={10.1007/BF02650179},
url={https://doi.org/10.1007/BF02650179}
}

@article{Banuls_2020,
   title={Simulating lattice gauge theories within quantum technologies},
   volume={74},
   ISSN={1434-6079},
   url={http://dx.doi.org/10.1140/epjd/e2020-100571-8},
   DOI={10.1140/epjd/e2020-100571-8},
   number={8},
   journal={The European Physical Journal D},
   publisher={Springer Science and Business Media LLC},
   author={Bañuls, Mari Carmen and Blatt, Rainer and Catani, Jacopo and Celi, Alessio and Cirac, Juan Ignacio and Dalmonte, Marcello and Fallani, Leonardo and Jansen, Karl and Lewenstein, Maciej and Montangero, Simone and Muschik, Christine A. and Reznik, Benni and Rico, Enrique and Tagliacozzo, Luca and Van Acoleyen, Karel and Verstraete, Frank and Wiese, Uwe-Jens and Wingate, Matthew and Zakrzewski, Jakub and Zoller, Peter},
   year={2020}}

@article{Hussain_2020,
   title={Optimal control of traffic signals using quantum annealing},
   volume={19},
   ISSN={1573-1332},
   url={http://dx.doi.org/10.1007/s11128-020-02815-1},
   DOI={10.1007/s11128-020-02815-1},
   number={9},
   journal={Quantum Information Processing},
   publisher={Springer Science and Business Media LLC},
   author={Hussain, Hasham and Javaid, Muhammad Bin and Khan, Faisal Shah and Dalal, Archismita and Khalique, Aeysha},
   year={2020}}

@article{2021_Bayer,
author = {Bayerstadler, Andreas and Becquin, Guillaume and Binder, Julia and Botter, Thierry and Ehm, Hans and Ehmer, Thomas and Erdmann, Marvin and Gaus, Norbert and Harbach, Philipp and Hess, Maximilian and Klepsch, Johannes and Leib, Martin and Luber, Sebastian and Luckow, Andre and Mansky, Maximilian and Mauerer, Wolfgang and Neukart, Florian and Niedermeier, Christoph and Palackal, Lilly and Winter, Fabian},
year = {2021},
pages = {},
title = {Industry quantum computing applications},
volume = {8},
journal = {EPJ Quantum Technology},
doi = {10.1140/epjqt/s40507-021-00114-x}
}

@article{bova_2021,
author = {Bova, Francesco and Goldfarb, Avi and Melko, Roger},
year = {2021},
pages = {},
title = {Commercial applications of quantum computing},
volume = {8},
journal = {EPJ Quantum Technology},
doi = {10.1140/epjqt/s40507-021-00091-1}
}

@article{Preskill_2018,
   title={Quantum Computing in the NISQ era and beyond},
   volume={2},
   ISSN={2521-327X},
   url={http://dx.doi.org/10.22331/q-2018-08-06-79},
   DOI={10.22331/q-2018-08-06-79},
   journal={Quantum},
   publisher={Verein zur Forderung des Open Access Publizierens in den Quantenwissenschaften},
   author={Preskill, John},
   year={2018}, pages={79} }

@article{Arute2019-mb,
  title     = "Quantum supremacy using a programmable superconducting processor",
  author    = "Arute, Frank and Arya, Kunal and Babbush, Ryan and Bacon, Dave
               and Bardin, Joseph C and Barends, Rami and Biswas, Rupak and
               Boixo, Sergio and Brandao, Fernando G S L and Buell, David A and
               Burkett, Brian and Chen, Yu and Chen, Zijun and Chiaro, Ben and
               Collins, Roberto and Courtney, William and Dunsworth, Andrew and
               Farhi, Edward and Foxen, Brooks and Fowler, Austin and Gidney,
               Craig and Giustina, Marissa and Graff, Rob and Guerin, Keith and
               Habegger, Steve and Harrigan, Matthew P and Hartmann, Michael J
               and Ho, Alan and Hoffmann, Markus and Huang, Trent and Humble,
               Travis S and Isakov, Sergei V and Jeffrey, Evan and Jiang, Zhang
               and Kafri, Dvir and Kechedzhi, Kostyantyn and Kelly, Julian and
               Klimov, Paul V and Knysh, Sergey and Korotkov, Alexander and
               Kostritsa, Fedor and Landhuis, David and Lindmark, Mike and
               Lucero, Erik and Lyakh, Dmitry and Mandr{\`a}, Salvatore and
               McClean, Jarrod R and McEwen, Matthew and Megrant, Anthony and
               Mi, Xiao and Michielsen, Kristel and Mohseni, Masoud and Mutus,
               Josh and Naaman, Ofer and Neeley, Matthew and Neill, Charles and
               Niu, Murphy Yuezhen and Ostby, Eric and Petukhov, Andre and
               Platt, John C and Quintana, Chris and Rieffel, Eleanor G and
               Roushan, Pedram and Rubin, Nicholas C and Sank, Daniel and
               Satzinger, Kevin J and Smelyanskiy, Vadim and Sung, Kevin J and
               Trevithick, Matthew D and Vainsencher, Amit and Villalonga,
               Benjamin and White, Theodore and Yao, Z Jamie and Yeh, Ping and
               Zalcman, Adam and Neven, Hartmut and Martinis, John M",
  journal   = "Nature",
  publisher = "Springer Science and Business Media LLC",
  volume    =  574,
  number    =  7779,
  pages     = "505--510",
  year      =  2019,
  language  = "en"
}

@article{Barison_2021,
   title={An efficient quantum algorithm for the time evolution of parameterized circuits},
   volume={5},
   ISSN={2521-327X},
   url={http://dx.doi.org/10.22331/q-2021-07-28-512},
   DOI={10.22331/q-2021-07-28-512},
   journal={Quantum},
   publisher={Verein zur Forderung des Open Access Publizierens in den Quantenwissenschaften},
   author={Barison, Stefano and Vicentini, Filippo and Carleo, Giuseppe},
   year={2021}, pages={512} }

@article{Brown_2010,
   title={Using Quantum Computers for Quantum Simulation},
   volume={12},
   ISSN={1099-4300},
   url={http://dx.doi.org/10.3390/e12112268},
   DOI={10.3390/e12112268},
   number={11},
   journal={Entropy},
   publisher={MDPI AG},
   author={Brown, Katherine L. and Munro, William J. and Kendon, Vivien M.},
   year={2010}, pages={2268–2307} }

@article{McArdle_2020,
   title={Quantum computational chemistry},
   volume={92},
   ISSN={1539-0756},
   url={http://dx.doi.org/10.1103/RevModPhys.92.015003},
   DOI={10.1103/revmodphys.92.015003},
   number={1},
   journal={Reviews of Modern Physics},
   publisher={American Physical Society (APS)},
   author={McArdle, Sam and Endo, Suguru and Aspuru-Guzik, Alán and Benjamin, Simon C. and Yuan, Xiao},
   year={2020}}

@article{Cao_2019,
   title={Quantum Chemistry in the Age of Quantum Computing},
   volume={119},
   ISSN={1520-6890},
   url={http://dx.doi.org/10.1021/acs.chemrev.8b00803},
   DOI={10.1021/acs.chemrev.8b00803},
   number={19},
   journal={Chemical Reviews},
   publisher={American Chemical Society (ACS)},
   author={Cao, Yudong and Romero, Jonathan and Olson, Jonathan P. and Degroote, Matthias and Johnson, Peter D. and Kieferová, Mária and Kivlichan, Ian D. and Menke, Tim and Peropadre, Borja and Sawaya, Nicolas P. D. and Sim, Sukin and Veis, Libor and Aspuru-Guzik, Alán},
   year={2019}, pages={10856–10915} }

@article{Bauer_2020,
   title={Quantum Algorithms for Quantum Chemistry and Quantum Materials Science},
   volume={120},
   ISSN={1520-6890},
   url={http://dx.doi.org/10.1021/acs.chemrev.9b00829},
   DOI={10.1021/acs.chemrev.9b00829},
   number={22},
   journal={Chemical Reviews},
   publisher={American Chemical Society (ACS)},
   author={Bauer, Bela and Bravyi, Sergey and Motta, Mario and Chan, Garnet Kin-Lic},
   year={2020}, pages={12685–12717} }

@article{Kandala_2017,
   title={Hardware-efficient variational quantum eigensolver for small molecules and quantum magnets},
   volume={549},
   ISSN={1476-4687},
   url={http://dx.doi.org/10.1038/nature23879},
   DOI={10.1038/nature23879},
   number={7671},
   journal={Nature},
   publisher={Springer Science and Business Media LLC},
   author={Kandala, Abhinav and Mezzacapo, Antonio and Temme, Kristan and Takita, Maika and Brink, Markus and Chow, Jerry M. and Gambetta, Jay M.},
   year={2017}, pages={242–246} }

@article{Bharti_2022,
   title={Noisy intermediate-scale quantum algorithms},
   volume={94},
   ISSN={1539-0756},
   url={http://dx.doi.org/10.1103/RevModPhys.94.015004},
   DOI={10.1103/revmodphys.94.015004},
   number={1},
   journal={Reviews of Modern Physics},
   publisher={American Physical Society (APS)},
   author={Bharti, Kishor and Cervera-Lierta, Alba and Kyaw, Thi Ha and Haug, Tobias and Alperin-Lea, Sumner and Anand, Abhinav and Degroote, Matthias and Heimonen, Hermanni and Kottmann, Jakob S. and Menke, Tim and Mok, Wai-Keong and Sim, Sukin and Kwek, Leong-Chuan and Aspuru-Guzik, Alán},
   year={2022}}

@article{Cerezo_2021,
   title={Variational quantum algorithms},
   volume={3},
   ISSN={2522-5820},
   url={http://dx.doi.org/10.1038/s42254-021-00348-9},
   DOI={10.1038/s42254-021-00348-9},
   number={9},
   journal={Nature Reviews Physics},
   publisher={Springer Science and Business Media LLC},
   author={Cerezo, M. and Arrasmith, Andrew and Babbush, Ryan and Benjamin, Simon C. and Endo, Suguru and Fujii, Keisuke and McClean, Jarrod R. and Mitarai, Kosuke and Yuan, Xiao and Cincio, Lukasz and Coles, Patrick J.},
   year={2021}, pages={625–644} }

@misc{fedorov2021vqe,
      title={VQE Method: A Short Survey and Recent Developments}, 
      author={Dmitry A. Fedorov and Bo Peng and Niranjan Govind and Yuri Alexeev},
      year={2021},
      eprint={2103.08505},
      archivePrefix={arXiv},
      primaryClass={quant-ph}
}

@article{Tilly_2022,
   title={The Variational Quantum Eigensolver: A review of methods and best practices},
   volume={986},
   ISSN={0370-1573},
   url={http://dx.doi.org/10.1016/j.physrep.2022.08.003},
   DOI={10.1016/j.physrep.2022.08.003},
   journal={Physics Reports},
   publisher={Elsevier BV},
   author={Tilly, Jules and Chen, Hongxiang and Cao, Shuxiang and Picozzi, Dario and Setia, Kanav and Li, Ying and Grant, Edward and Wossnig, Leonard and Rungger, Ivan and Booth, George H. and Tennyson, Jonathan},
   year={2022}, pages={1–128} }

@misc{benfenati2021improved,
      title={Improved accuracy on noisy devices by non-unitary Variational Quantum Eigensolver for chemistry applications}, 
      author={Francesco Benfenati and Guglielmo Mazzola and Chiara Capecci and Panagiotis Kl. Barkoutsos and Pauline J. Ollitrault and Ivano Tavernelli and Leonardo Guidoni},
      year={2021},
      eprint={2101.09316},
      archivePrefix={arXiv},
      primaryClass={quant-ph}
}

@article{Barkoutsos_2018,
   title={Quantum algorithms for electronic structure calculations: Particle-hole Hamiltonian and optimized wave-function expansions},
   volume={98},
   ISSN={2469-9934},
   url={http://dx.doi.org/10.1103/PhysRevA.98.022322},
   DOI={10.1103/physreva.98.022322},
   number={2},
   journal={Physical Review A},
   publisher={American Physical Society (APS)},
   author={Barkoutsos, Panagiotis Kl. and Gonthier, Jerome F. and Sokolov, Igor and Moll, Nikolaj and Salis, Gian and Fuhrer, Andreas and Ganzhorn, Marc and Egger, Daniel J. and Troyer, Matthias and Mezzacapo, Antonio and Filipp, Stefan and Tavernelli, Ivano},
   year={2018}}

@article{Whitfield_2011,
   title={Simulation of electronic structure Hamiltonians using quantum computers},
   volume={109},
   ISSN={1362-3028},
   url={http://dx.doi.org/10.1080/00268976.2011.552441},
   DOI={10.1080/00268976.2011.552441},
   number={5},
   journal={Molecular Physics},
   publisher={Informa UK Limited},
   author={Whitfield, James D. and Biamonte, Jacob and Aspuru-Guzik, Alán},
   year={2011}, pages={735–750} }

@article{Grimsley_2019,
   title={An adaptive variational algorithm for exact molecular simulations on a quantum computer},
   volume={10},
   ISSN={2041-1723},
   url={http://dx.doi.org/10.1038/s41467-019-10988-2},
   DOI={10.1038/s41467-019-10988-2},
   number={1},
   journal={Nature Communications},
   publisher={Springer Science and Business Media LLC},
   author={Grimsley, Harper R. and Economou, Sophia E. and Barnes, Edwin and Mayhall, Nicholas J.},
   year={2019}}

@article{Yordanov_2021,
   title={Qubit-excitation-based adaptive variational quantum eigensolver},
   volume={4},
   ISSN={2399-3650},
   url={http://dx.doi.org/10.1038/s42005-021-00730-0},
   DOI={10.1038/s42005-021-00730-0},
   number={1},
   journal={Communications Physics},
   publisher={Springer Science and Business Media LLC},
   author={Yordanov, Yordan S. and Armaos, V. and Barnes, Crispin H. W. and Arvidsson-Shukur, David R. M.},
   year={2021}}

@misc{rattew2020domainagnostic,
      title={A Domain-agnostic, Noise-resistant, Hardware-efficient Evolutionary Variational Quantum Eigensolver}, 
      author={Arthur G. Rattew and Shaohan Hu and Marco Pistoia and Richard Chen and Steve Wood},
      year={2020},
      eprint={1910.09694},
      archivePrefix={arXiv},
      primaryClass={quant-ph}
}

@article{Tang_2021,
   title={Qubit-ADAPT-VQE: An Adaptive Algorithm for Constructing Hardware-Efficient Ansätze on a Quantum Processor},
   volume={2},
   ISSN={2691-3399},
   url={http://dx.doi.org/10.1103/PRXQuantum.2.020310},
   DOI={10.1103/prxquantum.2.020310},
   number={2},
   journal={PRX Quantum},
   publisher={American Physical Society (APS)},
   author={Tang, Ho Lun and Shkolnikov, V.O. and Barron, George S. and Grimsley, Harper R. and Mayhall, Nicholas J. and Barnes, Edwin and Economou, Sophia E.},
   year={2021} }

@article{ratini_2022,
author = {Ratini, Leonardo and Capecci, Chiara and Benfenati, Francesco and Guidoni, Leonardo},
title = {Wave Function Adapted Hamiltonians for Quantum Computing},
journal = {Journal of Chemical Theory and Computation},
volume = {18},
number = {2},
pages = {899-909},
year = {2022},
doi = {10.1021/acs.jctc.1c01170},
    note ={PMID: 35041784},
URL = {https://doi.org/10.1021/acs.jctc.1c01170},
eprint={https://doi.org/10.1021/acs.jctc.1c01170}
}

@article{Tkachenko_2021,
   title={Correlation-Informed Permutation of Qubits for Reducing Ansatz Depth in the Variational Quantum Eigensolver},
   volume={2},
   ISSN={2691-3399},
   url={http://dx.doi.org/10.1103/PRXQuantum.2.020337},
   DOI={10.1103/prxquantum.2.020337},
   number={2},
   journal={PRX Quantum},
   publisher={American Physical Society (APS)},
   author={Tkachenko, Nikolay V. and Sud, James and Zhang, Yu and Tretiak, Sergei and Anisimov, Petr M. and Arrasmith, Andrew T. and Coles, Patrick J. and Cincio, Lukasz and Dub, Pavel A.},
   year={2021} }

@article{Ganzhorn_2019,
   title={Gate-Efficient Simulation of Molecular Eigenstates on a Quantum Computer},
   volume={11},
   ISSN={2331-7019},
   url={http://dx.doi.org/10.1103/PhysRevApplied.11.044092},
   DOI={10.1103/physrevapplied.11.044092},
   number={4},
   journal={Physical Review Applied},
   publisher={American Physical Society (APS)},
   author={Ganzhorn, M. and Egger, D.J. and Barkoutsos, P. and Ollitrault, P. and Salis, G. and Moll, N. and Roth, M. and Fuhrer, A. and Mueller, P. and Woerner, S. and Tavernelli, I. and Filipp, S.},
   year={2019} }

@article{White1992,
  title = {Density matrix formulation for quantum renormalization groups},
  author = {White, Steven R.},
  journal = {Phys. Rev. Lett.},
  volume = {69},
  issue = {19},
  pages = {2863--2866},
  numpages = {0},
  year = {1992},
  publisher = {American Physical Society},
  doi = {10.1103/PhysRevLett.69.2863},
  url = {https://link.aps.org/doi/10.1103/PhysRevLett.69.2863}
}

@article{White93,
  title = {Density-matrix algorithms for quantum renormalization groups},
  author = {White, Steven R.},
  journal = {Phys. Rev. B},
  volume = {48},
  issue = {14},
  pages = {10345--10356},
  numpages = {0},
  year = {1993},
  publisher = {American Physical Society},
  doi = {10.1103/PhysRevB.48.10345},
  url = {https://link.aps.org/doi/10.1103/PhysRevB.48.10345}
}

@article{Schollwock2005,
  title = {The density-matrix renormalization group},
  author = {Schollw\"ock, U.},
  journal = {Rev. Mod. Phys.},
  volume = {77},
  issue = {1},
  pages = {259--315},
  numpages = {0},
  year = {2005},
  publisher = {American Physical Society},
  doi = {10.1103/RevModPhys.77.259},
  url = {https://link.aps.org/doi/10.1103/RevModPhys.77.259}
}

@article{SCHOLLWOCK2011,
title = {The density-matrix renormalization group in the age of matrix product states},
journal = {Annals of Physics},
volume = {326},
number = {1},
pages = {96-192},
year = {2011},
note = {January 2011 Special Issue},
issn = {0003-4916},
doi = {https://doi.org/10.1016/j.aop.2010.09.012},
url = {https://www.sciencedirect.com/science/article/pii/S0003491610001752},
author = {Ulrich Schollwöck}
}

@article{Baiardi_2020,
   title={The density matrix renormalization group in chemistry and molecular physics: Recent developments and new challenges},
   volume={152},
   ISSN={1089-7690},
   url={http://dx.doi.org/10.1063/1.5129672},
   DOI={10.1063/1.5129672},
   number={4},
   journal={The Journal of Chemical Physics},
   publisher={AIP Publishing},
   author={Baiardi, Alberto and Reiher, Markus},
   year={2020}}

@article{Liu_2022,
   title={Layer VQE: A Variational Approach for Combinatorial Optimization on Noisy Quantum Computers},
   volume={3},
   ISSN={2689-1808},
   url={http://dx.doi.org/10.1109/TQE.2021.3140190},
   DOI={10.1109/tqe.2021.3140190},
   journal={IEEE Transactions on Quantum Engineering},
   publisher={Institute of Electrical and Electronics Engineers (IEEE)},
   author={Liu, Xiaoyuan and Angone, Anthony and Shaydulin, Ruslan and Safro, Ilya and Alexeev, Yuri and Cincio, Lukasz},
   year={2022},
   pages={1–20} }

@article{Hoffmann_1988,
    author = {Hoffmann, Mark R. and Simons, Jack},
    title = "{A unitary multiconfigurational coupled‐cluster method: Theory and applications}",
    journal = {The Journal of Chemical Physics},
    volume = {88},
    number = {2},
    pages = {993-1002},
    year = {1988},
    issn = {0021-9606},
    doi = {10.1063/1.454125},
    url = {https://doi.org/10.1063/1.454125},
    eprint = {https://pubs.aip.org/aip/jcp/article-pdf/88/2/993/11189139/993\_1\_online.pdf},
}

@article{Cooper_2010,
    author = {Cooper, Bridgette and Knowles, Peter J.},
    title = "{Benchmark studies of variational, unitary and extended coupled cluster methods}",
    journal = {The Journal of Chemical Physics},
    volume = {133},
    number = {23},
    pages = {234102},
    year = {2010},
    issn = {0021-9606},
    doi = {10.1063/1.3520564},
    url = {https://doi.org/10.1063/1.3520564},
    eprint = {https://pubs.aip.org/aip/jcp/article-pdf/doi/10.1063/1.3520564/15434716/234102\_1\_online.pdf},
}

@article{Evangelista_2011,
    author = {Evangelista, Francesco A.},
    title = "{Alternative single-reference coupled cluster approaches for multireference problems: The simpler, the better}",
    journal = {The Journal of Chemical Physics},
    volume = {134},
    number = {22},
    pages = {224102},
    year = {2011},
    issn = {0021-9606},
    doi = {10.1063/1.3598471},
    url = {https://doi.org/10.1063/1.3598471},
    eprint = {https://pubs.aip.org/aip/jcp/article-pdf/doi/10.1063/1.3598471/15437394/224102\_1\_online.pdf},
}

@misc{romero2018strategies,
      title="{Strategies for quantum computing molecular energies using the unitary coupled cluster ansatz}", 
      author={Jonathan Romero and Ryan Babbush and Jarrod R. McClean and Cornelius Hempel and Peter Love and Alán Aspuru-Guzik},
      year={2018},
      eprint={1701.02691},
      archivePrefix={arXiv},
      primaryClass={quant-ph}
}

@article{ITensor,
	title={{The ITensor Software Library for Tensor Network Calculations}},
	author={Matthew Fishman and Steven R. White and E. Miles Stoudenmire},
	journal={SciPost Phys. Codebases},
	pages={4},
	year={2022},
	publisher={SciPost},
	doi={10.21468/SciPostPhysCodeb.4},
	url={https://scipost.org/10.21468/SciPostPhysCodeb.4},
}

@article{ITensor-r0.3,
	title={{Codebase release 0.3 for ITensor}},
	author={Matthew Fishman and Steven R. White and E. Miles Stoudenmire},
	journal={SciPost Phys. Codebases},
	pages={4-r0.3},
	year={2022},
	publisher={SciPost},
	doi={10.21468/SciPostPhysCodeb.4-r0.3},
	url={https://scipost.org/10.21468/SciPostPhysCodeb.4-r0.3},
}

@misc{Qiskit,
    author = {{Qiskit contributors}},
    title = {Qiskit: An Open-source Framework for Quantum Computing},
    year = {2023},
    doi = {10.5281/zenodo.2573505}
}

@article{vonneumann1996,
   author = {John von Neumann},
   doi = {10.1007/978-3-642-61409-5},
   journal = {Mathematische Grundlagen der Quantenmechanik},
   publisher = {Springer Berlin Heidelberg},
   title = {Mathematische Grundlagen der Quantenmechanik},
   year = {1996},
}

@article{Magoulas2023,
   author = {Ilias Magoulas and Francesco A. Evangelista},
   doi = {10.1021/acs.jpca.3c02781},
   issn = {1089-5639},
   issue = {31},
   journal = {The Journal of Physical Chemistry A},
   month = {8},
   pages = {6567-6576},
   title = {Unitary Coupled Cluster: Seizing the Quantum Moment},
   volume = {127},
   year = {2023},
}

@Book{Flet87,
  Title                    = {Practical Methods of Optimization},
  Author                   = {Roger Fletcher},
  Publisher                = {John Wiley \& Sons},
  Year                     = {1987},

  Address                  = {New York, NY, USA},
  Edition                  = {Second}
}

@article{Ding_2020,
   title={Concept of Orbital Entanglement and Correlation in Quantum Chemistry},
   volume={17},
   ISSN={1549-9626},
   url={http://dx.doi.org/10.1021/acs.jctc.0c00559},
   DOI={10.1021/acs.jctc.0c00559},
   number={1},
   journal={Journal of Chemical Theory and Computation},
   publisher={American Chemical Society (ACS)},
   author={Ding, Lexin and Mardazad, Sam and Das, Sreetama and Szalay, Szilárd and Schollwöck, Ulrich and Zimborás, Zoltán and Schilling, Christian},
   year={2020},
   month=dec, pages={79–95} }

@article{Zhang_2021,
   title={Mutual information-assisted adaptive variational quantum eigensolver},
   volume={6},
   ISSN={2058-9565},
   url={http://dx.doi.org/10.1088/2058-9565/abdca4},
   DOI={10.1088/2058-9565/abdca4},
   number={3},
   journal={Quantum Science and Technology},
   publisher={IOP Publishing},
   author={Zhang, Zi-Jian and Kyaw, Thi Ha and Kottmann, Jakob S and Degroote, Matthias and Aspuru-Guzik, Alán},
   year={2021},
   month=jul, pages={035001} }

@misc{larocca2024review,
      title={A Review of Barren Plateaus in Variational Quantum Computing}, 
      author={Martin Larocca and Supanut Thanasilp and Samson Wang and Kunal Sharma and Jacob Biamonte and Patrick J. Coles and Lukasz Cincio and Jarrod R. McClean and Zoë Holmes and M. Cerezo},
      year={2024},
      eprint={2405.00781},
      archivePrefix={arXiv},
      primaryClass={quant-ph}
}

@article{Wiersema2020,
  title = {Exploring Entanglement and Optimization within the Hamiltonian Variational Ansatz},
  author = {Wiersema, Roeland and Zhou, Cunlu and de Sereville, Yvette and Carrasquilla, Juan Felipe and Kim, Yong Baek and Yuen, Henry},
  journal = {PRX Quantum},
  volume = {1},
  issue = {2},
  pages = {020319},
  numpages = {14},
  year = {2020},
  publisher = {American Physical Society},
  doi = {10.1103/PRXQuantum.1.020319},
  url = {https://link.aps.org/doi/10.1103/PRXQuantum.1.020319}
}

@Article{WeiHo2019,
	title={{Efficient variational simulation of non-trivial quantum states}},
	author={Wen Wei Ho and Timothy H. Hsieh},
	journal={SciPost Phys.},
	volume={6},
	pages={029},
	year={2019},
	publisher={SciPost},
	doi={10.21468/SciPostPhys.6.3.029},
	url={https://scipost.org/10.21468/SciPostPhys.6.3.029},
}

@article{Cade2020,
  title = {Strategies for solving the Fermi-Hubbard model on near-term quantum computers},
  author = {Cade, Chris and Mineh, Lana and Montanaro, Ashley and Stanisic, Stasja},
  journal = {Phys. Rev. B},
  volume = {102},
  issue = {23},
  pages = {235122},
  numpages = {25},
  year = {2020},
  publisher = {American Physical Society},
  doi = {10.1103/PhysRevB.102.235122},
  url = {https://link.aps.org/doi/10.1103/PhysRevB.102.235122}
}

@misc{bosse2021probinggroundstateproperties,
      title={Probing ground state properties of the kagome antiferromagnetic Heisenberg model using the Variational Quantum Eigensolver}, 
      author={Jan Lukas Bosse and Ashley Montanaro},
      year={2021},
      eprint={2108.08086},
      archivePrefix={arXiv},
      primaryClass={quant-ph},
      url={https://arxiv.org/abs/2108.08086}, 
}

@article{Kattem_lle_2022,
   title={Variational quantum eigensolver for the Heisenberg antiferromagnet on the kagome lattice},
   volume={106},
   ISSN={2469-9969},
   url={http://dx.doi.org/10.1103/PhysRevB.106.214429},
   DOI={10.1103/physrevb.106.214429},
   number={21},
   journal={Physical Review B},
   publisher={American Physical Society (APS)},
   author={Kattemölle, Joris and van Wezel, Jasper},
   year={2022},
   month=dec }

@Article{Kokail2019,
author={Kokail, C.
and Maier, C.
and van Bijnen, R.
and Brydges, T.
and Joshi, M. K.
and Jurcevic, P.
and Muschik, C. A.
and Silvi, P.
and Blatt, R.
and Roos, C. F.
and Zoller, P.},
title={Self-verifying variational quantum simulation of lattice models},
journal={Nature},
year={2019},
day={01},
volume={569},
number={7756},
pages={355-360},
issn={1476-4687},
doi={10.1038/s41586-019-1177-4},
url={https://doi.org/10.1038/s41586-019-1177-4}
}

\appendix
\counterwithin{figure}{section}
\renewcommand{\thefigure}{\thesection\arabic{figure}}
\counterwithin{table}{section}
\renewcommand{\thetable}{\thesection\arabic{table}}
\begin{appendices}
\section{Results collection}
\begin{figure*}[!htb]
    \centering
    \begin{subfigure}[b]{.30\linewidth}
    \includegraphics[width=\linewidth]{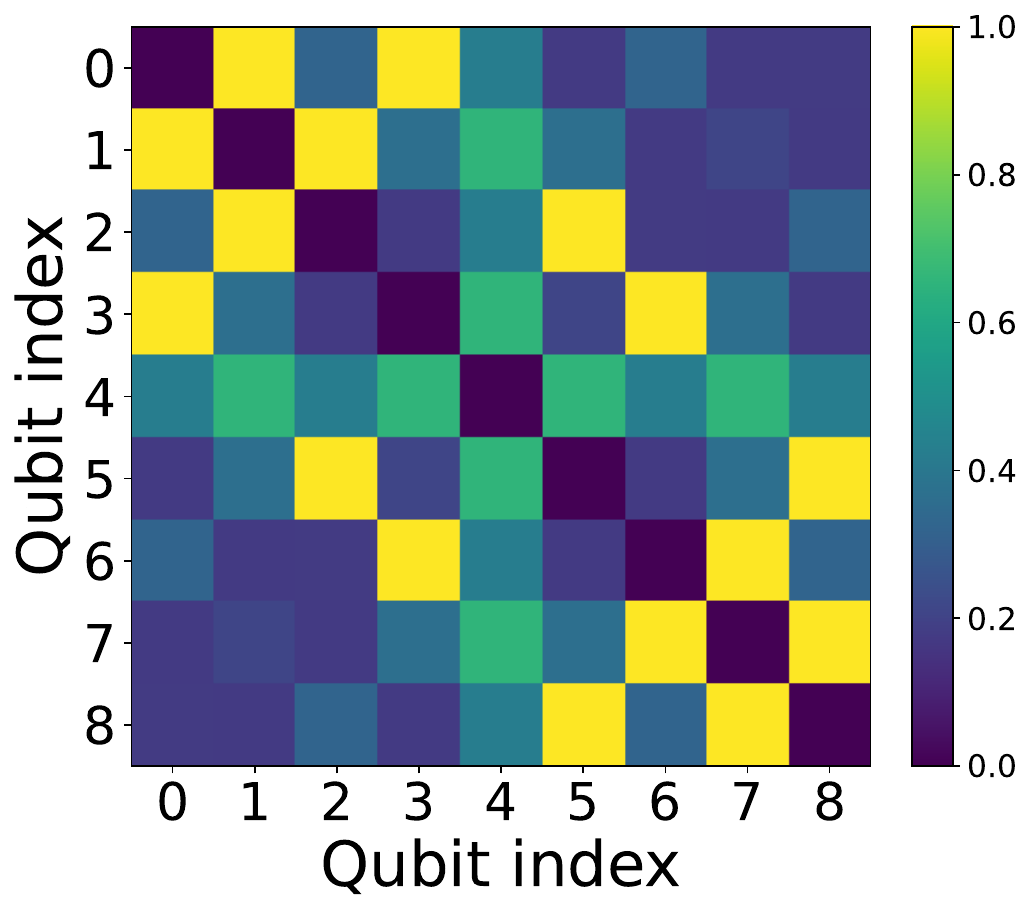}
    \caption{}\label{appendix:fig_3x3}
    \end{subfigure}
    \begin{subfigure}[b]{.30\linewidth}
    \includegraphics[width=\linewidth]{images/3x4_mi_map.pdf}
    \caption{}\label{appendix:fig_3x4}
    \end{subfigure}
    \begin{subfigure}[b]{.30\linewidth}
    \includegraphics[width=\linewidth]{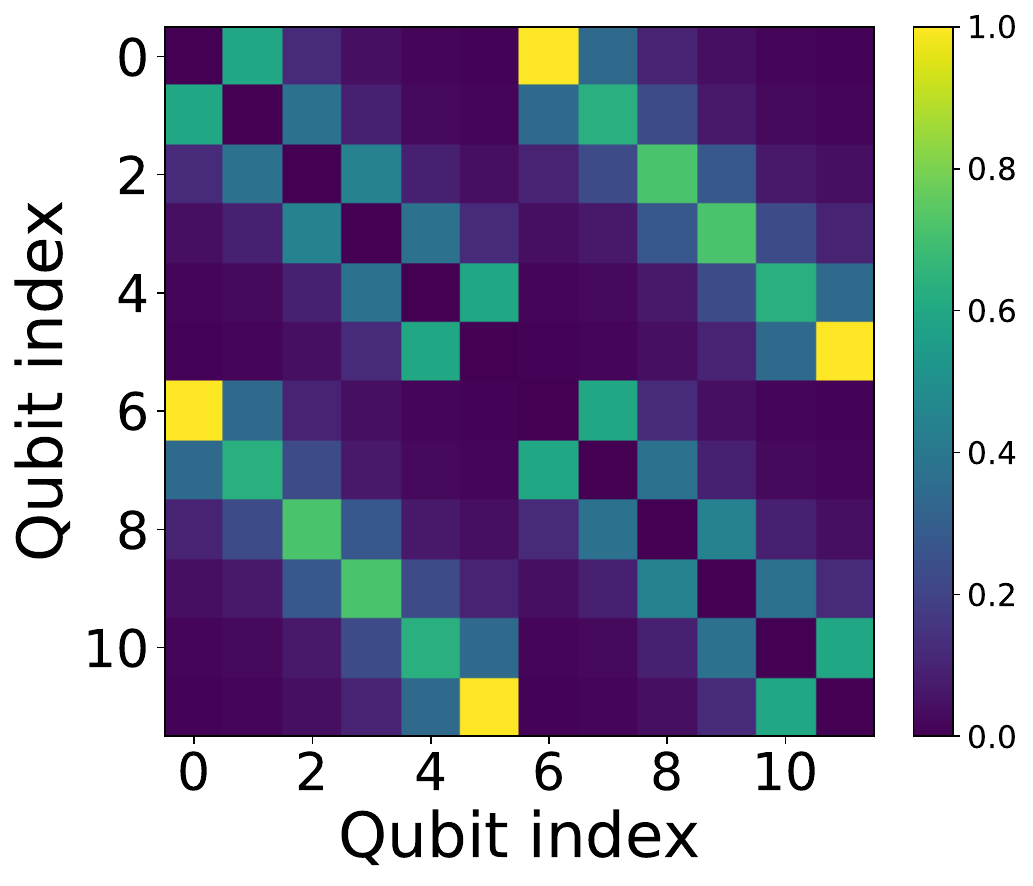}
    \caption{}\label{appendix:fig_2x6}
    \end{subfigure}
    \begin{subfigure}[b]{.30\linewidth}
    \includegraphics[width=\linewidth]{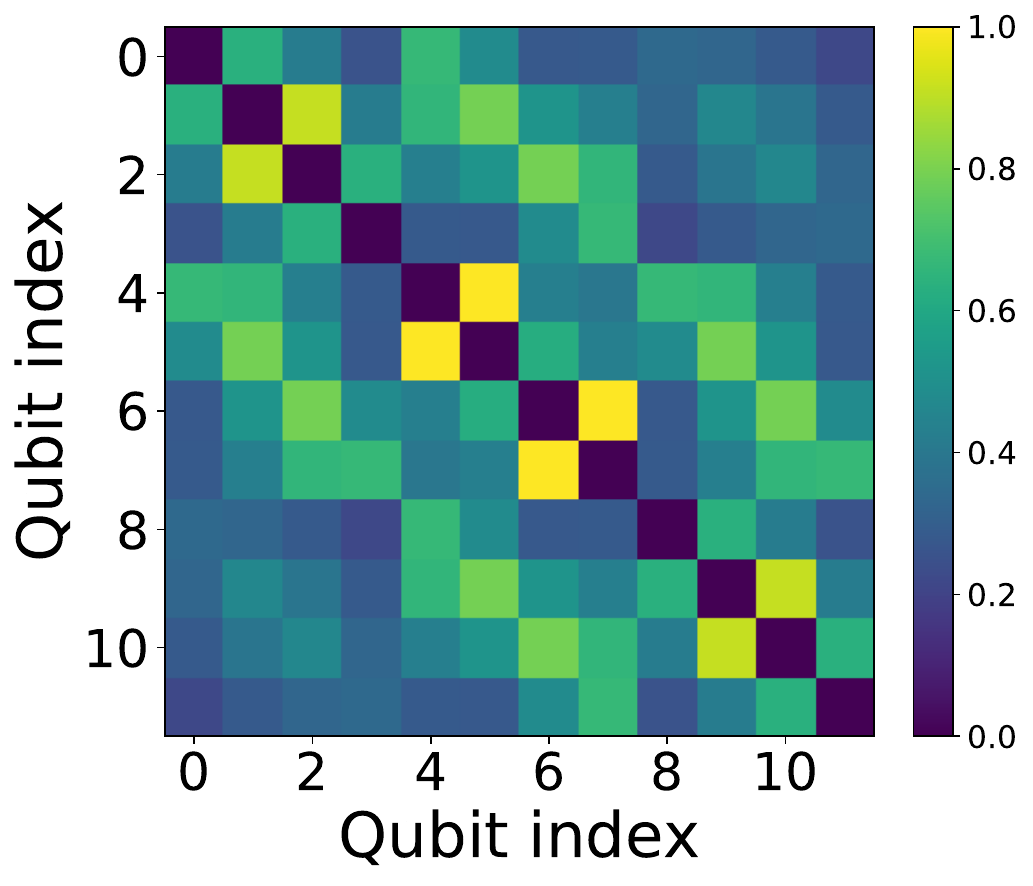}
    \caption{}\label{appendix:fig_3x4h}
    \end{subfigure}
    \begin{subfigure}[b]{.30\linewidth}
    \includegraphics[width=\linewidth]{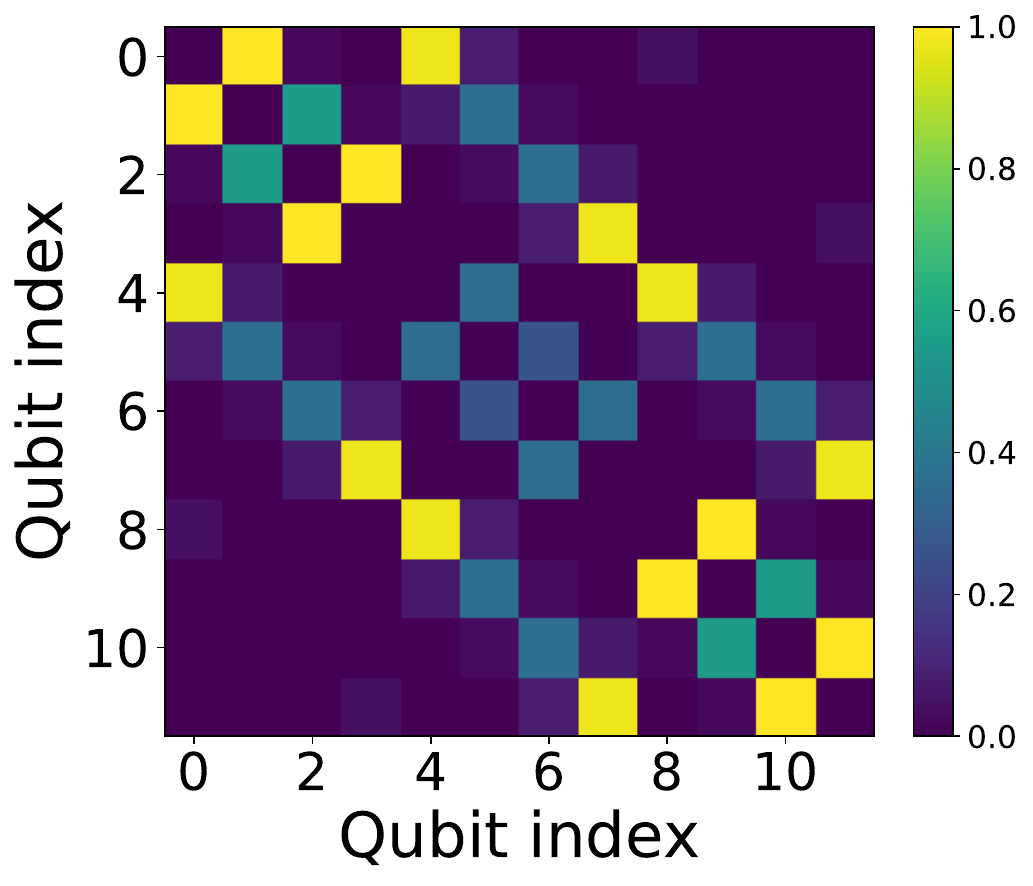}
    \caption{}\label{appendix:fig_3x4_0_1}
    \end{subfigure}
    \begin{subfigure}[b]{.30\linewidth}
    \includegraphics[width=\linewidth]{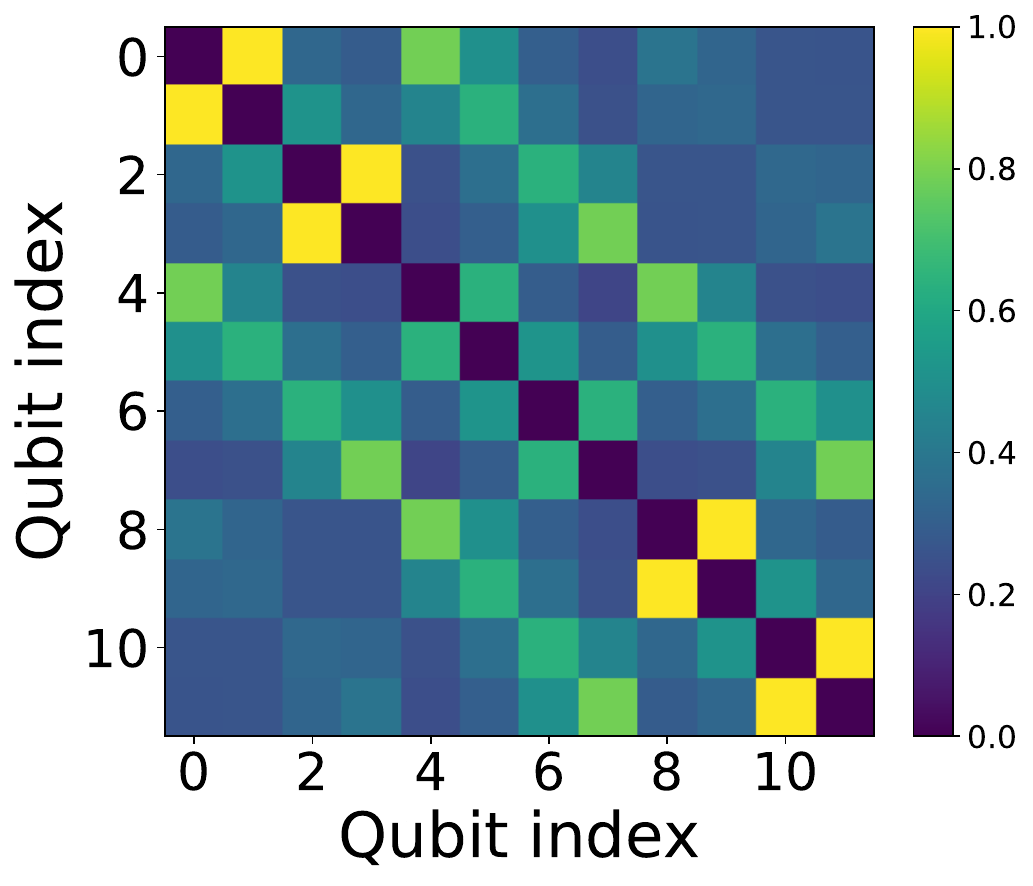}
    \caption{}\label{appendix:fig_3x4_0_66}
    \end{subfigure}
    \caption{QMI matrices obtained by the tested systems. Maps (a)-(b)-(c) refer to isotropic Heisenberg systems, (d) to isotropic Heisenberg with external magnetic field $h=2$, (e)-(f) to anisotropic Heisenberg with $\Delta = 1/10$ and $\Delta=2/3$, respectively.}
\label{appendix:fig_QMIs}
\end{figure*}

\begin{figure*}[!htb]
\centering
    \begin{subfigure}[b]{.450\linewidth}
    \includegraphics[width=\linewidth]{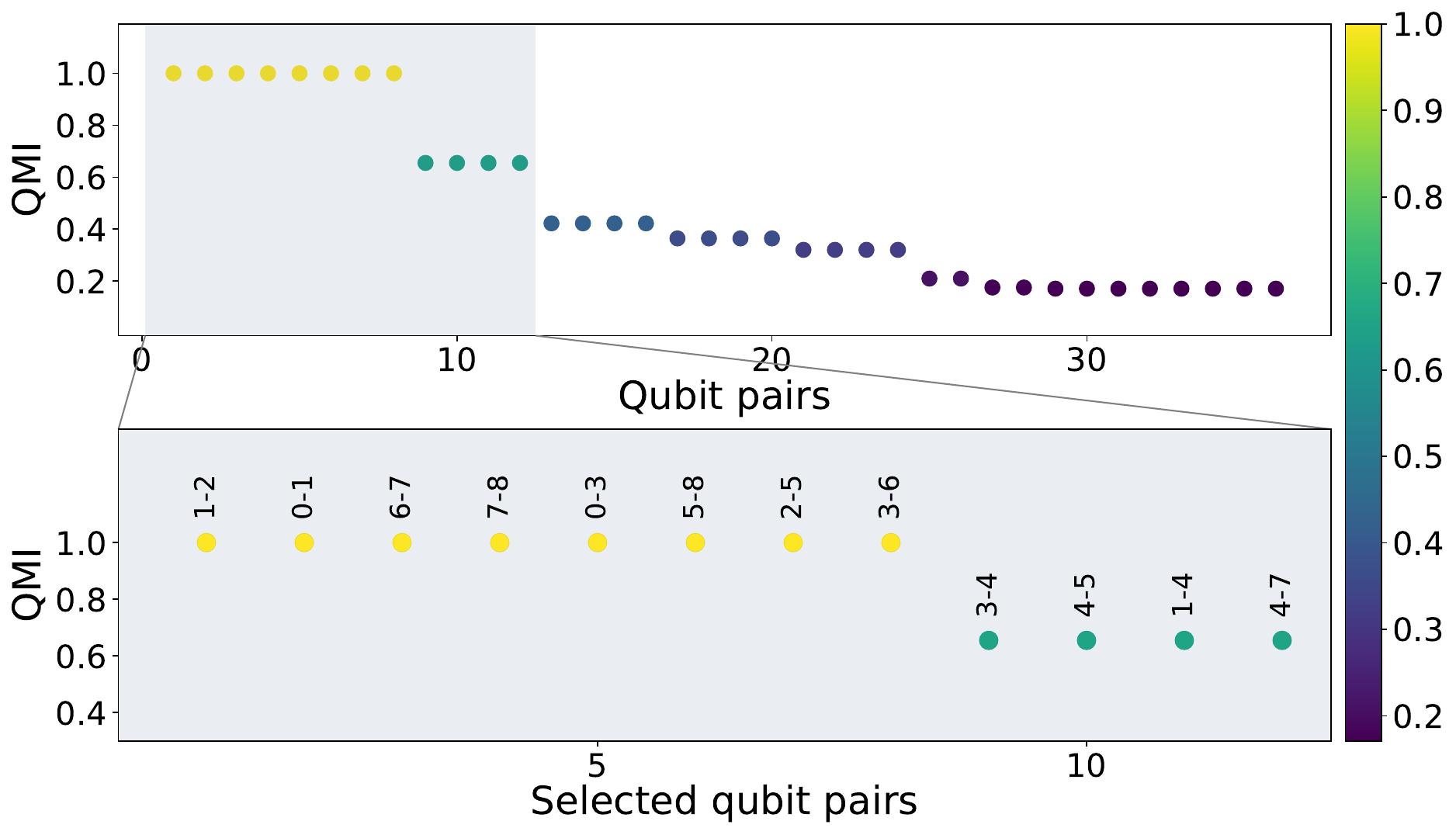}
    \caption{$3\times3$ isotropic Heisenberg Hamiltonian QMI value pairing selection.}\label{appendix:fig_sel3x3}
    \end{subfigure}\hfill
    \begin{subfigure}[b]{.450\linewidth}
    \includegraphics[width=\linewidth]{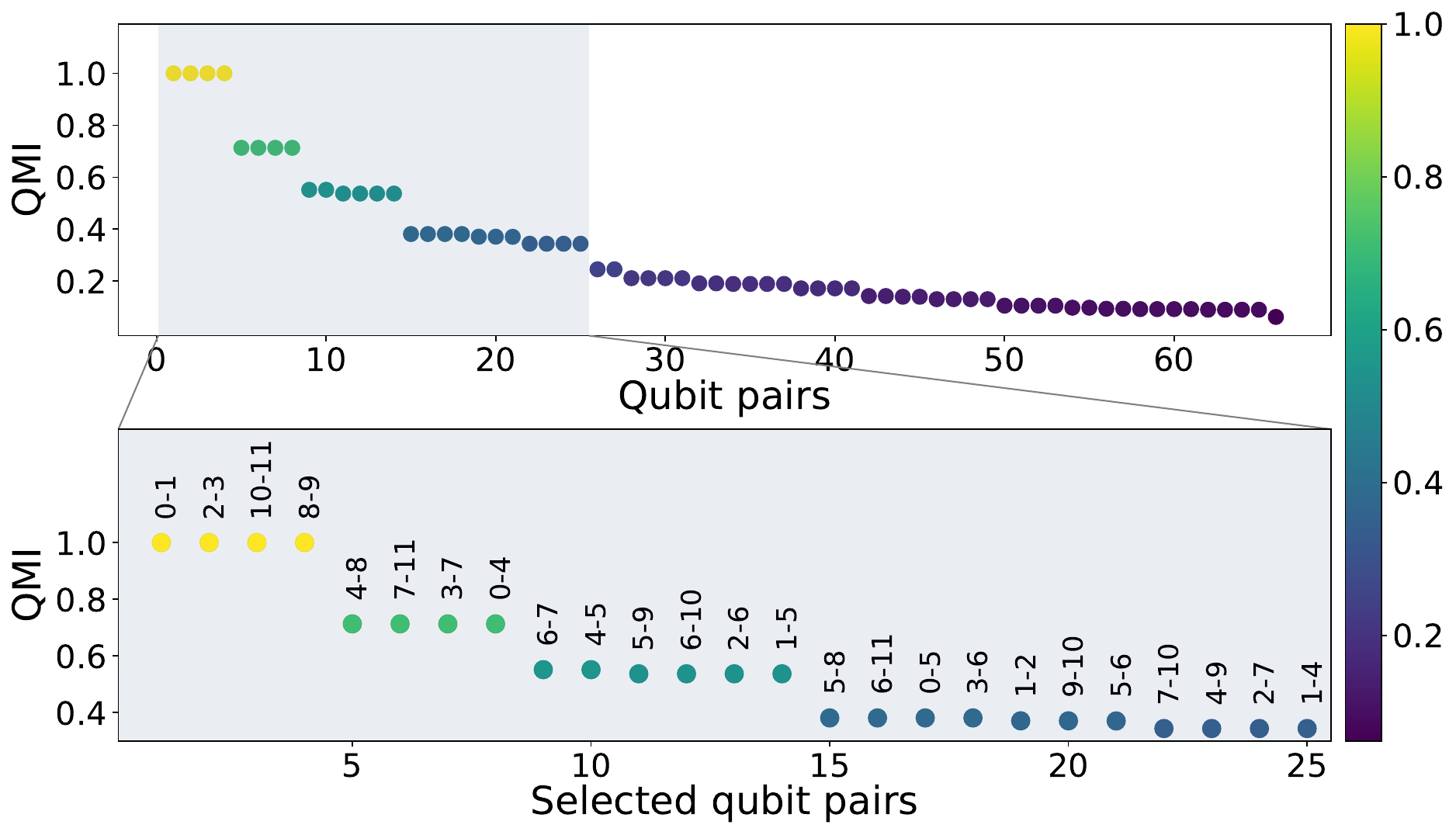}
    \caption{$3\times4$ isotropic Heisenberg Hamiltonian QMI value pairing selection.}\label{appendix:fig_sel3x4}
    \end{subfigure}
    \begin{subfigure}[b]{.450\linewidth}
    \includegraphics[width=\linewidth]{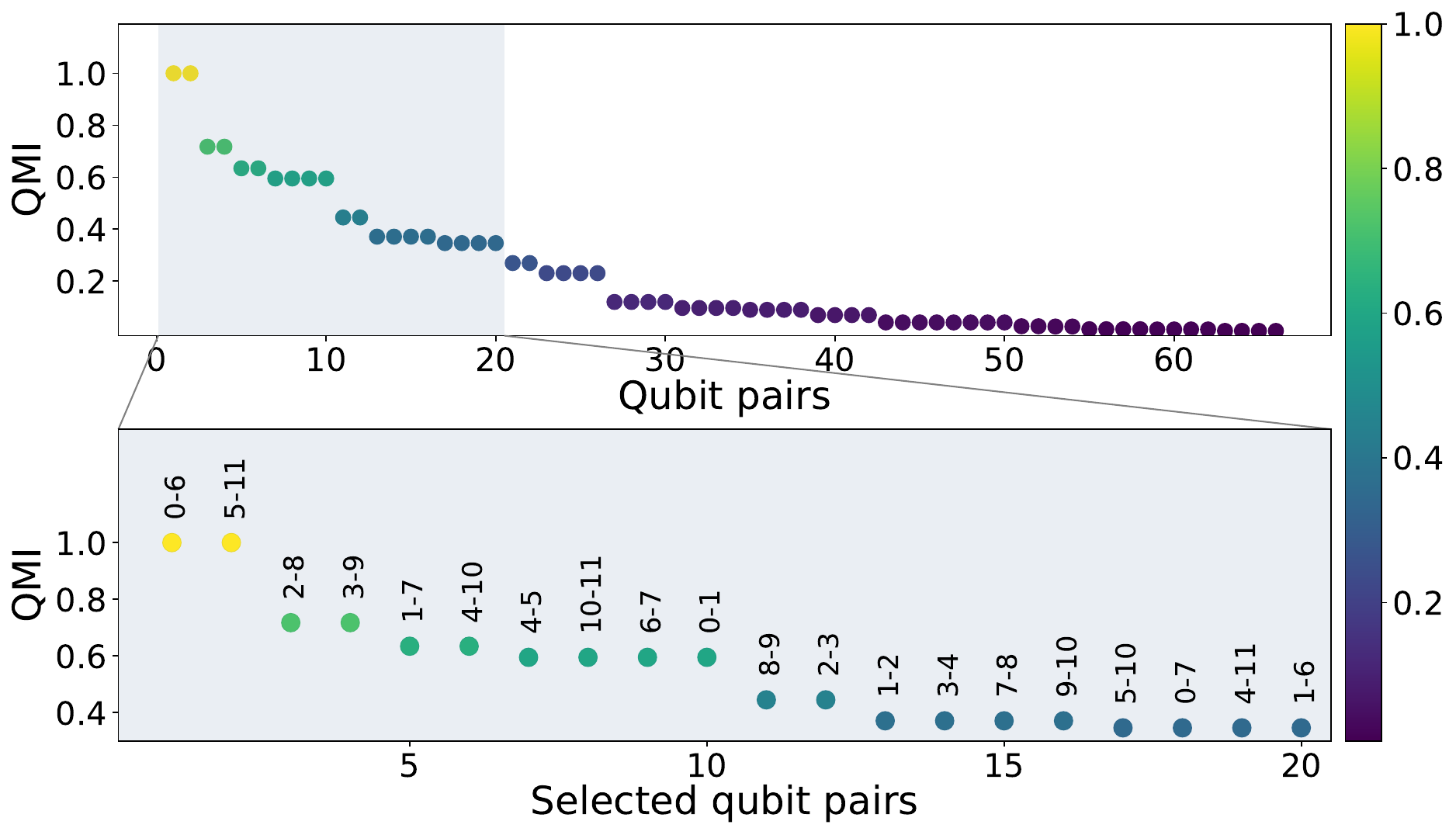}
    \caption{$2\times6$ isotropic Heisenberg Hamiltonian QMI pairings  selection.}\label{appendix:fig_sel2x6}
    \end{subfigure}\hfill
    \begin{subfigure}[b]{.450\linewidth}
    \includegraphics[width=\linewidth]{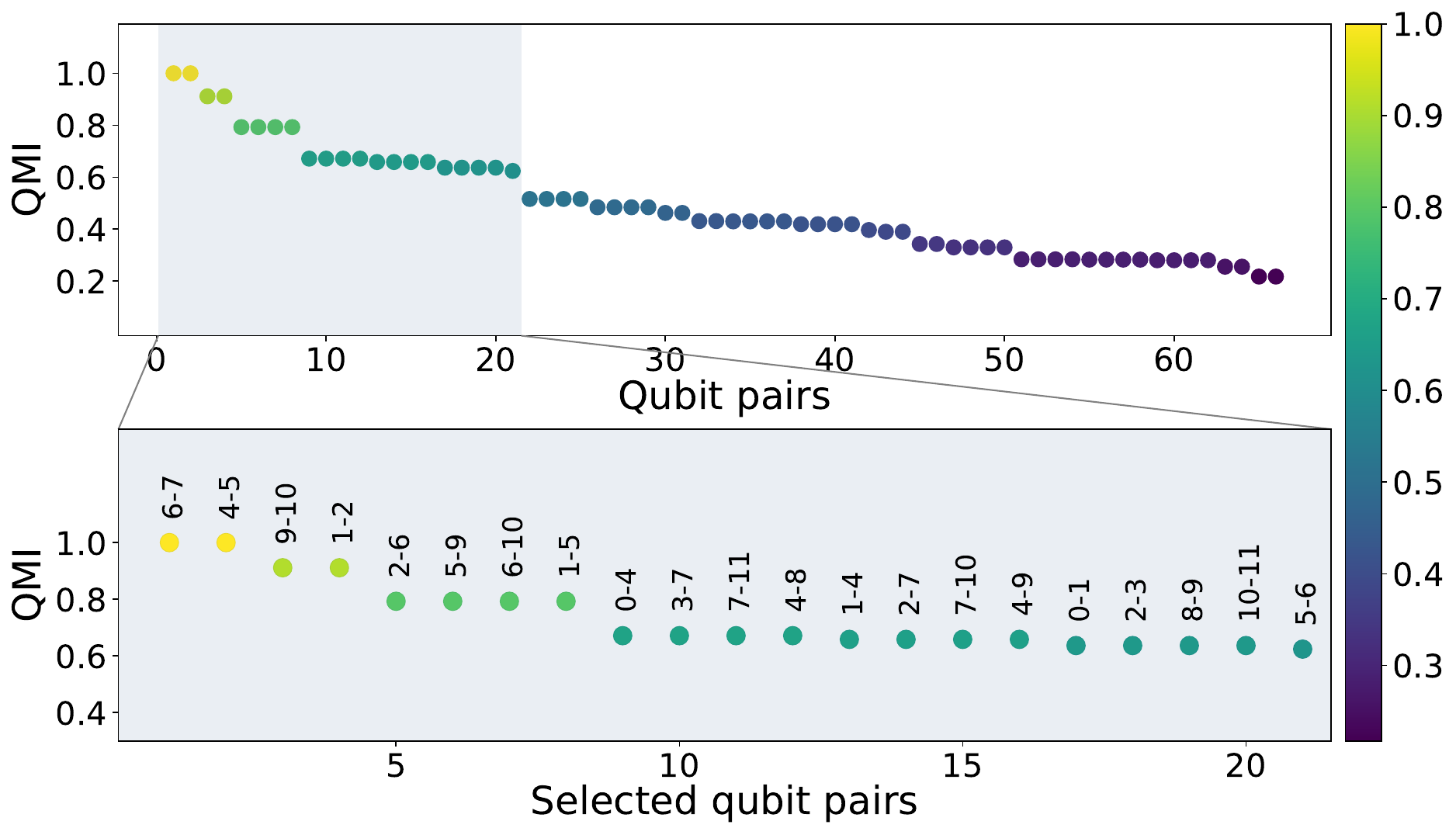}
    \caption{$3\times4$ isotropic Heisenberg Hamiltonian  with external magnetic field $h=2$ QMI entangling pairings selection.}\label{appendix:fig_sel3x4h}
    \end{subfigure}
    \begin{subfigure}[b]{.450\linewidth}
    \includegraphics[width=\linewidth]{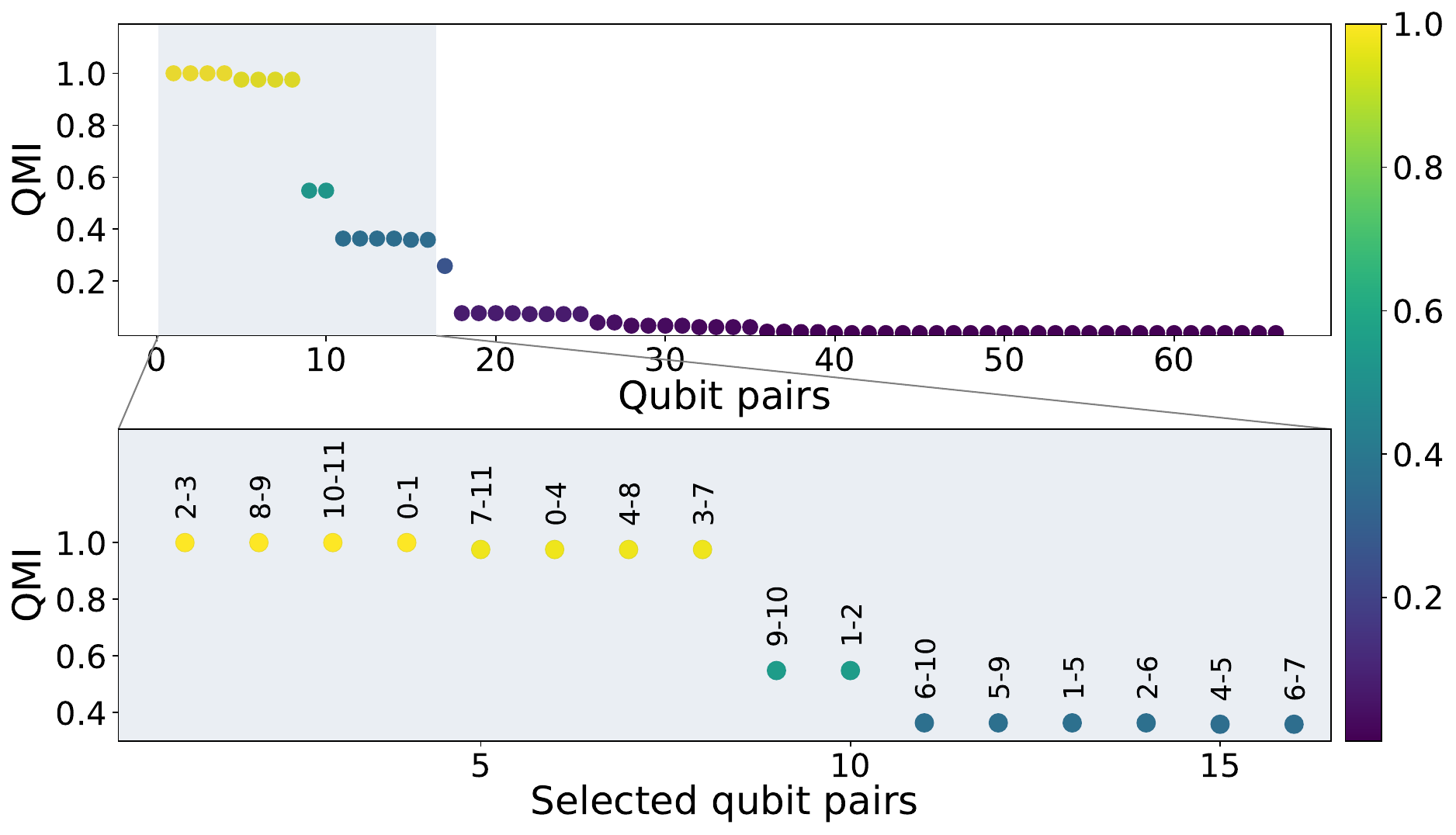}
    \caption{Anisotropic $3\times4$ Heisenberg Hamiltonian with $\Delta=1/10$ QMI value pairing selection.}\label{appendix:fig_sel3x4_0_1}
    \end{subfigure}\hfill
    \begin{subfigure}[b]{.450\linewidth}
    \includegraphics[width=\linewidth]{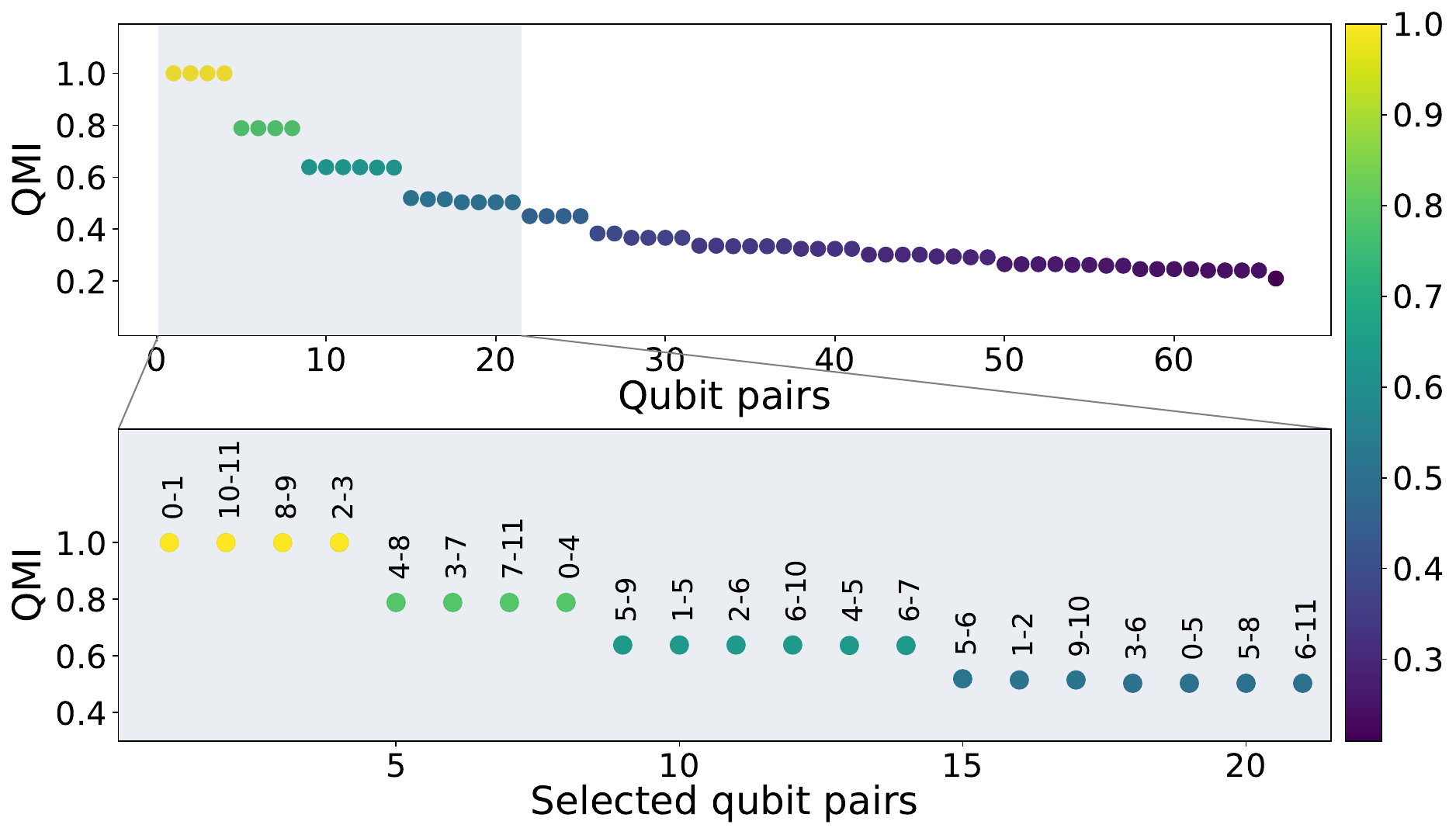}
    \caption{Anisotropic $3\times4$ Heisenberg Hamiltonian with $\Delta=2/3$ QMI value pairing selection.}\label{appendix:fig_sel3x4_0_66}
    \end{subfigure}
    \caption{Qubit-Pairs selection plot. For each image, the higher plot shows all the qubit-pairings in the system. The lower plot is a close-up on the couplers that have been selected by the first phase of the Multi-QIDA layers-building procedure. The qubit-pairs obtained from Algorithm \ref{algo:sel_pseudocode} are shown in Table \ref{appendix:tab_multi_qida_conf}.}
\label{appendix:fig_pairs_collected}
\end{figure*}

\begin{table*}[!htb]
    \centering
    \fontsize{8pt}{8pt}
    \setlength\tabcolsep{2pt}
    \begin{tabular}{|c||c|c|c|c|}
    \hline
    $System$&$Layer$ $1$&$Layer$ $2$&$Layer$ $2$&$Layer$ $4$\\
    \hline\hline
        $3\times3$&\makecell{$[0,1],[0,3],[1,2],[2,5],$\\$[5,8],[7,8],[6,7],[3,6]$}&$[1,4],[3,4],[4,5],[4,7]$&-&-\\\hline
         $2\times6$&\makecell{$[0,6],[1,7],[2,8],[3,9],$\\$[4,10],[5,11](\text{\textbf{*}})$}&$[0,1],[4,5],[6,7],[10,11]$&$[2,3],[8,9]$&$[1,2],[3,4],[7,8],[9,10]$\\\hline
         $3\times4$&$[0,1],[2,3],[8,9],[10,11]$&$[0,4],[3,7],[4,8],[7,11]$&\makecell{$[1,5],[2,6],[4,5],[5,9],$\\$[6,7],[6,10]$}&$[1,2],[5,6],[9,10]$\\\hline
         $3\times4,h=2$&$[1,2],[4,5],[6,7],[9,10]$&$[1,5],[2,6],[5,9],[6,10]$&\makecell{$[0,1],[0,4],[2,3],[3,7],$\\$[4,8],[7,11],[8,9],[10,11]$}&-\\\hline
         $3\times4, \Delta=1/10$&\makecell{$[0,1],[0,4],[2,3],[3,7],$\\$[4,8],[7,11],[8,9],[10,11]$}&$[1,2],[9,10]$&\makecell{$[1,5],[2,6],[4,5],[5,9],$\\$[6,7],[6,10]$}&-\\\hline
         $3\times4, \Delta=2/3$&$[0,1],[2,3],[8,9],[10,11]$&$[0,4],[3,7],[4,8],[7,11]$&\makecell{$[1,5],[2,6],[4,5],[6,7],$\\$[5,9],[6,10]$}&$[1,2],[5,6],[9,10]$\\
         \hline
    \end{tabular}
    
    \caption{Entangling map for each Multi-QIDA layer divided by system configurations. (\textbf{*}) in this configuration, the grouping optimization is used allowing the contraction of 3 layers into a unique layer. The selection has been done using a finesse ratio set to $\mu = 0.1$. Upon these layers, each Multi-QIDA configuration is completed with an additional ladder layer. In $(QIDA)^{CX}$ configuration, every layer after the first one is applied in $V$-shape i.e. only the first half of the full additional layers is shown. }
    \label{appendix:tab_multi_qida_conf}
\end{table*}

\begin{table}[!htb]
\fontsize{10pt}{10pt}
    \centering
    \scalebox{.82}{
    \setlength\tabcolsep{2pt}
    \begin{tabular}{|c||c|c|c|c|c|c|}
    \hline
        $\empty$&$3\times3$&$2\times6$&$3\times4$&\makecell{$3\times4$\\$h=2$}&\makecell{$3\times4$\\$\Delta=2/3$}&\makecell{$3\times4$\\$ \Delta=1/10$}\\
        \hline\hline
        $(L)_{4}^{CX}$&$32$&$44$&$44$&$44$&$44$&$44$\\
        $(L)_{5}^{CX}$&$40$&$55$&$55$&$55$&$55$&$55$\\
        $(L)_{6}^{CX}$&$-$&$-$&$66$&$-$&$66$&$-$\\\hline
        $(QIDA)^{CX}$&$32$&$48$&$52$&$50$&$52$&$46$\\
        $(QIDA)^{SO4}$&$40$&$54$&$56$&$54$&$56$&$54$\\
        \hline
    \end{tabular}}
    \caption{Number of CNOTs for every system and configuration.$(QIDA)^{CX}$ and $(QIDA)^{SO4}$ represent QIDA configuration with CNOTs and \textbf{SO}(4) gates,respectively, while $(L)^{CX}_d$ describe ladder ansatz with depth $d$ and CNOT as entangling gates.}
    \label{appendix:tab_CNOT_numbers}
\end{table}

\begin{figure*}[!htb]
    \centering
    \begin{subfigure}[b]{.480\linewidth}
    \includegraphics[width=\linewidth]{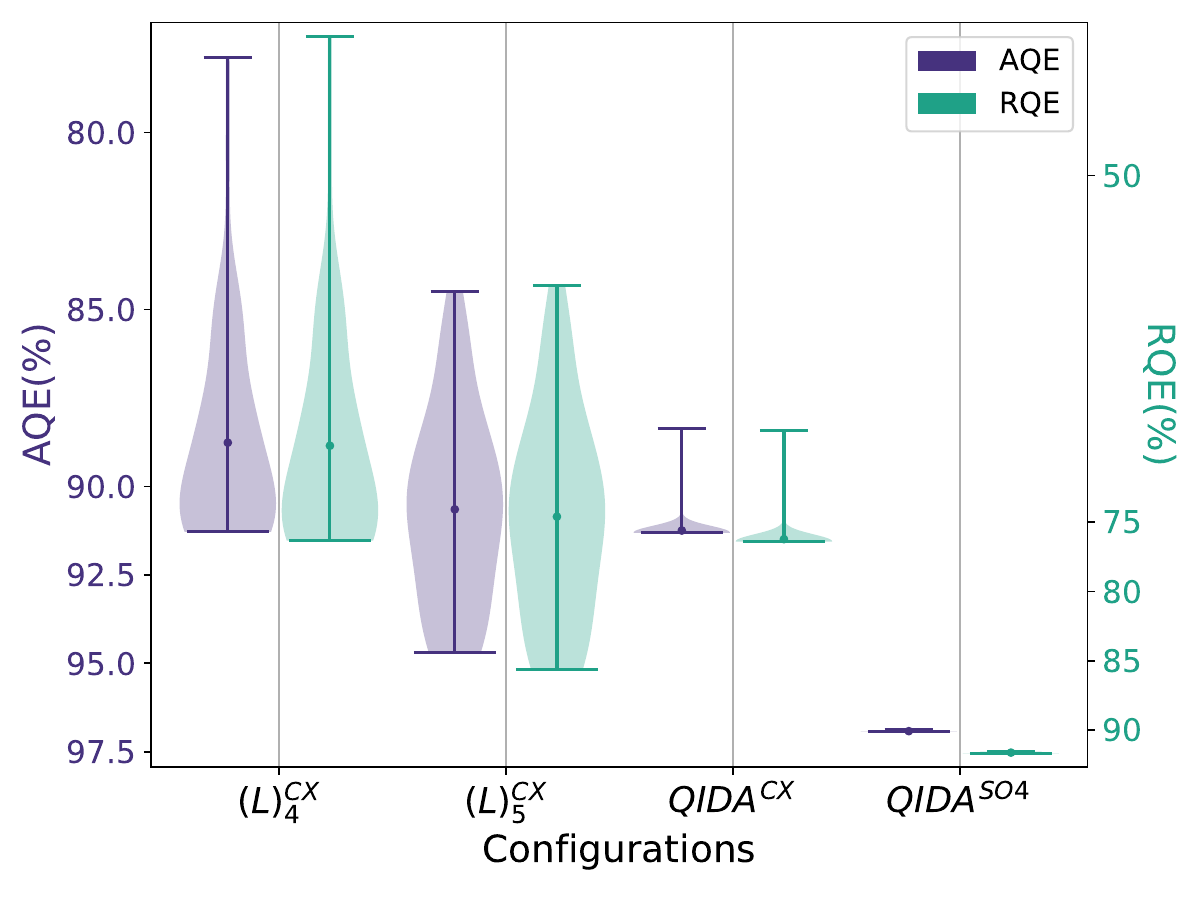}
    \caption{$3\times3$ Isotropic Heisenberg model system results.\\
    HEA: 32,40. Multi-QIDA: 32,40.}\label{appendix:fig_3x3_violin}
    \end{subfigure}\hfill
    \begin{subfigure}[b]{.480\linewidth}
    \includegraphics[width=\linewidth]{images/3x4_bigger.pdf}
    \caption{$3\times4$ Isotropic Heisenberg model system results.\\
    HEA: 44,55,66. Multi-QIDA: 52,56. }\label{appendix:fig_3x4_violin}
    \end{subfigure}
    \begin{subfigure}[b]{.480\linewidth}
    \includegraphics[width=\linewidth]{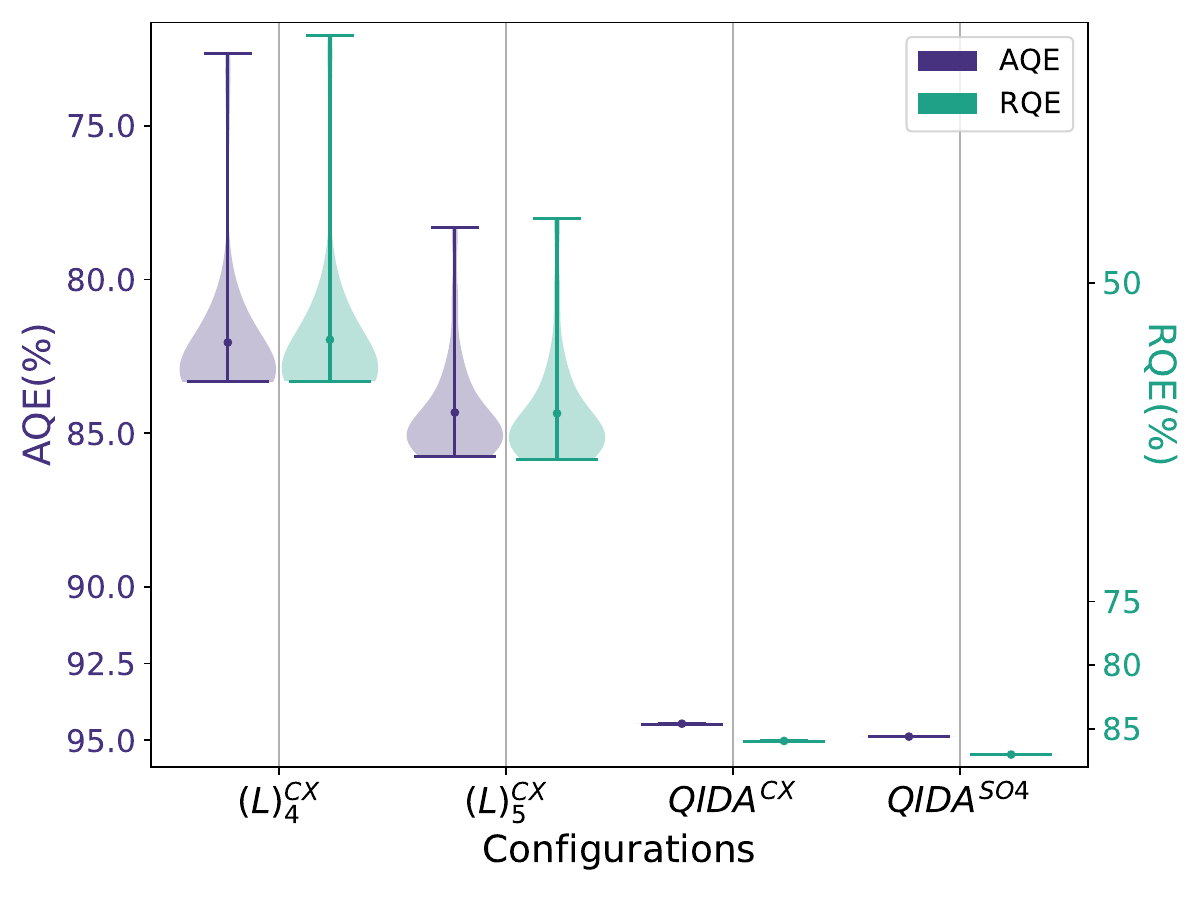}
    \caption{$2\times6$ Isotropic Heisenberg model system results.\\
    HEA: 44,55. Multi-QIDA: 48,54.}\label{appendix:fig_2x6_violin}
    \end{subfigure}\hfill
    \begin{subfigure}[b]{.480\linewidth}
    \includegraphics[width=\linewidth]{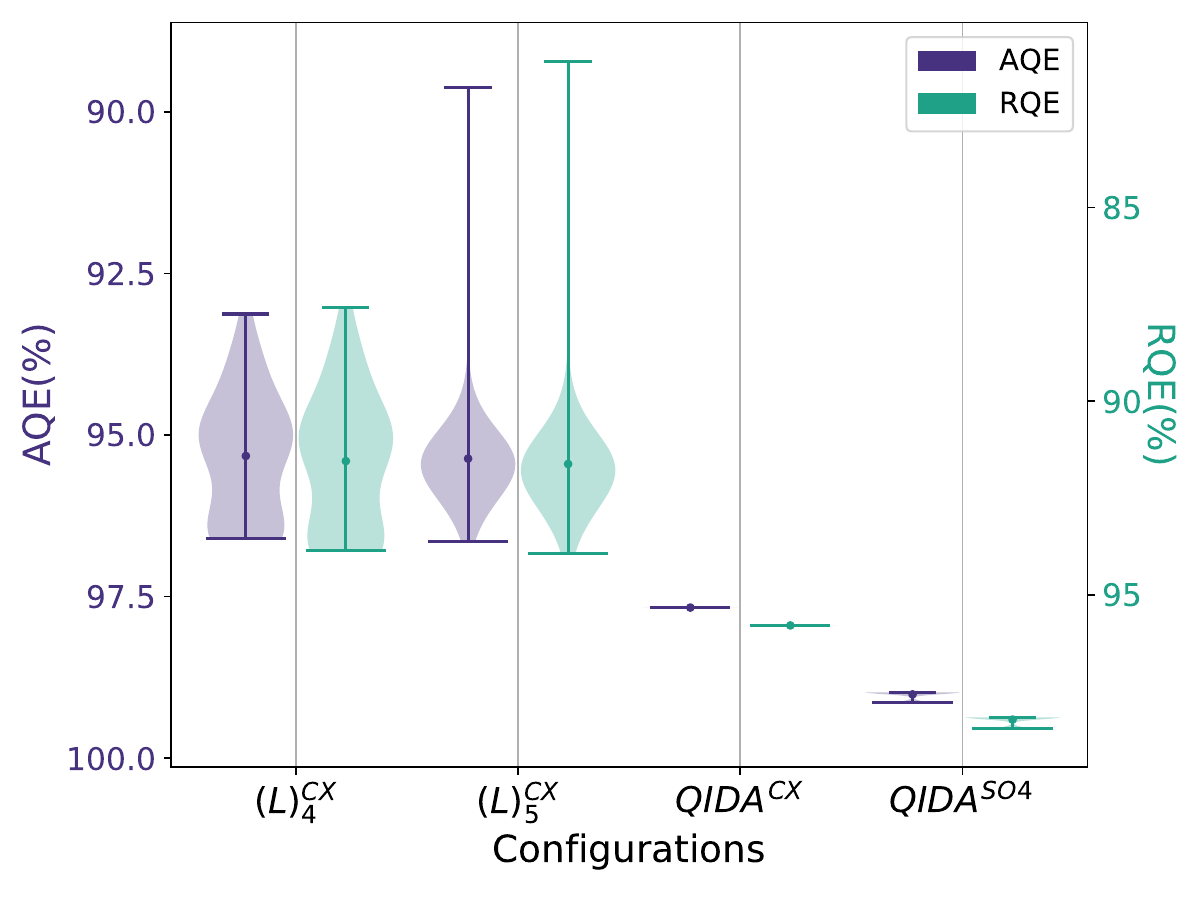}
    \caption{$3\times3$ Isotropic Heisenberg model with External magnetic field $h=2$ results.\\
    HEA: 44,55. Multi-QIDA: 50,54.}\label{appendix:fig_3x4h_violin}
    \end{subfigure}
    \begin{subfigure}[b]{.480\linewidth}
    \includegraphics[width=\linewidth]{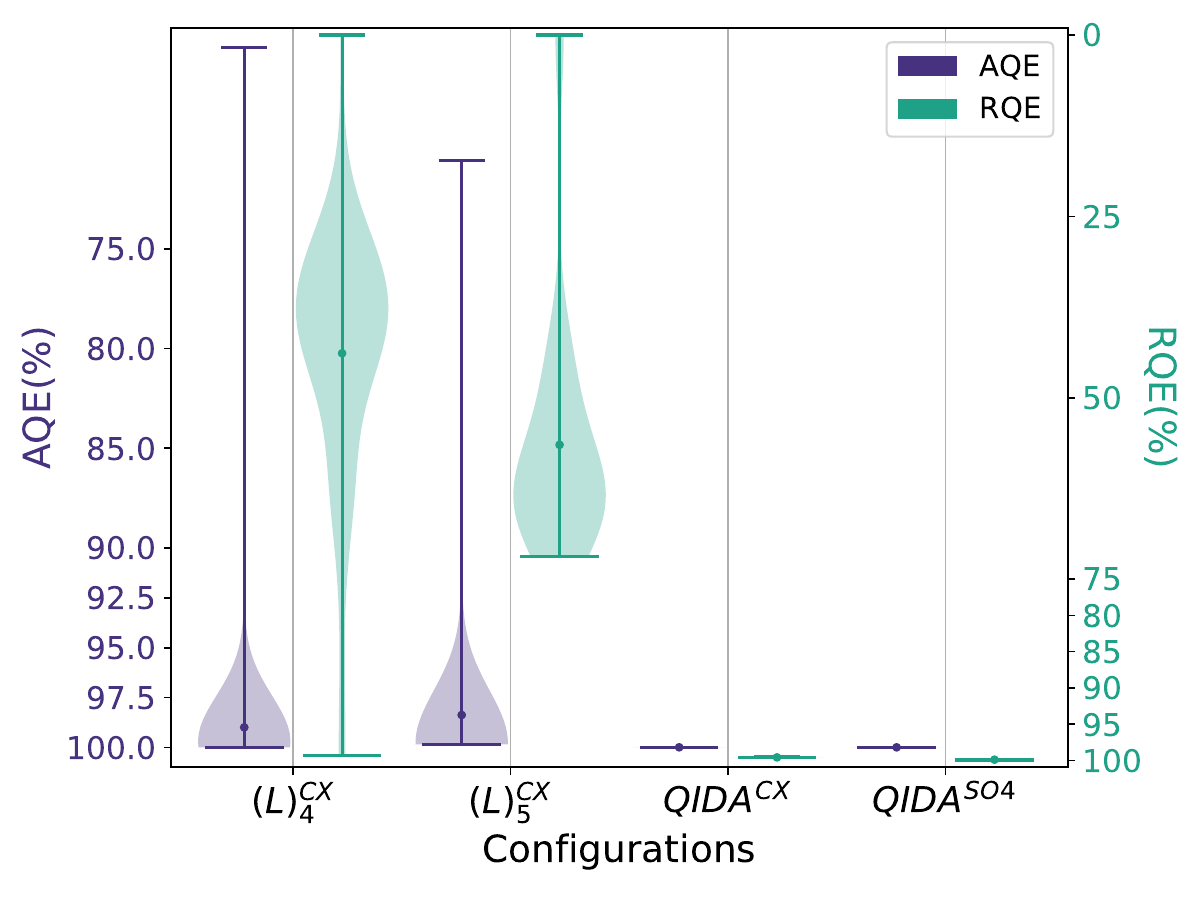}
    \caption{$3\times4$ Anisotropic Heisenberg model with $\Delta=1/10$ results.\\
    HEA: 44,55. Multi-QIDA: 46,54.}\label{appendix:fig_3x4_0_1_violin}
    \end{subfigure}\hfill
    \begin{subfigure}[b]{.480\linewidth}
    \includegraphics[width=\linewidth]{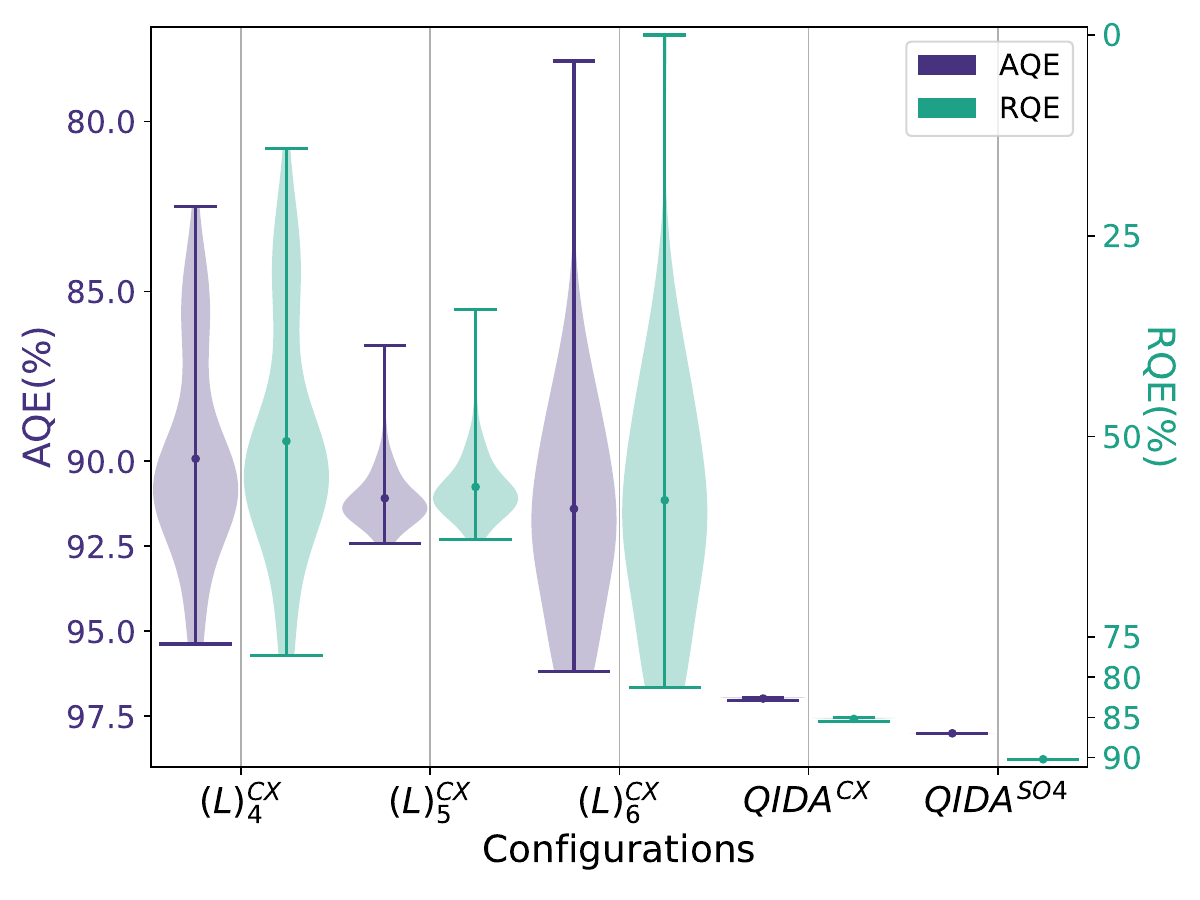}
    \caption{$3\times4$ Anisotropic Heisenberg model with $\Delta=2/3$ results.\\
    HEA: 44,55,66. Multi-QIDA: 52,56.}\label{appendix:fig_3x4_0_66_violin}
    \end{subfigure}
    \caption{Comparison between $AQE$ and $RQE$ for different systems and ansatz configuration. The number of $CNOT$s are showed in the caption for both hardware-efficient ans\"atze (HEA) and for multi-QIDA ans\"atze}
\label{appendix:fig_violins}
\end{figure*}

\begin{table}[!htb]
\scalebox{0.84}{
    \fontsize{10pt}{10pt}
    \setlength\tabcolsep{2pt}
    \centering
    \begin{tabular}{|c||c|c|c|c|c|c|}
    \hline
        $\empty$&$3\times3$&$2\times6$&$3\times4$&\makecell{$3\times4$\\$h=2$}&\makecell{$3\times4$\\$\Delta=2/3$}&\makecell{$3\times4$\\$ \Delta=1/10$}\\
        \hline\hline
        $(L)_{4}^{CX}$&$211$&$393$&$338$&$506$&$426$&$480$\\
        $(L)_{5}^{CX}$&$314$&$456$&$427$&$730$&$623$&$791$\\
        $(L)_{6}^{CX}$&-&-&$538$&-&$666$&-\\\hline
        $(QIDA)^{CX}$&$274$&$799$&$825$&$682$&$1082$&$1578$\\
        $(QIDA)^{SO4}$&$638$&$1011$&$1073$&$1592$&$1161$&$1115$\\
        \hline
    \end{tabular}}
    \caption{Average number of iterations required for convergence.}
    \label{appendix:tab_conv_time}
\end{table}

\begin{figure*}[!htb]
    \centering
    \begin{subfigure}[b]{.450\linewidth}
    \includegraphics[width=\linewidth]{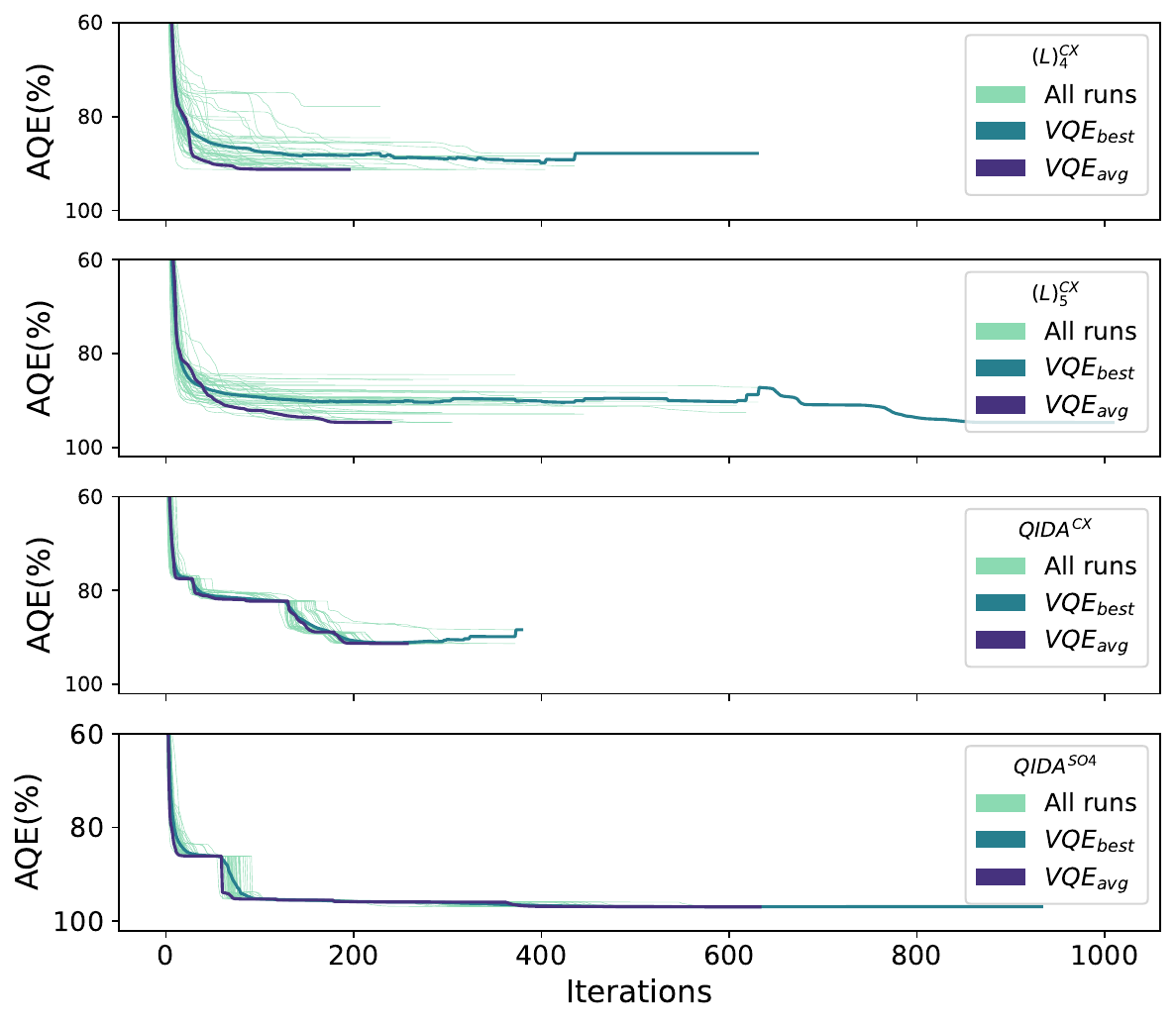}
    \caption{$3\times3$ isotropic Heisenberg Hamiltonian. From the top to the bottom: $(L)_{4}^{CX}$, $(L)_5^{CX}$, $QIDA^{CX}$, and $QIDA^{SO4}$}\label{appendix:fig_3x3_traj}
    \end{subfigure}\hfill
    \begin{subfigure}[b]{.450\linewidth}
    \includegraphics[width=\linewidth]{images/3x4_i_traj_slight_new.pdf}
    \caption{$3\times4$ isotropic Heisenberg Hamiltonian. From the top to the bottom: $(L)_{4}^{CX}$, $(L)_5^{CX}$, $(L)_6^{CX}$, $QIDA^{CX}$, and $QIDA^{SO4}$}\label{appendix:fig_3x4_traj}
    \end{subfigure}
    \begin{subfigure}[b]{.450\linewidth}
    \includegraphics[width=\linewidth]{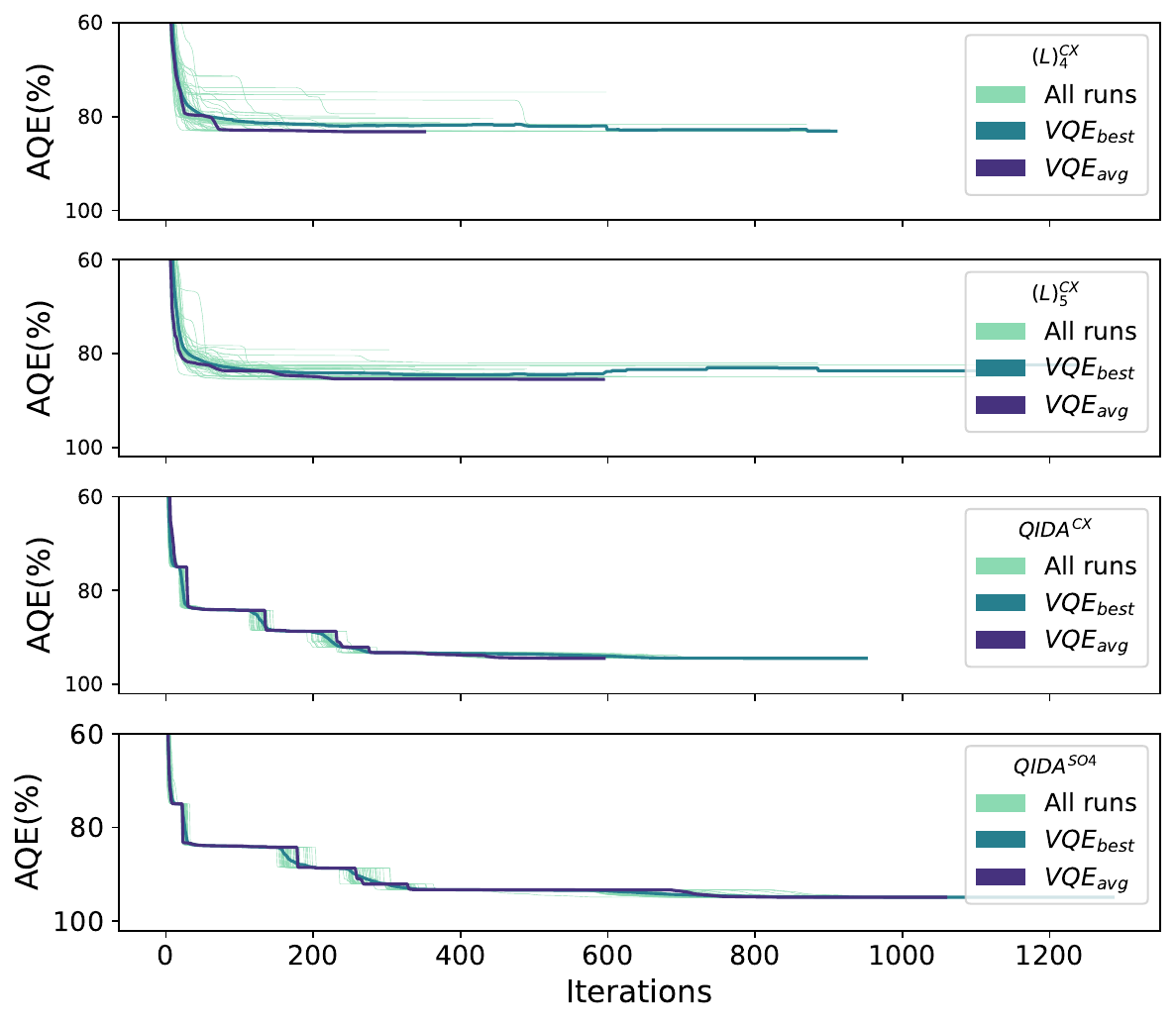}
    \caption{$2\times6$ isotropic Heisenberg Hamiltonian. From the top to the bottom: $(L)_{4}^{CX}$, $(L)_5^{CX}$, $QIDA^{CX}$, and $QIDA^{SO4}$}\label{appendix:fig_2x6_traj}
    \end{subfigure}\hfill
    \begin{subfigure}[b]{.450\linewidth}
    \includegraphics[width=\linewidth]{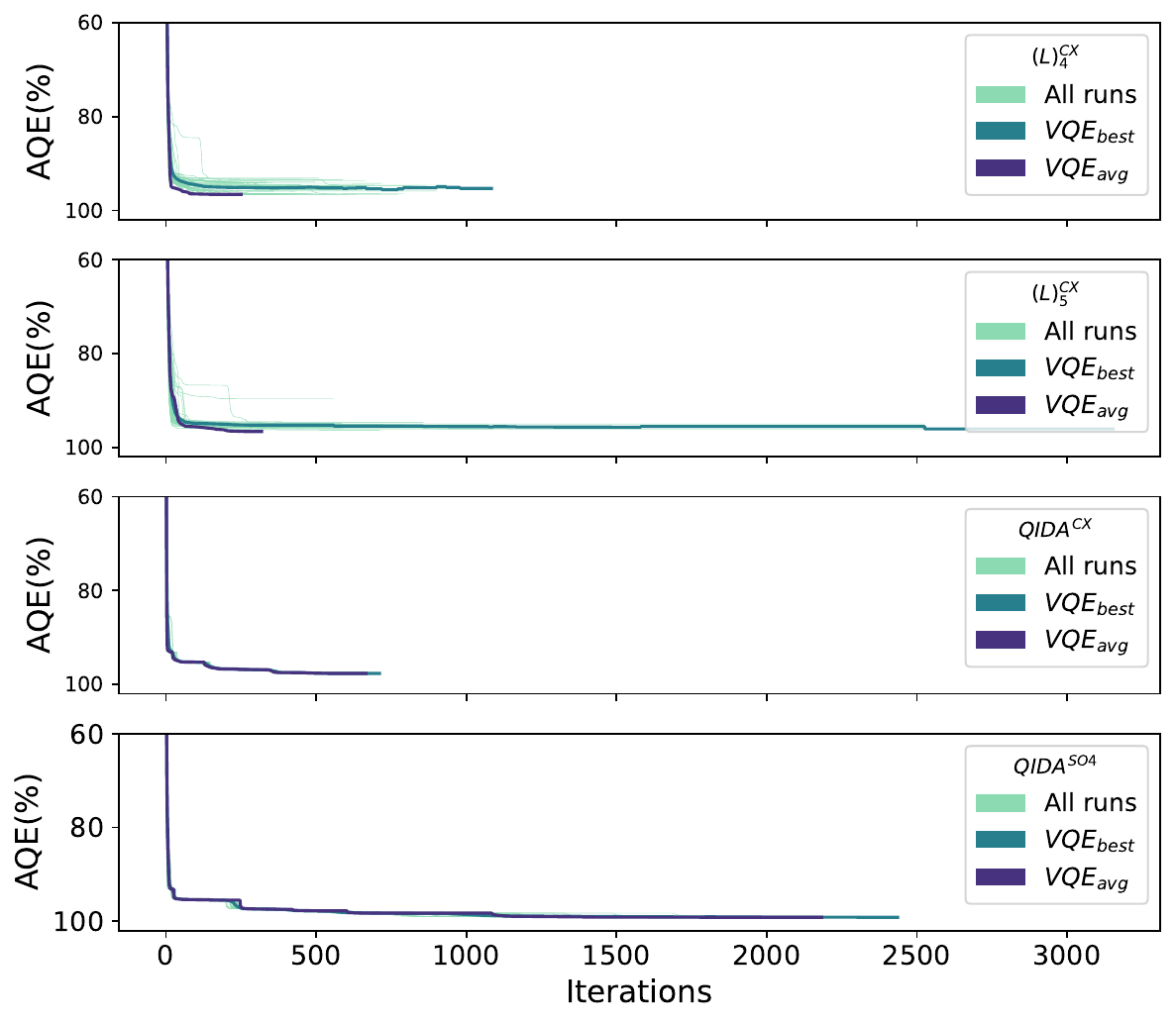}
    \caption{$3\times4$ isotropic Heisenberg Hamiltonian with an external magnetic field $h=2$. From the top to the bottom: $(L)_{4}^{CX}$, $(L)_5^{CX}$, $QIDA^{CX}$, and $QIDA^{SO4}$}\label{appendix:fig_3x4h_traj}
    \end{subfigure}
    \begin{subfigure}[b]{.450\linewidth}
    \includegraphics[width=\linewidth]{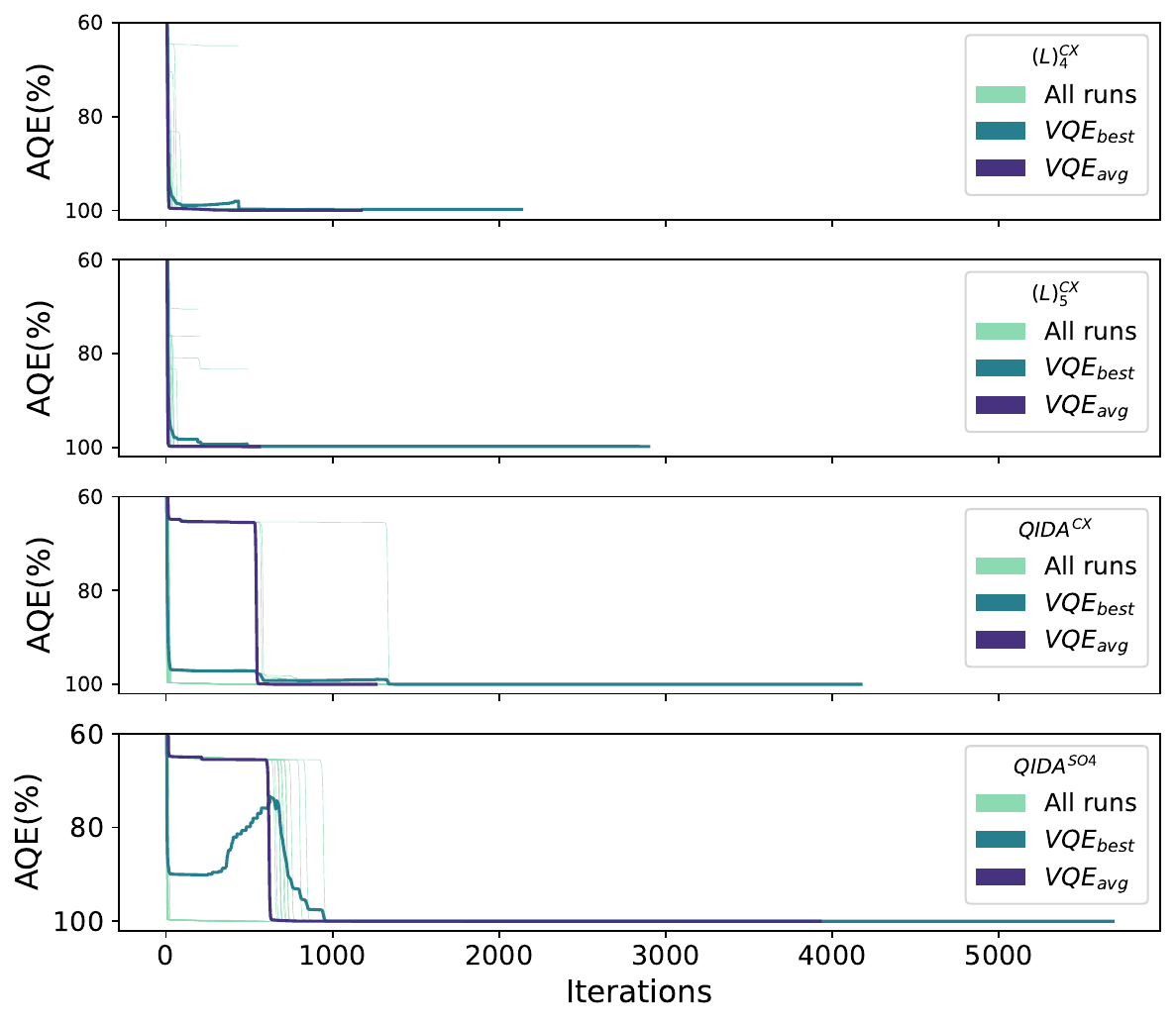}
    \caption{$3\times4$ anisotropic Heisenberg Hamiltonian with $\Delta=1/10$. From the top to the bottom: $(L)_{4}^{CX}$, $(L)_5^{CX}$, $QIDA^{CX}$, and $QIDA^{SO4}$}\label{appendix:fig_3x4_0_1_traj}
    \end{subfigure}\hfill
    \begin{subfigure}[b]{.450\linewidth}
    \includegraphics[width=\linewidth]{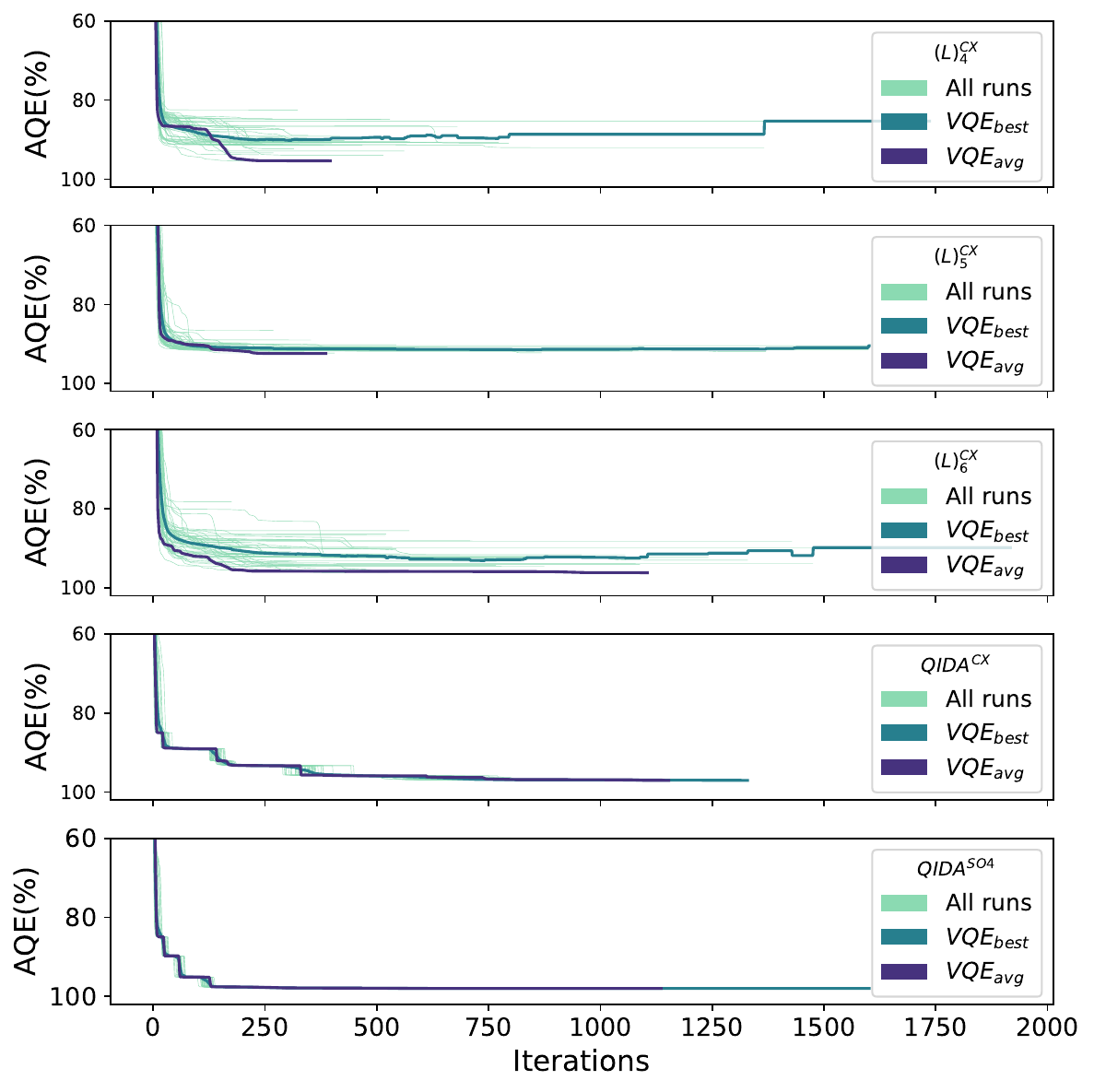}
    \caption{$3\times4$ anisotropic Heisenberg Hamiltonian with $\Delta=2/3$. From the top to the bottom: $(L)_{4}^{CX}$, $(L)_5^{CX}$, $(L)_6^{CX}$, $QIDA^{CX}$, and $QIDA^{SO4}$}\label{appendix:fig_3x4_0_66_traj}
    \end{subfigure}
    \caption{ Comparison between different systems of the optimization trajectories. The number of iterations corresponds to the number of evaluations required for the optimizator to converge.}
\label{appendix:fig_trajectories}
\end{figure*}
\end{appendices}

\end{document}